\documentclass[lineno]{jfm}
\usepackage{graphicx}
\usepackage{epstopdf, epsfig}
\usepackage{psfrag}
\usepackage{mathrsfs}
\usepackage{bm}
\usepackage{natbib}
\usepackage{color,xcolor}
\usepackage{amsmath,graphicx}
\usepackage{verbatim}

\ifCUPmtlplainloaded \else
  \checkfont{eurm10}
  \iffontfound
    \IfFileExists{upmath.sty}
      {\typeout{^^JFound AMS Euler Roman fonts on the system,
                   using the 'upmath' package.^^J}%
       \usepackage{upmath}}
      {\typeout{^^JFound AMS Euler Roman fonts on the system, but you
                   dont seem to have the}%
       \typeout{'upmath' package installed. JFM.cls can take advantage
                 of these fonts,^^Jif you use 'upmath' package.^^J}%
      }
  \else
  \fi
\fi

\ifCUPmtlplainloaded \else
  \checkfont{msam10}
  \iffontfound
    \IfFileExists{amssymb.sty}
      {\typeout{^^JFound AMS Symbol fonts on the system, using the
                'amssymb' package.^^J}%
       \usepackage{amssymb}%
         
         \let\geq=\geqslant
      }{}
  \fi
\fi

\newcommand{\ri}{\mathop{\rm i}\nolimits}
\newcommand{\re}{\mathop{\rm e}\nolimits}

\title[Boundary conditions aiming at eliminating PDI]{Asymptotic-analysis-inspired boundary conditions aiming at eliminating  polymer diffusive instability}
\author{
      Ming Dong\aff{1}\corresp{\email{dongming@imech.ac.cn}}
      \and Dongdong Wan\aff{2}
}

 \affiliation{
   \aff{1}State Key Laboratory of Nonlinear Mechanics, Institute of Mechanics, Chinese Academy of Sciences, Beijing 100190, PR China
   \aff{2}Department of Mechanical Engineering, National University of Singapore, 9 Engineering Drive 1, 117575 Singapore
}

\begin{document}

\maketitle
\begin{abstract}
The recent discovery of polymer diffusive instability (PDI) by Beneitez et al. (Phys. Rev. Fluids, 2023, 8: L101901) poses  challenges in implementing artificial conformation diffusion (ACD) in transition simulations of viscoelastic wall-shear flows. In this paper, we demonstrate  that the unstable PDI is primarily induced by the conformation boundary conditions additionally introduced in the ACD equation system, which could be eliminated if  a {\color{black}new} set of  conformation conditions is adopted. To address this issue,   we begin with  an asymptotic analysis of the PDI within the near-wall thin diffusive layer, which  simplifies the complexity of the instability system by reducing the number of the controlling parameters from five to {\color{black}zero}. Then, based on this simplified model, {\color{black}we construct a stable asymptotic solution that minimises the perturbations in the wall sublayer. From the near-wall behaviour of this solution, we derive a new set of conformation boundary conditions, prescribing a Neumann-type condition for its streamwise stretching component, $c_{11}$, and Dirichlet-type conditions for all the other conformation components.} These  boundary conditions are subsequently  validated within the original ACD instability system, incorporating both the Oldroyd-B and FENE-P constitutive models. Finally, we perform direct numerical simulations based on the traditional and the new conformation conditions, demonstrating the effectiveness of the latter in eliminating the unstable PDI. {\color{black}Importantly, this improvement does not affect the calculations of other types of instabilities.}
Therefore, this work offers a promising approach for achieving reliable polymer-flow simulations with ACD, ensuring both numerical stability and accuracy.
\end{abstract}

\section{Introduction}
Viscoelastic shear flows exhibit distinct features compared to Newtonian flows, primarily due to the additional viscoelastic stresses arising from the stretching and relaxation of polymers, even in dilute solutions \citep{Sanchez2022}. In addition to the Newtonian turbulence that is dominated by large inertial forces, two viscoelastic turbulent states emerge: elastic turbulence (ET), which occurs at vanishingly small inertia \citep{Groisman2000,Steinberg2021}, and elasto-inertial turbulence (EIT), for which the effects of inertia and elasticity are comparable \citep{Samanta2013,Dubief2023}.
The origins of these two turbulent states are of particular interest, and studies aimed at identifying their transition routes are still ongoing. For a comprehensive review of this topic, one can refer to \cite{Couchman2024}.

{\color{black}In viscoelastic flows, experimental techniques do not allow for direct measurements of the polymer conformation tensor. Numerical simulations can compensate for this limitation, making them an efficient tool for investigating the transition routes to elastic turbulence (ET) or elasto-inertial turbulence (EIT).}  However, as highlighted by \cite{Alves2021}, the lack of diffusion in the conformation tensor equations poses significant numerical difficulties in simulating viscoelastic fluid flows, especially at high Weissenberg  numbers.  To achieve numerical stability, adding an artificial conformation diffusion (ACD) term to the conformation equation appears as an efficient approach. {\color{black} This term is expressed as $(ReSc)^{-1}\nabla^2\bf c$, where $Re$ denotes  the Reynolds number, $Sc$ is the Schmidt number (defined by the ratio of the solvent viscosity to diffusivity of polymers in solution), and $\bf c$ represents the polymer conformation tensor.}
Although  there is always a minimal degree of molecular diffusion of the polymer stress in reality, {\color{black}quantified by $Sc^{-1}\sim 10^{-6}$, this level is too small to stabilise the computations \citep{Dubief2023}}. Using a spectral method, {\color{black}\cite{Sureshkumar1995} demonstrated that numerical stability requires a diffusion coefficient of $(ReSc)^{-1}\sim 10^{-3}$, or in their case, $Sc\approx 0.2$. This study provided justification for the subsequent use of $Sc < 1$, although other simulations have employed different $Sc$ values \citep{Dubief2023}.}

Although the ACD has faced criticism, it remains a widely employed approach in polymer-flow simulations due to its simplicity and robustness. However, the recent identification of a new type of instability induced by ACD, referred to as polymer diffusive instability (PDI), complicates this situation.
In viscoelastic plane Couette flows, \cite{Beneitez2023} first reported the emergence of an unstable wall-mode PDI that arises at vanishing inertia and low diffusivity of the polymer stress. This instability was found to occur even at extremely small, non-zero diffusion levels, and was claimed to be robust against various boundary conditions applied to the polymer conformation equation. Direct numerical simulations indicated that PDI could trigger a self-sustaining chaotic state, offering a potential mechanism  for the origin of ET in numerical simulations with ACD. Subsequently, \cite{Couchman2024} conducted a systematic study of PDI across a wide parameter space, employing both Oldroyd-B and FENE-P constitutive models for plane Couette flows and plane Poiseuille flows. Their investigations also considered configurations with finite Reynolds numbers, suggesting that PDI may also lead to EIT in addition to ET in polymer fluids. Furthermore, through two-dimensional numerical simulations, \cite{Beneitez2024} established a transition route to ET and EIT based on the secondary instability of PDI. In recent studies, \cite{Lewy2024} particularly focused on the PDI in  highly concentrated polymeric solutions, {\color{black}and \cite{Beneitez2025} confirmed the emergence of PDI even when explicit ACD is not present.}
 Through asymptotic analysis by taking both the diffusivity and the solvent-to-total-viscosity ratio to be small, the instability system was significantly simplified, reducing the 10$^{th}$-order ordinary differential equation system to a system of two algebraic equations. This simplification enables an in-depth exploration of the PDI mechanism.

Although the nature of PDI -- whether it is physical or artificial -- remains a topic of debate, its discovery poses challenges to previous transition simulations that employed ACD. The PDI is fundamentally  different from numerical instability, as it persists even when the artificial diffusivity is {\color{black}very small}. One may expect that suppressing or avoiding the unstable PDI is beneficial in simulations of polymer fluids incorporating ACD.
It is important to emphasise that the inclusion of polymer diffusion increases the order of the   differential system, requiring additional boundary conditions for the polymer conformation at the wall.
Considering three representative sets of boundary conditions {\color{black}(the detailed forms of which are presented in $\S$\ref{sec:instability})}, \cite{Beneitez2023} demonstrated the presence of unstable zones of PDI for all examined boundary conditions. However, variations in the neutral curves were observed depending on the choice of the conformation boundary conditions. \cite{Lewy2024} also reported that replacing the conformation boundary conditions from the vanishing diffusion to the Neumann conditions leads to a suppression of PDI in the high-concentration limit.
These observations raise a pertinent question: is there a boundary condition under which the unstable zone of PDI could be eliminated? If such a boundary condition exists, it could potentially offer a valid approach for simulating the transition process of polymer fluids with the introduction of ACD.
Addressing this question is inherently challenging, as it necessitates exploring the elimination of PDI across the entire parameter space. Nevertheless, since the generic instability mechanism can be captured through its asymptotic structure (as has been done in \cite{Lewy2024}), we can develop the required conformation boundary conditions  inspired by the asymptotic analysis. In particular, the perturbation behaviours observed in the near-wall region can offer valuable insights for selecting the boundary conditions. This is the primary focus of this paper.

The rest of this paper is organised as follows. In $\S$\ref{sec:physical_model}, we present the mathematical framework  of the linear viscoelastic instability in parallel wall-shear flows based on Oldroyd-B constitutive model, along with conformation diffusion. This is followed by an exploration of its asymptotic structure in  $\S$\ref{sec:asymptotics}, where we identify {\color{black}four} regimes with reducing the  complexity from regime I to regime {\color{black}IV}. Based on the simplified equation system in regime {\color{black}IV}, we propose a set of {\color{black}new} boundary conditions to eliminate PDI in $\S$\ref{sec:proper_III}, and assess its applicability in the other asymptotic regimes  in $\S$\ref{sec:proper_II} and the original system  in $\S$\ref{sec:proper_original}. In $\S$\ref{sec:proper_FENE_P}, we further validate the applicability of the new boundary conditions within the FENE-P constitutive model. Then, the new boundary conditions are integrated in direct numerical simulation (DNS) code, showing the effectiveness in eliminating PDI for a representative configuration in $\S$\ref{sec:DNS}. Finally, we conclude with remarks in $\S$\ref{sec:conclusion}.

\section{Physical model and mathematical descriptions}
\label{sec:physical_model}
\subsection{Formulation}
Following  \cite{Couchman2024}, we consider two types of parallel wall-shear flows of polymer solutions: a plane Poiseuille flow (PPF) driven by a pressure gradient and a plane Couette flow (PCF) driven by two counter-moving parallel plates. By selecting the half-channel width $H$ as the unit length and the maximum velocity $U_{max}$ (the centreline velocity for PPF and the wall velocity for PCF) as the reference velocity, we define two dimensionless parameters, the Reynolds number $Re$ and the Weissenberg number $Wi$:
\begin{equation}
Re=\frac{\rho U_{max}H}{\mu},\quad Wi=\frac{\bar\lambda U_{max}}{H},
\end{equation}
where $\rho$ denotes the density, and $\mu$ and $\bar\lambda$ denote the total viscosity and the relaxation time of the polymer molecules (to their equilibrium states), respectively. Here, the total viscosity is the sum of the solvent viscosity $\mu_s$ and the polymer viscosity $\mu_p$ \citep{Bird1987}, i.e., $\mu=\mu_s+\mu_p$, and a viscosity ratio $\beta$ is defined as
\begin{equation}
\beta=\mu_s/\mu\in[0,1].
\end{equation}
The dimensionless Cartesian coordinate system $(x,y,z)$ is employed, with $x$, $y$ and $z$ denoting the streamwise, wall-normal and spanwise directions, respectively. The velocity field $\bm u=(u,v,w)$ is normalised by $U_{max}$, the time $t$ is normalised by $H/U_{max}$, and the pressure is normalised by $\rho U_{max}^2/Re$. The conformation tensor $\bm c$ and stress ${\bm \tau}$ are normalised by $k_BT/H_0$ and $\mu_p U_{max}/H$, respectively, where $k_B$ is the Boltzmann constant, $T$ is temperature and $H_0$ is the spring constant in the elastic dumbbell model of the polymer.

The dimensionless  equations governings the viscoelastic fluid with artificial conformation diffusion  read \citep{Couchman2024}
\refstepcounter{equation}
$$
Re\Big(\frac{\partial \bm u}{\partial t}+(\bm u\cdot\nabla)\bm u\Big)=-\nabla p+\beta\nabla^2\bm u+(1-\beta)\nabla\cdot \bm \tau,\quad \nabla\cdot \bm u=0,\eqno{(\theequation a,b)}
$$
$$
\frac{\partial \bm c}{\partial t}+(\bm u\cdot \nabla)\bm c+\tau-\bm c\cdot \nabla\bm u-(\nabla \bm u)^T\cdot \bm c=\epsilon \nabla^2\bm c,
\label{eq:GE}\eqno{(\theequation c)}
$$
where $\epsilon$ denote the artificial diffusivity introduced in the polymer stress equation. It is related to the Schmidt number through $Sc=(\epsilon Re)^{-1} $. Notably, when  $Re=0$, {\color{black}one can avoid introducing the Schmidt number and directly use $\epsilon$ to denote the conformation diffusivity.} For representative demonstration, the following asymptotic analysis will be based on the Oldroyd-B constitutive model, for which $\bm \tau$ and $\bm c$ are related by
\begin{equation}
\bm \tau=\frac{\bm c-\bm I}{Wi},
\label{eq:OB}
\end{equation}
with $\bm I$ denoting the unity matrix. We would like to emphasise that the  boundary conditions to be developed in this paper will also be applied to the widely-used FENE-P constitutive model, because the asymptotic structures for both constitutive models are the same. This will be confirmed in $\S$\ref{sec:proper_FENE_P}.

\subsection{Linear stability theory (LST)}
\label{sec:instability}
To analyse the linear instability, we decompose the flow field $\bm\phi=(\bm u,p,\bm c)$ into a steady base flow $\bm\Phi=(\bm U,P,\bm C)$ and a harmonic perturbation $\hat{\bm\phi}=(\hat{\bm u},\hat p,\hat{\bm c})$,
\begin{equation}
\bm\phi=\bm\Phi(y)+\hat \epsilon\hat{\bm\phi}(y)\re^{\ri(kx+\gamma z-\omega t)}+c.c.,
\label{eq:decomposition}
\end{equation}
where $\hat \epsilon\ll 1$ denotes the perturbation amplitude, $k$ and $\gamma$ denote the streamwise and spanwise wavenumbers, $\omega$ is the complex frequency with its imaginary part denoting the growth rate, and $c.c.$ denotes the complex conjugate. In this paper, we focus on two-dimensional perturbations, for which $\gamma=0$. For parallel wall-shear flows, $\bm U=(U(y),0,0)$ with $y\in[-1,1]$, and
the base flow profile is
\begin{equation}
U=\left\{\begin{array}{ll}
           1-y^2& \mbox{for a PPF}; \\
           y&\mbox{for a PCF}.
         \end{array}
\right.
\label{eq:base_velocity}
\end{equation}
{\color{black}For convenience, we introduce the wall shear $\lambda:=U_y(0)$, with $\lambda=2$ and 1 for PPF and  PCF, respectively.}
 The non-zero elements of $\bm C$ for the Oldroyd-B constitutive model are
\begin{equation}
C_{11}=1+2Wi^2U'^2,\quad C_{12}=C_{21}=WiU',\quad C_{22}=C_{33}=1,
\label{eq:base_conformation}
\end{equation}
where,  throughout this paper, a prime denotes the derivative with respect to its argument.

Substituting (\ref{eq:decomposition}) into the governing equation system (\ref{eq:GE}) and (\ref{eq:OB}), and retaining the $O(\hat\epsilon)$ terms, we obtain the following linear system:
\refstepcounter{equation}
$$ \ri k Re(U-c)\hat u+ReU'\hat v+\ri k\hat p={\beta}{}(\hat u''-k^2\hat u)+{1-\beta}(\ri k\hat \tau_{11}+\hat \tau_{12}'),
\eqno{(\theequation a)}
$$
$$
\ri k Re(U-c)\hat v+\hat p'={\beta}{}(\hat v''-k^2\hat v)+{1-\beta}{}(\ri k\hat \tau_{12}+\hat \tau_{22}'),\quad \ri k \hat u+\hat v'=0,
\eqno{(\theequation b,c)}
$$
$$
[\ri k (U-c)+\epsilon k^2]\hat c_{11}+\hat\tau_{11}+C_{11}'\hat v-2(\ri k C_{11}\hat u+C_{12}\hat u'+U'\hat c_{12})=\epsilon \hat c_{11}'',\eqno{(\theequation d)}
$$
$$
[\ri k (U-c)+\epsilon k^2]\hat c_{12}+\hat\tau_{12}+C_{12}'\hat v-(\ri k C_{12}\hat v+C_{22}\hat u'+U'\hat c_{22})=\epsilon \hat c_{12}'',
\eqno{(\theequation e)}
$$
$$
[\ri k (U-c)+\epsilon k^2]\hat c_{22}+\hat\tau_{22}-2(\ri k C_{12}\hat v+C_{22}\hat v')=\epsilon \hat c_{22}'',\eqno{(\theequation f)}
\label{eq:instability}
$$
$$
[\ri k (U-c)+\epsilon k^2]\hat c_{33}+\hat\tau_{33}=\epsilon \hat c_{33}'',\eqno{(\theequation g)}
$$
where  $c=\omega/k=c_r+\ri c_i$ and the Oldroyd-B constitutive model determines
\refstepcounter{equation}
$$
(\hat\tau_{11},\hat\tau_{12},\hat\tau_{22},\hat \tau_{33})=Wi^{-1}(\hat c_{11},\hat c_{12},\hat c_{22},\hat c_{33}).\eqno{(\theequation)}\label{eq:instability_OB}
$$
 Note that because $\hat c_{33}$ is decoupled from the Oldroyd-B instability system, we do not need to consider its impact in the following, and direct introduce the boundary condition $\hat c_{33}(\pm 1)=0$.

At both walls $y=\pm 1$, the no-slip, non-penetration conditions are applied,
\refstepcounter{equation}
$$
\hat u(\pm 1)=\hat v(\pm1)=0.\quad
\label{eq:BCuv}
\eqno{(\theequation)}
$$
Following \cite{Beneitez2023}, we consider three representative types of boundary conditions for the conformation tensor.
For the type-(i) boundary conditions (referred to as BC1), the diffusive  effect is neglected at the wall, leading to
\refstepcounter{equation}
$$
\Big[({-\ri k c+Wi^{-1}})\hat c_{11}-{2(C_{12}\hat u'+U'\hat c_{12})}\Big]_{y=\pm1}=0,
\label{eq:BC1}\eqno{(\theequation a)}
$$
$$
 \Big[({-\ri k c+Wi^{-1}})\hat c_{12}-{\hat u'}\Big]_{y=\pm1}=0,\quad\hat c_{22}(\pm 1)=0.
\eqno{(\theequation b, c)}
$$
For the type-(ii) boundary conditions (referred to as BC2), only the terms $\epsilon\partial_{yy}{\textbf C}_{ij}$ are set to be zero at the wall,
\refstepcounter{equation}
$$
\Big[({-\ri k c+Wi^{-1}}+k^2)\hat c_{11}-{2(C_{12}\hat u'+U'\hat c_{12})}\Big]_{y=\pm1}=0,
\label{eq:BC2}\eqno{(\theequation a)}
$$
$$
 \Big[({-\ri k c+Wi^{-1}}+k^2)\hat c_{12}-{\hat u'}\Big]_{y=\pm1}=0,\quad\hat c_{22}(\pm 1)=0.
\eqno{(\theequation b, c)}
$$
For the type-(iii) boundary conditions (referred to as BC3), the conformation tensor is assumed to show zero gradient at the wall ($\partial_y{\textbf C}_{ij}|_{y=\pm1}=0$):
\refstepcounter{equation}
$$
\hat c_{11}'(\pm 1)=\hat c_{12}'(\pm 1)=\hat c_{22}'(\pm 1)=0.
\label{eq:BC3}\eqno{(\theequation a,b, c)}
$$

The equation system (\ref{eq:instability}) and (\ref{eq:instability_OB}), together  with the boundary conditions (\ref{eq:BCuv}) and one of (\ref{eq:BC1}) to (\ref{eq:BC3}), constitutes an eigenvalue problem, where $c$ serves  as the eigenvalue. The controlling parameters for this system include $Wi$, $Re$, $k$, $\beta$ and $\epsilon$. In this paper, the eigenvalue system is solved using the following two numerical approaches:
\begin{itemize}
  \item Spectral Collocation (SC) method: This method is based on Chebyshev polynomials and is a global method that excels in calculating the eigenvalues across a wide spectrum. A detailed description can be found in \cite{Zhang2021}.
  \item Newton Iterative (NI) Method: The equations are discretized using the fourth-order compact Malik scheme \citep{Malik1990}, and the eigenvalue solution is obtained by Newton iterative method to minimise the determinant of the coefficient matrix to a sufficiently small value (e.g. $10^{-9}$). This local method focuses on a single eigenvalue based on an initial guess and is well-known for its high accuracy. Our previous eigenvalue calculations in viscoelastic pipe flows \citep{Dong2022} and Newtonian boundary-layer flows \citep{Ji2023,Song2023} employed this method.
   \end{itemize}

In this paper, unless otherwise specified, most results are obtained using the NI method, with initial values for iteration derived either from the SC method or from a continuation of existing eigenvalue solutions.

\section{Asymptotic analyses of PDI}
\label{sec:asymptotics}
\subsection{Guideline of the asymptotic analyses}
As a guideline for the subsequent asymptotic analyses,  this subsection outlines the scaling relations for the various asymptotic regimes along with their corresponding structures, while detailed mathematical derivations are provided in the following sections.
Based on the linear system (\ref{eq:instability}), \cite{Couchman2024} computed the neutral curves for PDI in both PPF and PCF, revealing the emergence of the thin diffusive layer in the $O(\sqrt{\epsilon})$ vicinity of the wall. In this layer, the  conformation diffusivity becomes the dominant influence. At the critical state where PDI emerges, a clear scaling relation of $k\sim \epsilon^{-1/2}$  is observed, with the critical Weissenberg number reaching several tens.
Notably, the lower-branch neutral curve of PDI  manifests even when $Wi$ approaches infinity, with the wavenumber decreases accordingly.
{\color{black}In light of these findings, we conduct an asymptotic analysis under the assumption that $\epsilon\ll1$  while taking $Wi=O(1)$,  referred to as regime I. As will be shown in $\S$\ref{sec:Numerical_regime_I}, the most unstable PDI in regime I appears in the limit $Wi\to \infty$ and $k\to 0$, with the scaling $\bar k=k\epsilon^{1/2}\sim Wi^{-3/2}$. We then derive the instability properties in these limits,  referred to as regime II. Furthermore, to simplify the asymptotic system, we consider the dilute limit -- similar to the centre mode analysis in \cite{Dong2022}. In this limit, the rescaled wavenumber, $\tilde k=k\epsilon^{1/2}Wi^{3/2}$, scales as $(1-\beta)^{-3/2}$, so that the resulting asymptotic system is characterized by a single controlling parameter, $\breve \sigma=\tilde k^{2/3}(1-\beta)$. This regime is referred to as regime III. Finally, we show that if the limit $\breve \sigma\to \infty$ is considered,  unstable PDI is still maintained (termed as regime IV); however, this asymptotic state exhibits a parameter-free character, which facilitates the derivation of analytical solutions.} In summary, this paper considers the following four regimes:
\begin{itemize}
        \item Regime I: $Wi=O(1)$, $1-\beta=O(1)$, $k\sim\epsilon^{-1/2}$ and $c-U(-1)\sim \epsilon^{1/2}$;
        \item Regime II: $Wi\gg 1$, $1-\beta=O(1)$, $k\epsilon^{1/2}\sim Wi^{-3/2}\ll 1$ and $c-U(-1)\sim\epsilon^{1/2}Wi^{1/2}$;
        \item Regime III: $Wi\gg 1$, $\beta\to 1$, $k\epsilon^{1/2}Wi^{3/2}\sim (1-\beta)^{-3/2}\gg 1$ and $c-U(-1)\sim \epsilon^{1/2}Wi^{1/2}\tilde k^\alpha$, where $\alpha=-1$ for BC1/BC2 and $=-1/3$ for BC3;
        \item {\color{black}Regime IV: $Wi\gg 1$, $\beta\to 1$, $k^{2/3}\epsilon^{1/3}Wi(1-\beta)\gg 1$, and $c-U(-1)\sim \epsilon^{1/2}Wi^{1/2}\tilde k^{-1}\breve \sigma^{1/2}$ for BC1/BC2 and $\sim\epsilon^{1/2}Wi^{1/2}\tilde k^{-1/3}\breve\sigma^{-1/2}$ for BC3.}
\end{itemize}

\begin{figure}
\begin{center}
  \includegraphics[width = 0.96\textwidth]{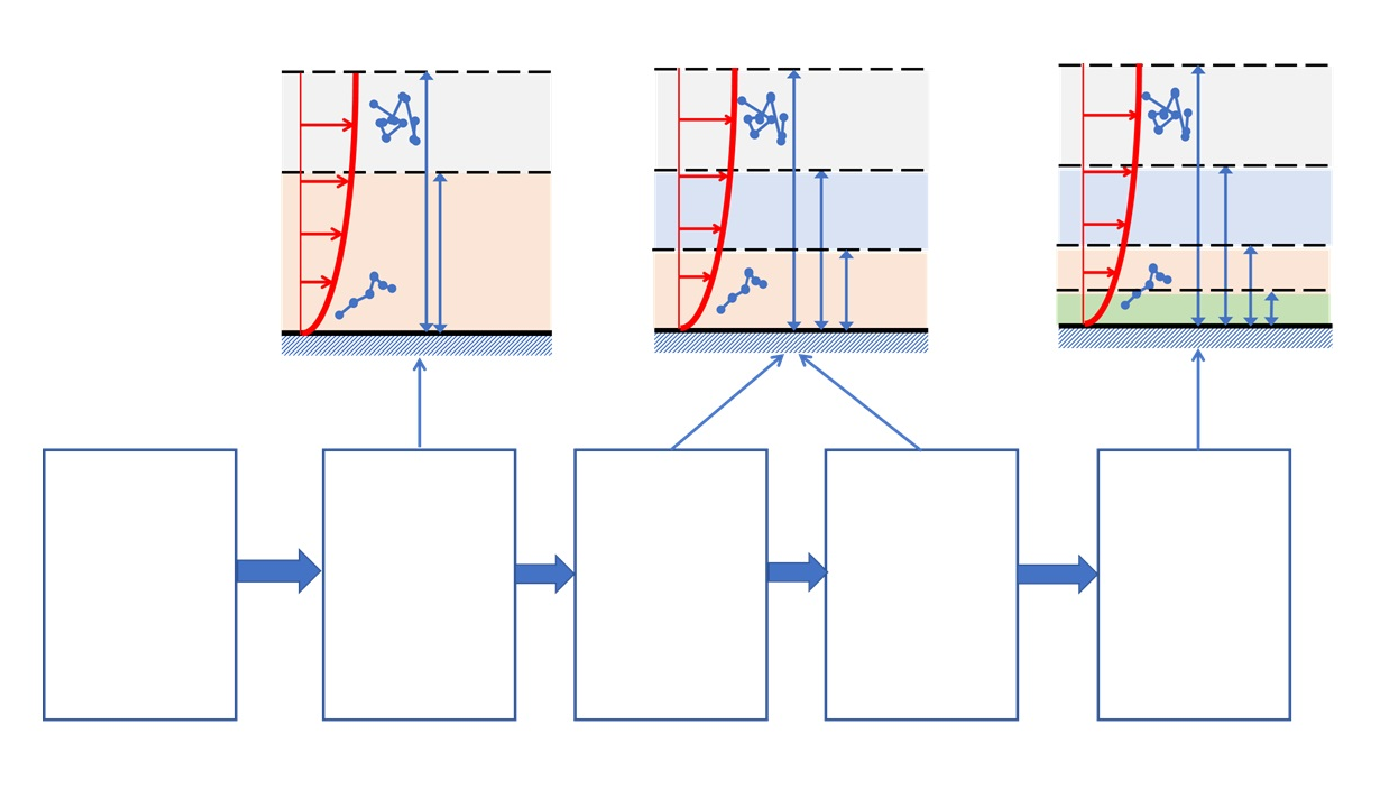}
      \put(-250,140){{\large (ii)}}\put(-250,175){{\large (i)}}
         \put(-307,195){{\large PPF profile}}
         \put(-207,195){{\large PPF profile}}
          \put(-102,195){{\large PPF profile}}
          \put(-150,150){{\large (ii-1)}}\put(-153,175){{\large (i)}}
          \put(-142,130){{\large (ii-2)}}
      \put(-208,180){\large(b)}  \put(-308,180){\large(a)} \put(-102,180){\large(c)}
      \put(-42,150){{\large (ii-1)}}\put(-46,175){{\large (i)}}
      \put(-35,135){{\large (iii)}}\put(-29,124){{\large (iv)}}
 \put(-349,78){\bf System } \put(-345,68){\bf (\ref{eq:instability}): }
  \put(-355,50){$(Re,Wi,\beta$}
    \put(-350,40){$k)=O(1)$}
    \put(-346,25){$\epsilon$ small}
  \put(-278,78){\bf Regime I }
     \put(-278,65){$Wi=O(1)$}
   \put(-272,45){$\epsilon\ll 1$}
      \put(-280,55){$1\!-\!\beta\!=\!O(1)$}
\put(-280,32){$k\!\sim\!\epsilon^{-\frac12}\!\gg\! 1$}
\put(-276,20){$Re\ll \epsilon^{-1}$}
 \put(-214,78){\bf Regime II}
   \put(-207,58){$Wi\gg1$}
   \put(-213,45){$1\!-\!\beta\!=\!O(1)$}
   \put(-214,30){$k\epsilon^{\frac12}\!\sim\! Wi^{-\frac32}$}
    \put(-149,78){\bf Regime III}
    \put(-142,58){$Wi\gg1$}
    \put(-140,45){$\beta\to 1$}
        \put(-147,30){$k\epsilon^{1/2}Wi^{3/2}$}
           \put(-149,22){$\sim(1-\beta)^{-\frac32}$}
               \put(-75,78){\bf Regime IV}
               \put(-60,58){$\breve\sigma=$}
               \put(-70,47){$k^{2/3}\epsilon^{1/3}\times$}
               \put(-75,36){$Wi(1-\beta)$}
               \put(-60,25){$\gg1$}
      \put(-358,9){5 controlling}
      \put(-356,0){parameters}
      \put(-282,9){3 controlling}
      \put(-280,0){parameters}
       \put(-215,9){2 controlling}
      \put(-213,0){parameters}
    \put(-150,9){1 controlling}
      \put(-148,0){parameter}
          \put(-79,9){No controlling}
      \put(-75,0){parameter}
  \caption{{\color{black}A summary of the reduction process from the instability system (\ref{eq:instability}) to regime IV. Inbox panels (a), (b) and (c) illustrate the asymptotic structures for  regimes I, II/III and IV, respectively. Layer (i) denotes the main layer; layer (ii) denotes the diffusive layer; layers (ii-1) and (ii-2) denote the upper and lower diffusive layer, respectively; layers (iii) and (iv) denote the bulk sublayer and the wall sublayer, respectively.} The elongated chains represent the polymer states in the high and low shear rates. }\label{fig:sketch}
\end{center}
\end{figure}
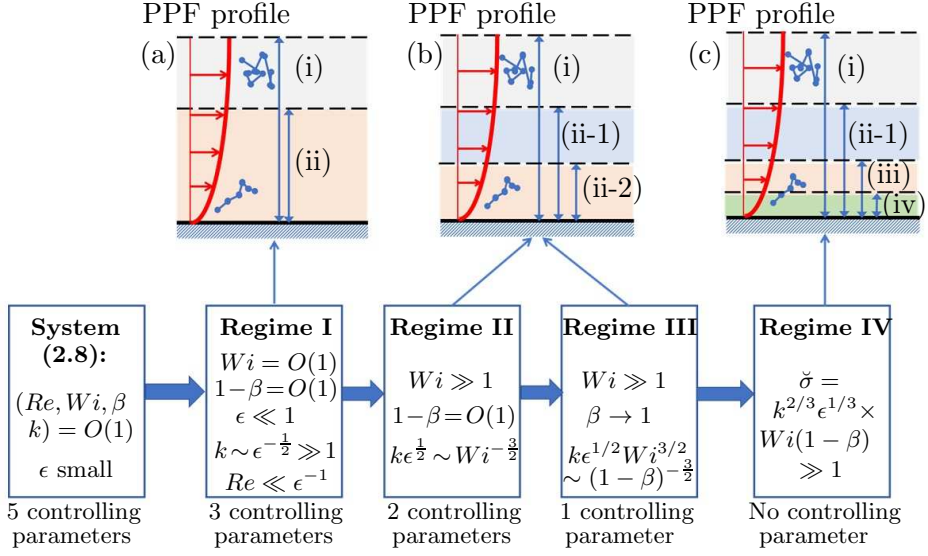

The asymptotic structures of these regimes are depicted in figure \ref{fig:sketch}.
In regime I, the instability primarily manifests within the diffusive layer (layer (ii)) in the $O(\sqrt{\epsilon})$ vicinity of the wall, while in regimes II and III, the diffusive layer  splits into two distinct layers: an upper diffusive layer (UDL, layer (ii-1)) and a lower diffusive layer (LDL, layer (ii-2)) {\color{black}(with the mathematical details illustrated in $\S$\ref{sec:regime_II})}.
 Remarkably, the thicknesses of the UDL and LDL are of different magnitudes for regimes II and III. For the former,  the thicknesses of the UDL and LDL are $O(Wi^{3/2}\epsilon^{-1/2})$ and $O(Wi^{1/2}\epsilon^{-1/2})$, respectively. In contrast, for the low-concentration regime III,  the rescaled wavenumber $\tilde k=k\epsilon^{1/2}Wi^{3/2}$   must be large, resulting in the modified thicknesses of $O(\tilde k^{-1}Wi^{3/2}\epsilon^{-1/2})$ and $O(\tilde k^{-1/3}Wi^{1/2}\epsilon^{-1/2})$, respectively.
Specifically, for regime III, the magnitude of the phase speed correction, $c-U(-1)$, may vary depending on the specific conformation boundary conditions imposed.
{\color{black}Regime IV is actually the asymptotic state of regime III by taking $\breve \sigma$ to be asymptotically large. In this limit, the LDL further splits into two sublayers: a bulk sublayer (layer (iii)) and a wall sublayer (layer (iv)).}
It is worth to note  that regime I converges to regime II as $Wi$ becomes large, regime II converges to regime III as $\beta$ approaches unity, and regime III converges to regime IV as $\breve\sigma$ becomes large.

\subsection{Regime I analysis}
\label{sec:regime_I}
\subsubsection{Formulation for regime I}
\label{sec:regime_I_formulation}
This subsection focuses on $Wi=O(1)$ and $k\sim\epsilon^{-1/2}$, and thus we introduce
\begin{equation}
\bar k=\epsilon^{1/2}k=O(1).
\end{equation}
Without loss of generality, we only consider the perturbations near the lower wall $y=-1$.
By balancing the unsteady terms with the convective terms in  (\ref{eq:instability}a,b), we can estimate the magnitude of the phase speed, which can be expressed as
\begin{equation}
c=U(-1)+\epsilon^{1/2}c_1+\cdots,
\label{eq:phase_speed}
\end{equation}
where $U(-1)=0$ for a PPF and $-1$ for a PCF.

In this regime, the perturbations are mainly concentrated in the near-wall diffusive layer, as depicted in figure \ref{fig:sketch}-(a). In this layer, we introduce a local coordinate
\begin{equation}
 Y=\epsilon^{-1/2}(y+1)=O(1),
\end{equation}
and the velocity and conformation tensor of the base flow  are approximated as
\begin{equation}
 U=\sqrt{\epsilon}\lambda Y+O(\epsilon),\quad (C_{11},C_{12},C_{22},C_{33})=(2Wi^2\lambda^2+1, Wi\lambda,1,1)+O(\sqrt{\epsilon}).
 \end{equation}
From a scaling estimate from the equation system  (\ref{eq:instability}) and (\ref{eq:instability_OB}), we can expand the perturbation field as
$$
\Big(\hat u,\hat v,\hat p\Big)=\Big(\hat u_0,\hat v_0,\epsilon^{-1/2}{\beta }{}\hat p_0\Big)+\cdots,\quad \Big(\hat c_{11},\hat c_{12},\hat c_{22}\Big)=\epsilon^{-1/2}Wi^2\Big(\hat c_{110},\hat c_{120},\hat c_{220}\Big)+\cdots.
\eqno{(\theequation a,b)}
\label{eq:expansion_regime_I}
$$
Substituting the above expansion  into the instability system (\ref{eq:instability}) and (\ref{eq:instability_OB}), and collecting the leading-order terms, we obtain
\refstepcounter{equation}
$$
\hat u_0''-\bar k^2\hat u_0+\sigma Wi(\ri\bar k\hat c_{110}+\hat c_{120}')-\ri\bar k\hat p_0=0,
\eqno{(\theequation a)}$$
 $$\hat v_0''-\bar k^2\hat v_0+ \sigma Wi(\ri\bar k\hat c_{120}+\hat c_{220}')-\hat p_0'=0,\quad  \ri\bar k\hat u_0+\hat v_0'=0,
\eqno{(\theequation b,c)}
$$
$$
 \hat c_{110}''= S\hat c_{110}-2\ri \bar k(2\lambda^2+Wi^{-2})\hat u_0-2Wi^{-1}\lambda\hat u_0'-2\lambda \hat c_{120},
\eqno{(\theequation d)}
\label{eq:equation_I}
$$
$$
\hat c_{120}''= S\hat c_{120}-\ri \bar k(2\lambda^2+Wi^{-2})\hat v_0-Wi^{-2}\hat u_0'-\lambda\hat c_{220},\eqno{(\theequation e)}$$
  $$\hat c_{220}''= S\hat c_{220}-2(\ri\bar k \lambda Wi^{-1}\hat v_0+Wi^{-2}\hat v_0'),\eqno{(\theequation f)}
$$
where  $S=\ri\bar k(\lambda Y-c_1)+Wi^{-1}+\bar k^2$, and $\sigma=(1-\beta)/\beta$. Here, the equation for $\hat c_{33}$ is omitted, as it does not influence the other perturbation quantities. We assume that $\epsilon Re \ll 1$, which allows us to neglect the inertia terms in the leading-order momentum equation. Considering that the perturbations should damp to match the main layer in the upper limit,  we  impose the attenuation conditions for the perturbation,
\begin{equation}
(\hat u_0,\hat v_0,\hat p_0,\hat c_{110},\hat c_{120},\hat c_{220})\to 0\quad\mbox{as }Y\to \infty.\label{eq:BC_upper}
\end{equation}
Following the boundary conditions from equations (\ref{eq:BCuv}) to (\ref{eq:BC3}), we derive the wall boundary conditions for the asymptotic system presented in  (\ref{eq:equation_I}):
 the BC1 conditions read
\refstepcounter{equation}
$$
 \hat u_0(0)=\hat v_0(0)=0,\quad(-\ri \bar kc_1+Wi^{-1})\hat c_{110}(0)=2\lambda\Big(\frac{\hat u_0'(0)}{Wi }+{ \hat c_{120}(0)}\Big),
 \eqno{(\theequation a,b,c)}
  $$
  $$ (-\ri \bar kc_1+Wi^{-1})\hat c_{120}(0)=\frac{\hat u_0'(0)}{Wi^2},\quad \hat c_{220}(0)=0;
\eqno{(\theequation d,e)}\label{eq:BC_I}
$$
the BC2 conditions read
\refstepcounter{equation}
$$
 \hat u_0(0)=\hat v_0(0)=0,\quad(-\ri \bar kc_1+Wi^{-1}+\tilde k^2)\hat c_{110}(0)=2\lambda\Big(\frac{\hat u_0'(0)}{Wi }+{ \hat c_{120}(0)}\Big),
 \eqno{(\theequation a,b,c)}
  $$
  $$ (-\ri \bar kc_1+Wi^{-1}+\tilde k^2)\hat c_{120}(0)=\frac{\hat u_0'(0)}{Wi^2},\quad \hat c_{220}(0)=0;
\eqno{(\theequation d,e)}\label{eq:BC_II}
$$
 the BC3 conditions read
\refstepcounter{equation}
$$
 \hat u_0(0)=\hat v_0(0)=\hat c_{11}'(0)=\hat c_{12}'(0)=\hat c_{22}'(0)=0.\label{eq:BC_III}\eqno{(\theequation a,b,c,d,e)}
$$

In this regime, the eigenvalue system discussed in $\S$\ref{sec:instability} is recast to (\ref{eq:equation_I}) with boundary conditions (\ref{eq:BC_upper}) and  one of (\ref{eq:BC_I}) to (\ref{eq:BC_III}), with $c_1$ being the eigenvalue. In this system, only three independent  controlling parameters ($Wi$, $\beta$ and $\bar k$) are present, indicating a simplification from the original system.
Such an asymptotic-analysis-inspired simplification was also achieved in \cite{Lewy2024}.
This system can be solved using the local NI numerical approach, with the initial value derived from solving (\ref{eq:instability}) and (\ref{eq:instability_OB}) using the global SC method. In the NI calculations, we select a computational domain of $Y\in[0,1000]$, and discritise the governing equations into 2001 grid points, with a clustering of points in the near-wall region. The results are confirmed to be of sufficient accuracy by  comparing with those obtained using a doubled computational domain and twice the number of grid points.

\subsubsection{Numerical results for regime I}
\label{sec:Numerical_regime_I}
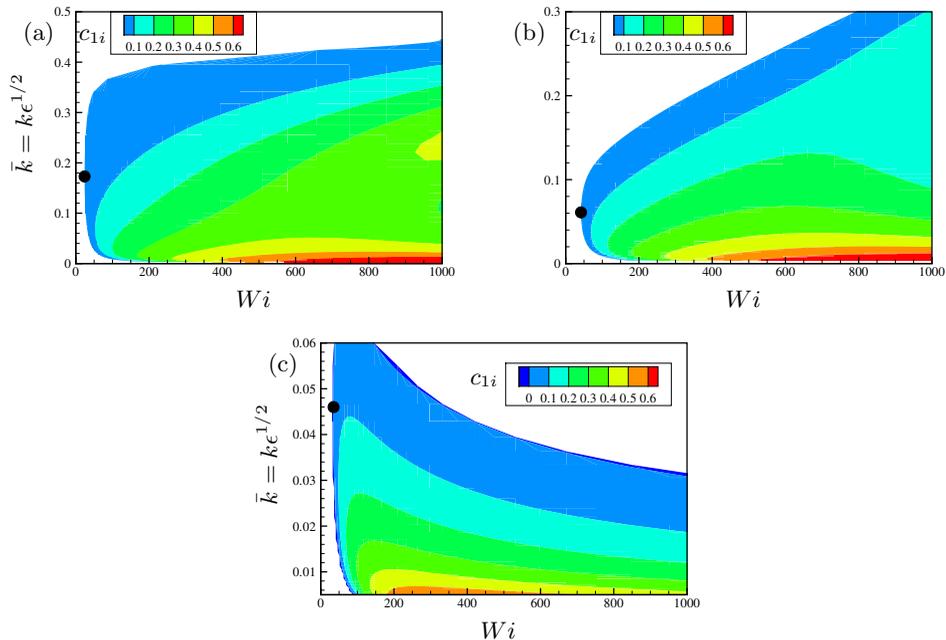
\begin{figure}
\begin{center}
  \includegraphics[width = 0.48\textwidth]{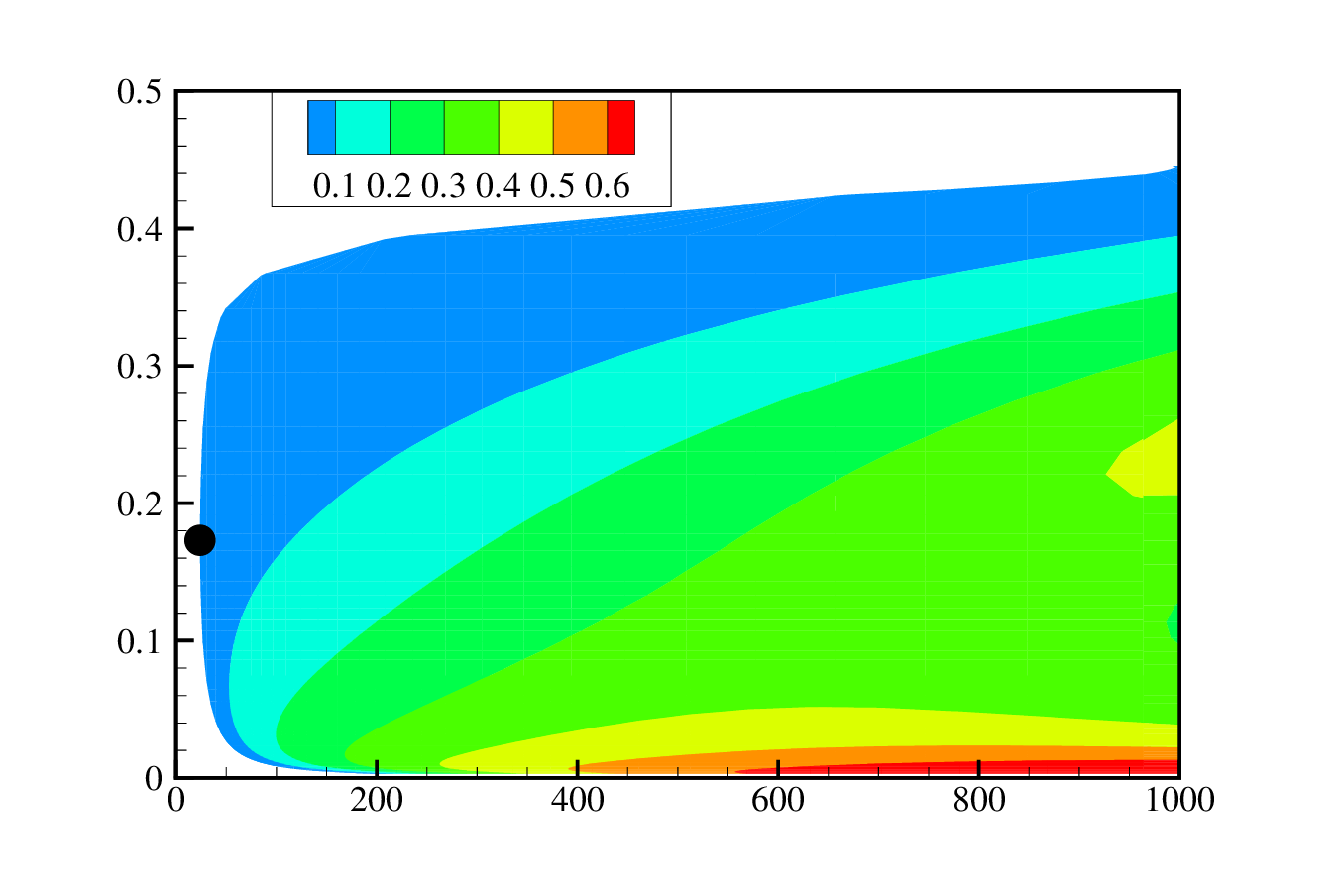}
  \put(-185,50){\rotatebox{90}{$\bar k= k\epsilon^{1/2}$}}
  \put(-180,100){(a)}
  \put(-100,0){$Wi$}\put(-159,100){$c_{1i}$}
    \includegraphics[width = 0.48\textwidth]{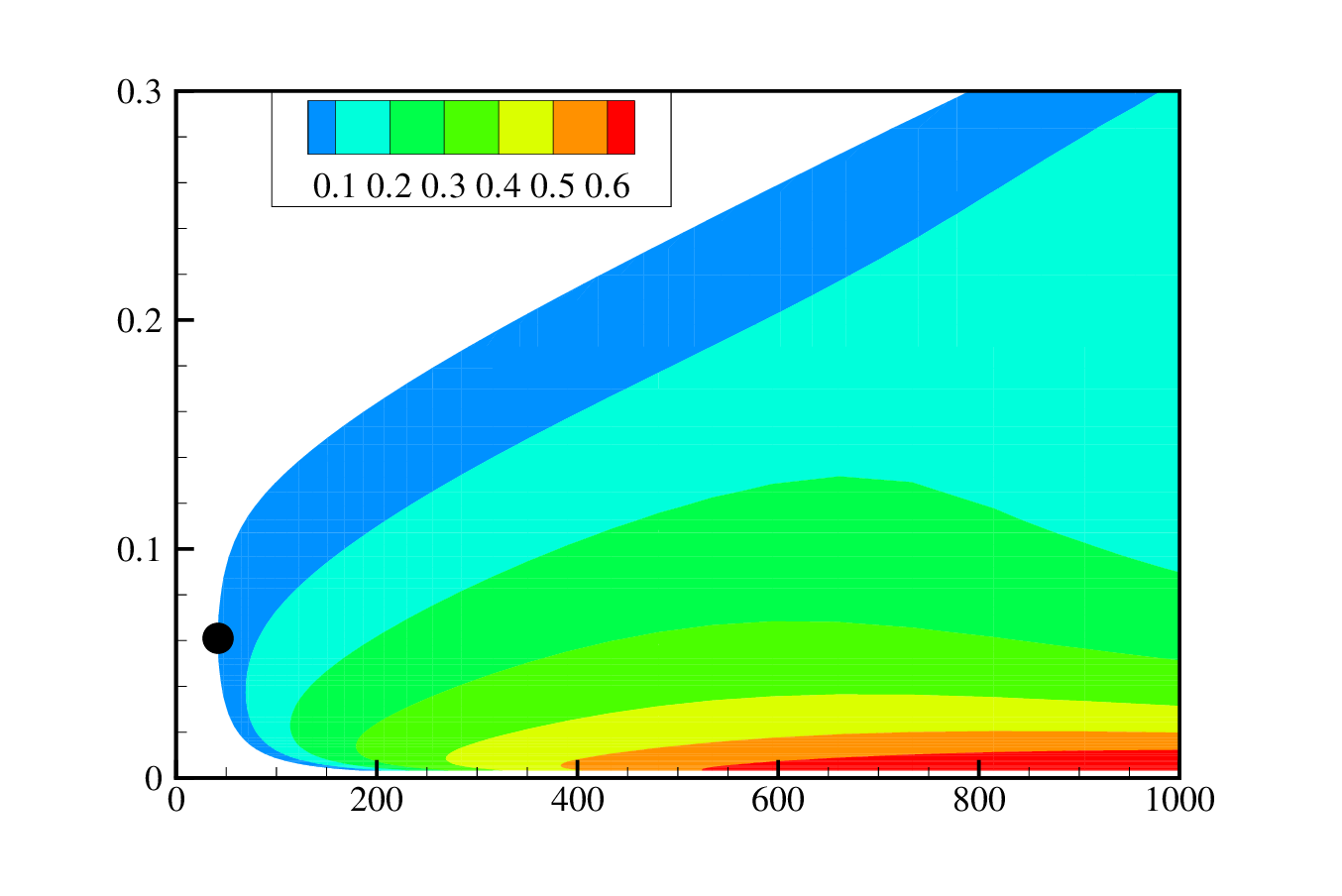}  \put(-180,100){(b)}\put(-159,100){$c_{1i}$}
  \put(-100,0){$Wi$}\\
    \includegraphics[width = 0.48\textwidth]{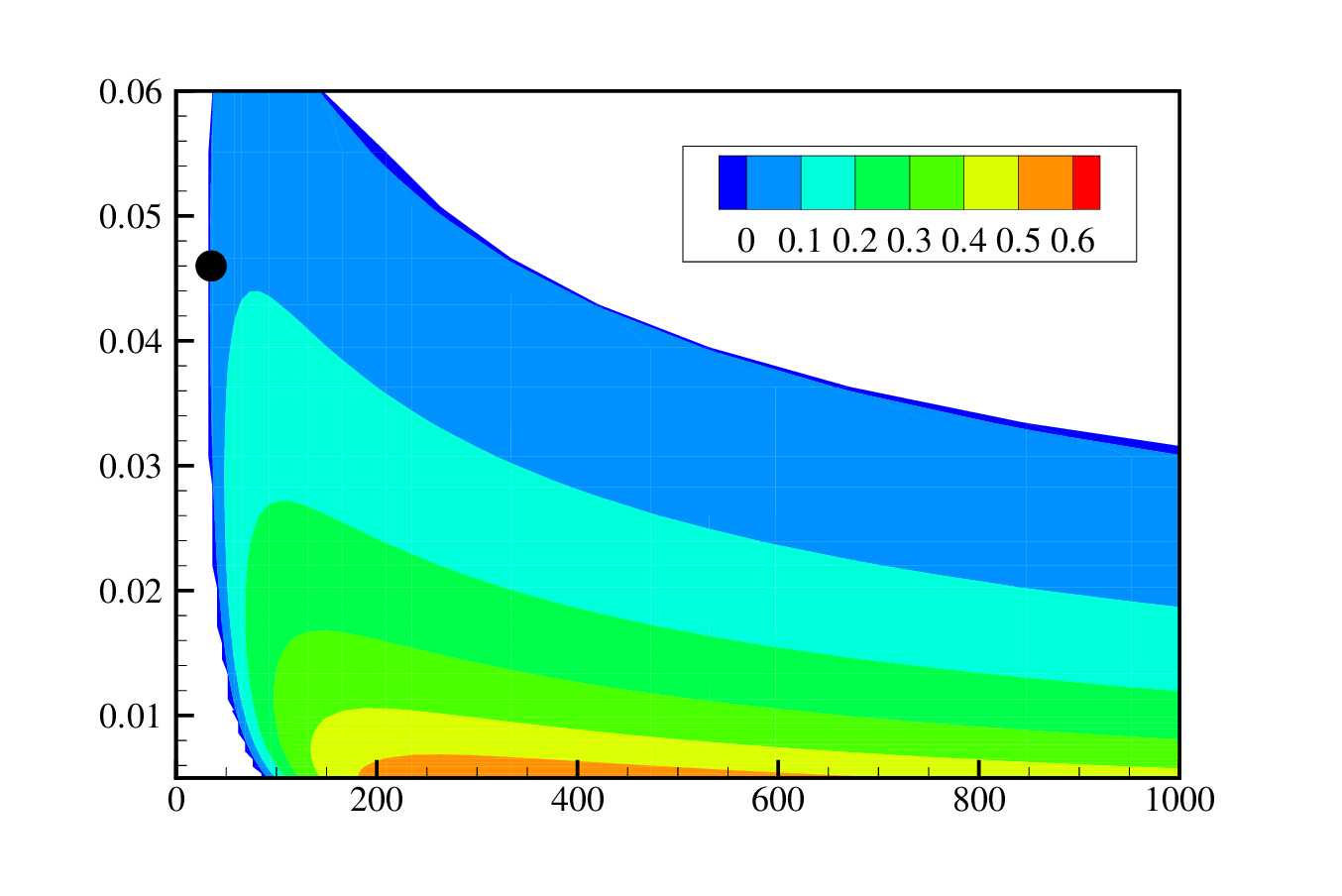}
 \put(-185,50){\rotatebox{90}{$\bar k= k\epsilon^{1/2}$}}
  \put(-180,100){(c)}
  \put(-100,0){$Wi$}\put(-104,95){$c_{1i}$}
  \caption{Contours of the rescaled growth rate $c_{1i}$ of the regime-I PDI for various wall boundary conditions for PPF with $\beta=0.9$, where the circles mark the critical point for the instability onset ($Wi_c$,$\bar k_c$). (a): BC1; (b): BC2; (c): BC3.}\label{fig:growth_contours}
\end{center}
\end{figure}
By selecting a viscoelastic PPF with  $\lambda=2$ and $\beta=0.9$, we calculate the rescaled growth rate $c_{1i}$ of PDI for the three wall boundary conditions represented in (\ref{eq:BC_I}) to (\ref{eq:BC_III}). The resulting contours in the $Wi$-$\bar k$   plane are depicted in figure \ref{fig:growth_contours}.
Although unstable zones emerge at moderate and high $Wi$ values for all considered boundary conditions, the overall shapes of these unstable zones vary significantly. The largest unstable zone is observed in panel (a), corresponding to  BC1. For each $Wi$ value exceeding the critical threshold $Wi_c$, two points with zero growth rates appear, referred to as the lower- and upper-branch neutral points. As $Wi$ increases, the most unstable state, indicated by the peak growth rate, occurs at a very low $\bar k$  value, in close proximity to the lower-branch neutral point. This reveals a regime of particular interest in the limit of   $Wi\gg 1$ and $\bar k\ll 1$.

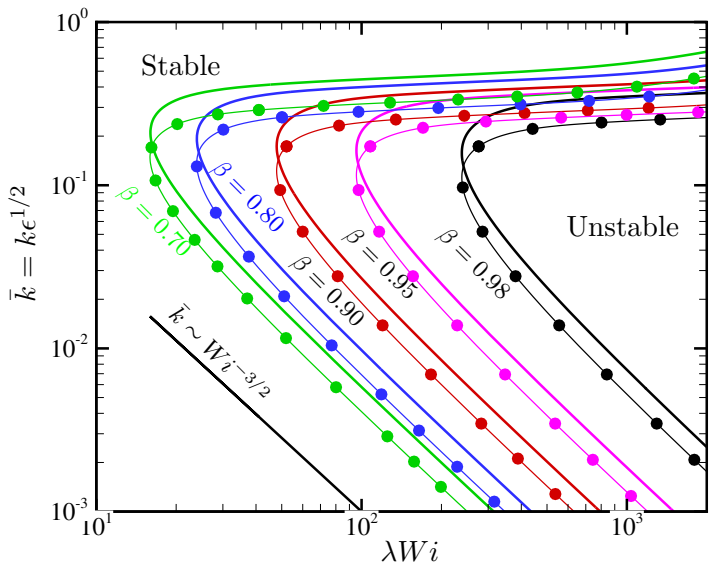
\begin{figure}
\begin{center}
  \includegraphics[width = 0.8\textwidth]{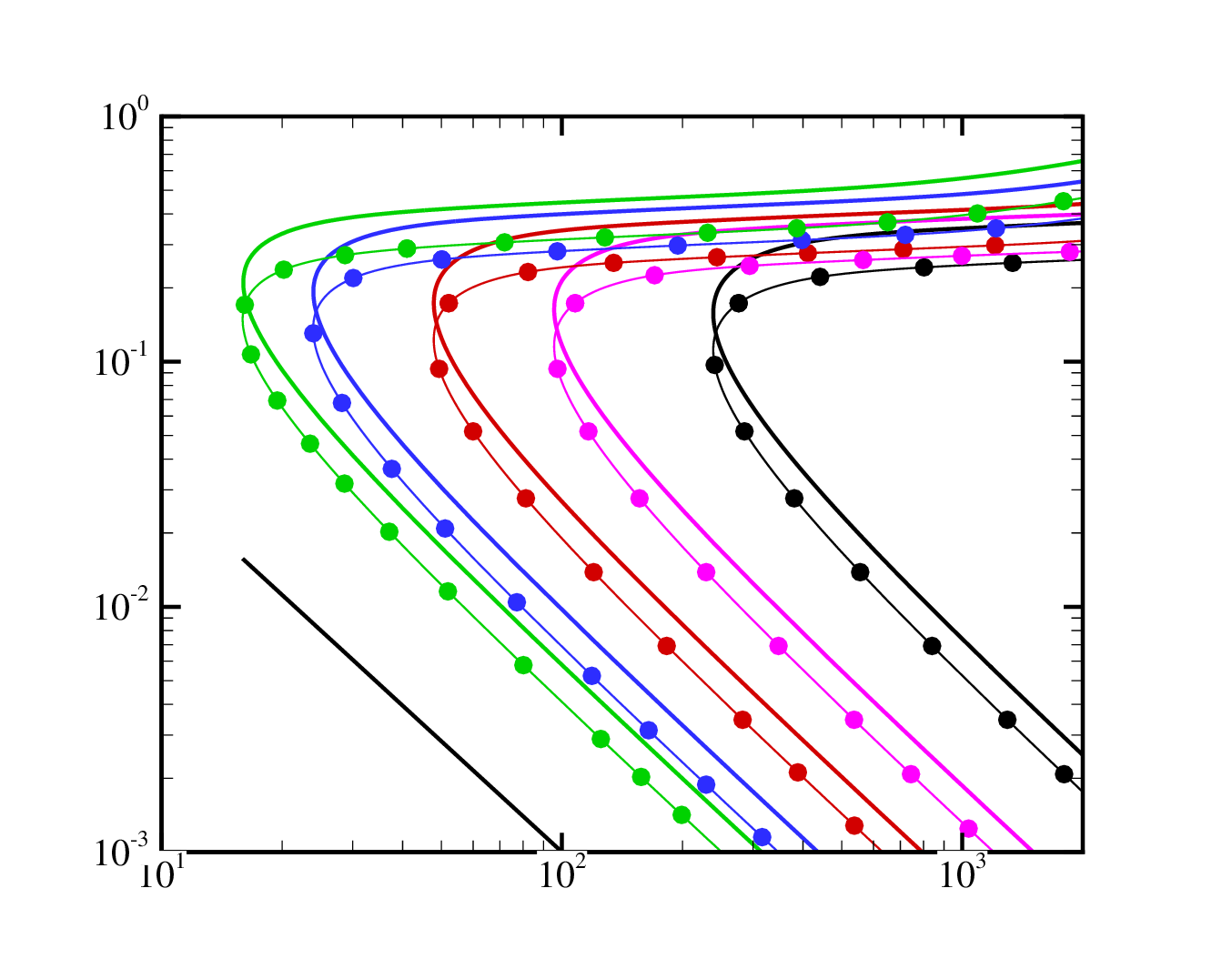}
  \put(-300,105){\rotatebox{90}{\large $\bar k= k\epsilon^{1/2}$}}
  \put(-195,120){\rotatebox{-45}{\color{black}$\beta=0.90$}}
   \put(-260,146){\rotatebox{-45}{\color{green}$\beta=0.70$}}
    \put(-225,155){\rotatebox{-45}{\color{blue}$\beta=0.80$}}
    \put(-175,130){\rotatebox{-45}{\color{black}$\beta=0.95$}}
     \put(-140,130){\rotatebox{-45}{\color{black}$\beta=0.98$}}
       \put(-240,100){\rotatebox{-45}{$\bar k\sim Wi^{-3/2}$}}
  \put(-160,8){\large$\lambda Wi$}
  \put(-90,130){\large Unstable}
  \put(-250,190){\large Stable}
  \caption{Neutral curves of regime-I PDI in the $\bar k$-$\lambda Wi$ plane for BC1 with representative $\beta$ values, where lines with and without are for the PCF ({\color{black}$\lambda=1$}) and PPF ({\color{black}$\lambda=2$}) flows, respectively. The fitting curve $\bar k\sim Wi^{-3/2}$ indicate the scaling law of the lower-branch neutral curve. }\label{fig:neutral_curve}
\end{center}
\end{figure}
In figure \ref{fig:neutral_curve}, we choose BC1 and plot the  neutral curves for representative $\beta$ values for both PPF and PCF.
Following \cite{Couchman2024}, the horizontal axis is rescaled by $\lambda Wi$.
For a given $\beta$ value, the neutral curves for PPF and PCF are quite similar, with the $\bar k$ band for PCF appearing in a slightly lower range. Reducing the polymer concentration (i.e. increasing $\beta$) leads to a shift of the unstable zone to higher $\lambda Wi$ values, while the wavenumber band of $\bar k$ remains almost unchanged. The critical values of $\lambda Wi_c$ and $\bar k_c$ are summarised in Table \ref{tab:cases}.
For each $\beta$ value, the critical values  of $\lambda Wi_c$  are almost identical for PCF and PPF cases,  aligning with the observation in \cite{Couchman2024}. However the critical values of $\bar k_c$ exhibit slight differences.
Remarkably, for the lower-branch neutral curve, the rescaled wavenumber $\bar k$   scales as $Wi^{-3/2}$. By further fitting the coefficient of the scaling relation ${\cal C}=\bar k/Wi^{-3/2}$, we obtain the coefficients for each $\beta$ value, as summarised in Table \ref{tab:cases2}. These coefficients increase remarkably as $\beta$ approaches unity.
\begin{table}
  \begin{center}
\def~{\hphantom{0}}
  \begin{tabular}{ccccccccccc}
  \vspace{.2cm}     $\beta$  & 0.70& 0.80&0.90&0.95&0.98& 0.70& 0.80&0.90&0.95&0.98  \\ \vspace{.2cm}
 $\lambda$& 2&2&2&2&2&1&1&1&1&1\\ \vspace{.2cm}
 $\bar k_c$&0.211&0.190&0.173&0.165&0.160&0.149&0.135&0.123&0.116&0.111\\ \vspace{.2cm}
 $\lambda Wi_c$&15.9&23.9&47.8&95.6&239&15.9 &23.9&47.8&95.6&239
\end{tabular}
  \caption{Critical values for various $\beta$ and $\lambda$ values.}
  \label{tab:cases}
  \end{center}
\end{table}
\begin{table}
  \begin{center}
\def~{\hphantom{0}}
  \begin{tabular}{cccccc}
  \vspace{.2cm}     $\beta$  & 0.70& 0.80&0.90&0.95&0.98  \\ \vspace{.2cm}
 $\lambda$& 2&2&2&2&2\\ \vspace{.2cm}
 ${\cal C}$&1.94&3.15&7.73&20.1&75.2
\end{tabular}
  \caption{Fitting of the scaling coefficient ${\cal C}=\bar k/Wi^{-3/2}$ for the lower-branch neutral curve of PDI instability in a viscoelastic PPF.}
  \label{tab:cases2}
  \end{center}
\end{table}

\begin{figure}
\begin{center}
  \includegraphics[width = 0.48\textwidth]{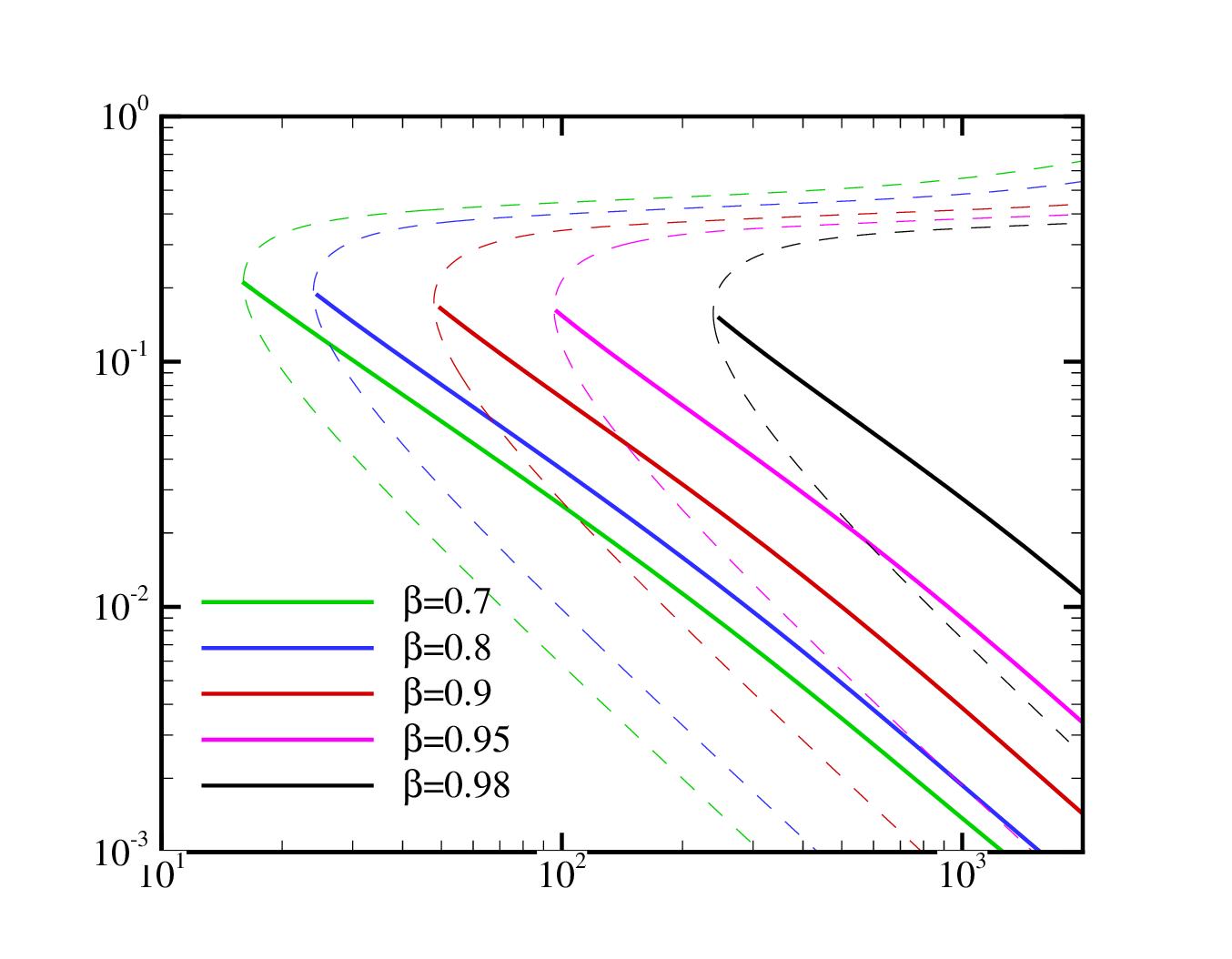}
  \put(-185,60){$ k\epsilon^{1/2}$}
  \put(-180,120){(a)}
  \put(-100,0){$\lambda Wi$}
    \includegraphics[width = 0.48\textwidth]{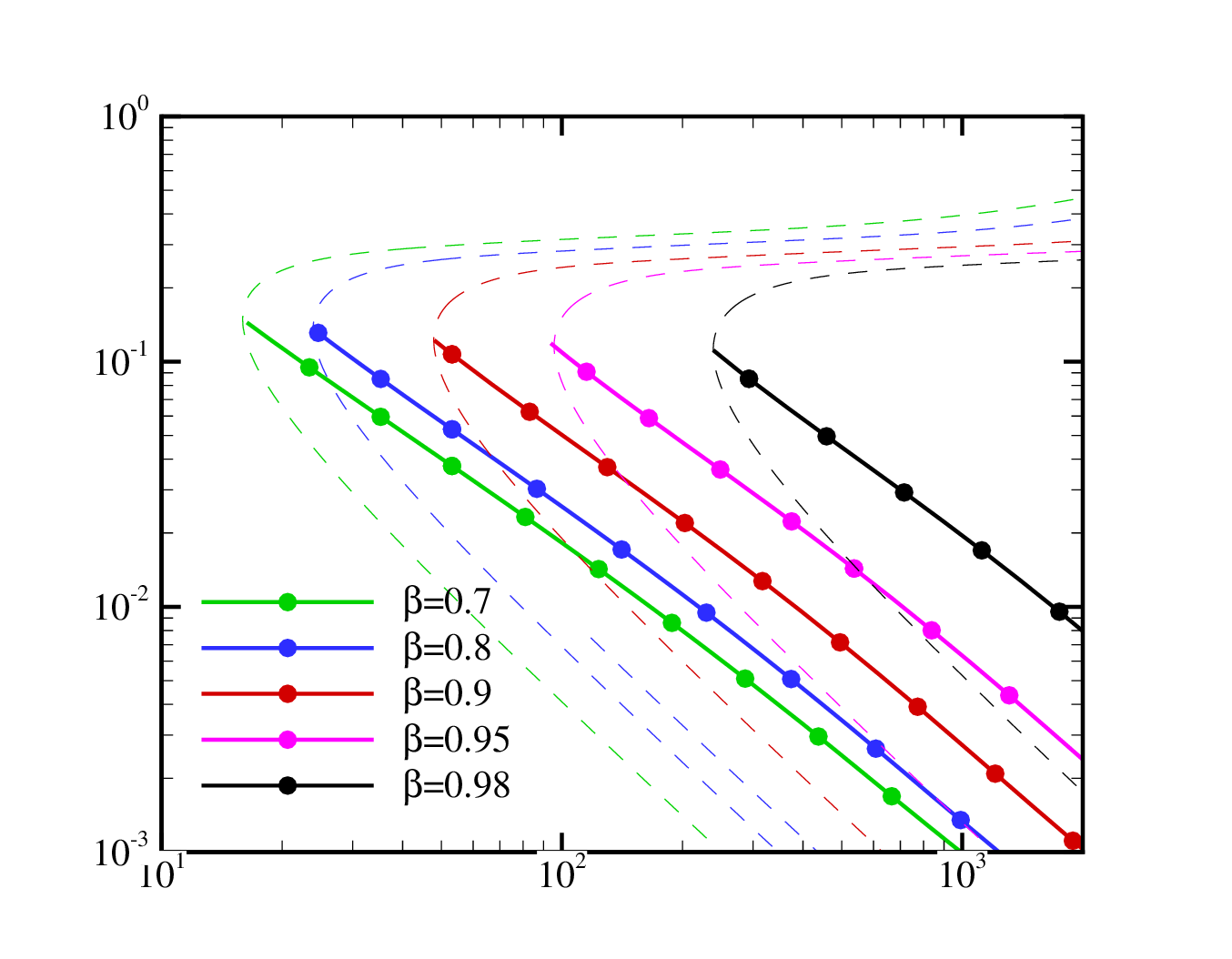}  \put(-180,120){(b)}
  \put(-100,0){$\lambda Wi$}\\
      \includegraphics[width = 0.48\textwidth]{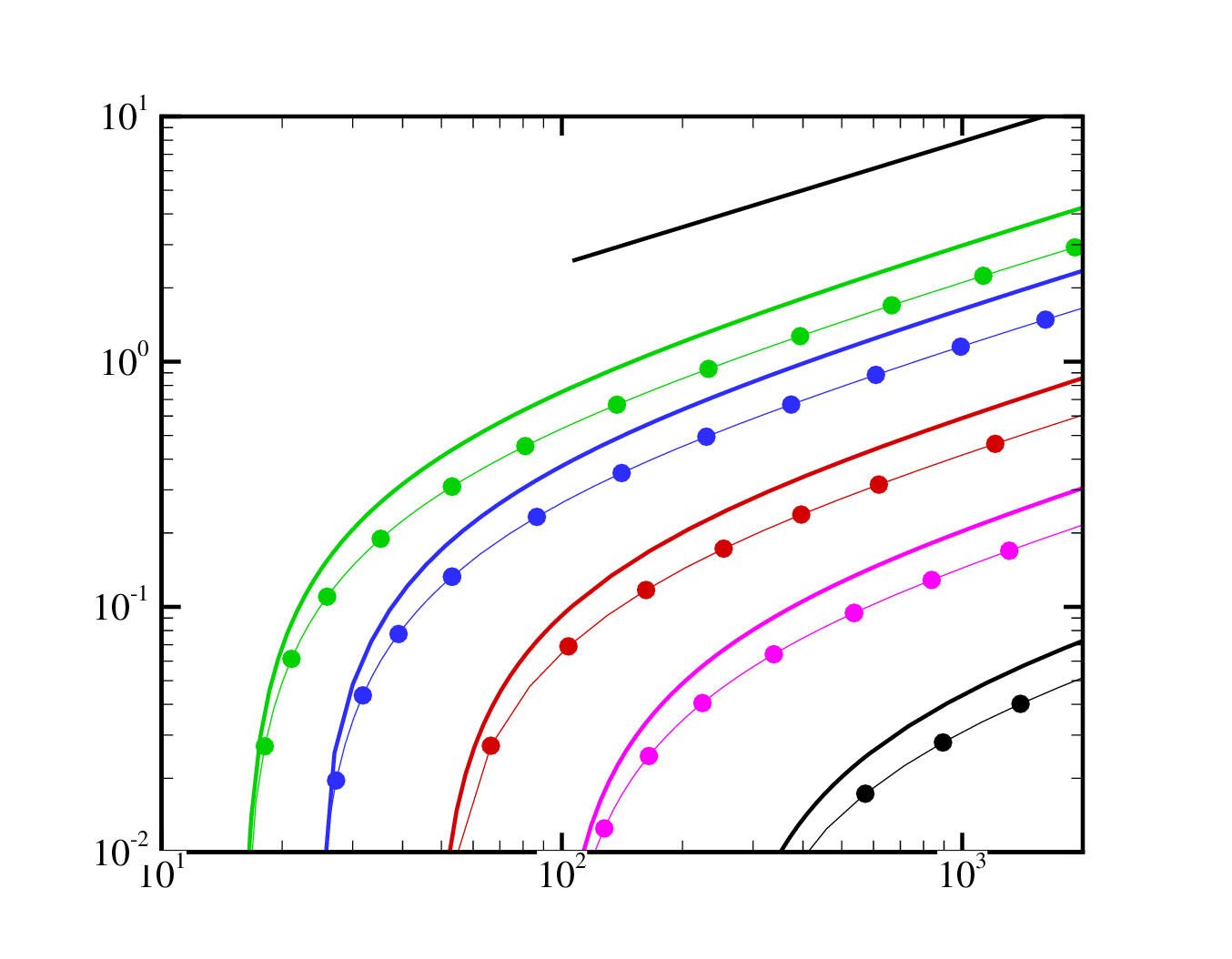}  \put(-180,120){(c)}
  \put(-100,0){$\lambda Wi$}  \put(-180,65){$ c_{1i}$} \put(-105,105){\rotatebox{18}{$\sim Wi^{1/2}$}}
  \caption{{\color{black}The properties of the most unstable regime-I PDI mode for different $\beta$ values under BC1}. (a) and (b): Rescaled wavenumber $\bar k=k\epsilon^{1/2}$ for PPF ($\lambda=2$) and PCF ($\lambda=1$), respectively; (c): rescaled growth rate $c_{1i}$. The light dashed lines in (a,b) indicate the neutral curves extracted from figure \ref{fig:neutral_curve}.}\label{fig:most_unstable}
\end{center}
\end{figure}
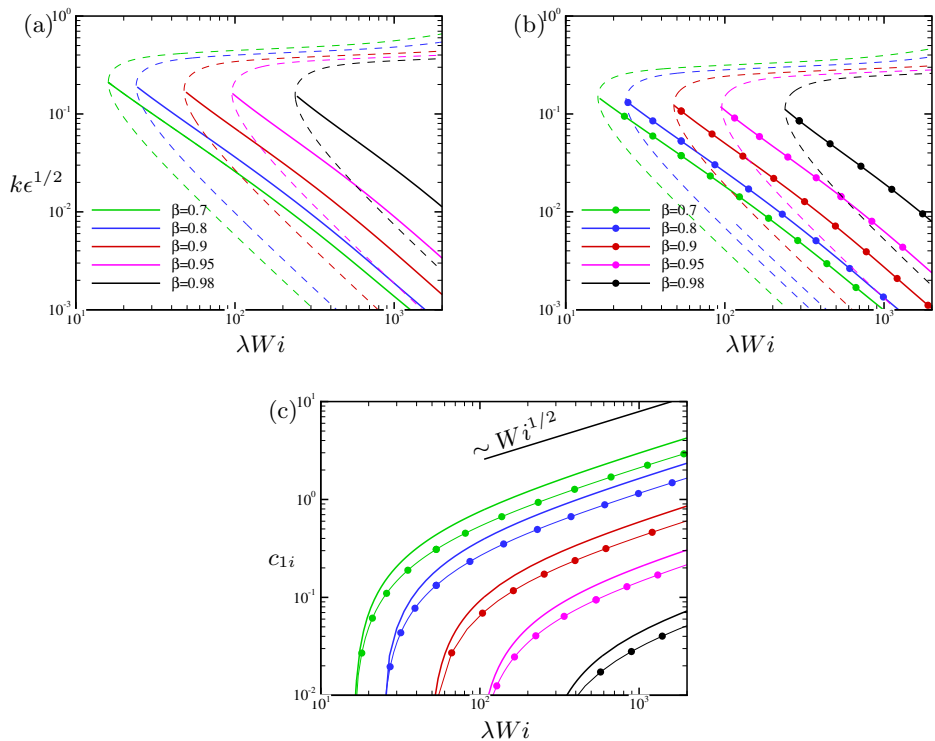
Figures \ref{fig:most_unstable}-(a) and (b) illustrate the parameter pairs $(\lambda Wi,\bar k)$ for the most unstable PDI mode in PPF and PCF, respectively. Analyzing the results across different polymer concentrations, a noticeable decrease in $\bar k$   with an increase in $Wi$ is evident, with only a limited discrepancy observed between PPF and PCF. Panel (c) shows the dependence of the rescaled growth rate $c_{1i}$ on $\lambda Wi$, where a scaling relation of  $c_{1i}\sim Wi^{1/2}$ is clearly identified in the high-$Wi$ limit for each curve.

The observations above motivate us to investigate the instability mechanism associated with the lower-branch PDI mode in the high-$Wi$ limit, as this regime also encompasses the most unstable state. Given that the instability properties of PPFs and PCFs are quite close, we will focus on PPFs for demonstration in the following. In particular, we will focus on the scaling relations  $\bar k\sim Wi^{-3/2}$ and $c_{1i}\sim Wi^{1/2}$ in the subsequent subsection.
\subsection{Regime II analysis}
\label{sec:regime_II}
 In regime II, we consider a large Weissenberg number with  $\bar k\sim Wi^{-3/2}$. Thus, for convenience, we further rescale the wavenumber as
\begin{equation}
\tilde k=Wi^{3/2}\bar k=Wi^{3/2}\epsilon^{1/2}k=O(1).
\end{equation}
Under this condition, the diffusive layer splits into two distinct regions: LDL and UDL, with the asymptotic structure illustrated in figure \ref{fig:sketch}-(b).

\subsubsection{Lower diffusive layer for regime II}
Balancing the conformation tensor and viscosity in the momentum equation (\ref{eq:equation_I}a), we obtain $\hat u_0''\sim \sigma Wi\ri \bar k\hat c_{110}$, which leads to  $\hat c_{110}\sim Y^{-2}Wi^{-1}\bar k^{-1}\hat u_0$. Considering the conformation tensor equation (\ref{eq:equation_I}d), we need to balance the diffusive term ($\hat c_{110}''$) with the convective term (e.g. $-4\ri\bar k\lambda^2\hat u_0$), which yields $\hat c_{110}\sim Y^2\bar k\hat u_0$. These two relations together determine the thickness of the diffusive layer, yielding $Y\sim Wi^{1/2}$. For convenience, we introduce a local coordinate
 \begin{equation}
 \tilde Y=Wi^{-1/2}Y=Wi^{-1/2}\epsilon^{-1/2}(1+y)=O(1).
 \end{equation}
 Simultaneously, we can derive the magnitude of the conformation tensor from the relations established above,  $\hat c_{110}\sim Wi^{-1/2}\hat u_0$. Similarly, we find that $\hat v_0\sim Wi^{-1}\hat u_0$, $\hat p\sim Wi^{-3/2}\hat u_0$, $\hat c_{120}=Wi^{-3/2}\hat u_0$ and $\hat c_{220}=Wi^{-5/2}\hat u_0$. We can then expand the perturbation field as follows:
\begin{equation}
(\hat u_0,\hat v_0,\hat p_0,\hat c_{110},\hat c_{120},\hat c_{220})=\Big(\tilde u_0,\frac{\tilde v_0}{Wi},\frac{\tilde p_0}{Wi^{3/2}},\frac{\tilde c_{110}}{Wi^{1/2}},\frac{\tilde c_{120}}{Wi^{3/2}},\frac{\tilde c_{220}}{Wi^{5/2}}\Big)+\cdots.
\label{eq:expansion_LDL}
\end{equation}
By balancing the inertia term in the conformation tensor equation, we find that the expression for the phase speed expansion should be modified to
\begin{equation}
c=U(-1)+\epsilon^{1/2}Wi^{1/2}\tilde c_1+\cdots.
\label{eq:phase_speed1}
\end{equation}
It can be observed that the scaling of the phase speed for the regime-II analysis is consistent with that of the most unstable mode depicted in figure \ref{fig:most_unstable}.

Substituting the above expansions  into the instability equations (\ref{eq:equation_I}), we obtain
\refstepcounter{equation}
$$
\tilde u_0''+\sigma(\ri\tilde k\tilde c_{110}+\tilde c_{120}')=0,\quad \tilde v_0''+ \sigma(\ri\tilde k\tilde c_{120}+\tilde c_{220}')-\tilde p_0'=0,
\eqno{(\theequation a,b)}
$$
$$
 \ri\tilde k\tilde u_0+\tilde v_0'=0,\quad \tilde c_{110}''=\tilde S\tilde c_{110}-2(2\ri \tilde k\lambda^2\tilde u_0+\lambda\tilde u_0'+\lambda \tilde c_{120}),
\eqno{(\theequation c,d)}
\label{eq:equation_II}
$$
$$
\tilde c_{120}''= \tilde S\tilde c_{120}-2\ri \tilde k\lambda^2\tilde v_0-\tilde u_0'-\lambda\tilde c_{220},\quad \tilde c_{220}''=\tilde S\tilde c_{220}-2(\ri\tilde k \lambda\tilde v_0+\tilde v_0'),\eqno{(\theequation e,f)}
$$
where $\tilde S=\ri\tilde k(\lambda \tilde Y-\tilde c_1)+1$. It is seen that the streamwise pressure gradient is absent in the leading-order streamwise momentum equation in this regime.
Under these scaling relations, the wall boundary conditions for both BC1 and BC2, as illustrated in (\ref{eq:BC_I}) and (\ref{eq:BC_II}), respectively, can be recast into a unified form,
\refstepcounter{equation}
$$
 \tilde u_0(0)=\tilde v_0(0)=0,\quad { \tilde S_0}\tilde c_{110}(0)={2\lambda[\tilde u_0'(0)+\tilde c_{120}(0)]},\quad { \tilde S_0}\tilde c_{120}(0)={\tilde u_0'}(0),\quad \tilde c_{220}=0,
 \label{eq:LB_BC}\eqno{(\theequation a,b,c,d,e)}
$$
where $\tilde S_0=\tilde S(0)=-\ri \tilde k\tilde c_1+1$. Therefore, we use BC12 to represent this type of boundary conditions. For BC3, we can express the wall boundary conditions  as
\refstepcounter{equation}
$$
\tilde u_0(0)=\tilde v_0(0)=\tilde c_{110}'(0)=\tilde c_{120}'(0)=\tilde c_{220}'(0)=0. \label{eq:LB_BCiii}
\eqno{(\theequation a,b,c,d,e)}
$$

This system does not guarantee  the attenuation conditions in the limit of $\tilde Y\to \infty$.  Therefore, we must consider a thicker layer (UDL) where the streamwise pressure gradient comes back to the leading order. By analyzing the balance of the momentum equation, we find that the thickness of the UDL should be comparable to the rescaled streamwise length scale, specifically, $(1+y)\sim \bar k\sim Wi^{-3/2}$.

\subsubsection{Upper diffusive layer for $1-\beta=O(1)$}
\label{sec:UDL}
Following the aforementioned scaling estimate, we introduce
\begin{equation}
\bar Y=Wi^{-1}\tilde Y=Wi^{-3/2}Y=Wi^{-3/2}\epsilon^{-1/2}(1+y)=O(1).
\end{equation}
From a scaling estimate, we expand the perturbation field in this layer as
\begin{equation}
(\hat u_0,\hat v_0,\hat p_0,\hat c_{110},\hat c_{120},\hat c_{220})=\Big(\bar u_0,\bar v_0,\frac{\bar p_0}{Wi^{1/2}},\frac{\bar c_{110}}{Wi^{3/2}},\frac{\bar c_{120}}{Wi^{3/2}},\frac{\bar c_{220}}{Wi^{5/2}}\Big)+\cdots.\label{eq:expansion_UDL}
\end{equation}
Substituting these equations to (\ref{eq:equation_I}), we obtain the leading-order governing equations in this layer:
\refstepcounter{equation}
$$
\sigma(\ri\tilde k\bar  c_{110}+\bar  c_{120}')-\ri\tilde k\bar  p_0=0,\quad  \ri\sigma\tilde k\bar  c_{120}-\bar  p_0'=0,
\eqno{(\theequation a,b)}
$$
$$
 \ri\tilde k\bar  u_0+\bar  v_0'=0,\quad \ri\bar k\lambda \bar Y\bar  c_{110}-2(2\ri \tilde k\lambda^2\bar  u_0+\lambda \bar  c_{120})=0,
\eqno{(\theequation c,d)}
\label{eq:equation_III}
$$
$$
\ri\bar k\lambda \bar Y\bar  c_{120}-2\ri \tilde k\lambda^2\bar  v_0=0,\quad \ri\bar k\lambda \bar Y\bar  c_{220}-2\ri\tilde k \lambda\bar  v_0=0.\eqno{(\theequation e,f)}
$$
The system can be reduced to
\begin{equation}
\bar Y^2\bar v_0''-2\bar Y\bar v'_0+(2-\tilde k^2\bar Y^2)\bar v_0=0,
\end{equation}
whose solution, by excluding the exponentially growing part, is expressed as
\begin{equation}
\bar v_0=d_0\bar Y\re^{-\tilde k\bar Y},
\end{equation}
with $d_0$ being a constant. We can therefore obtain the other perturbation quantities from (\ref{eq:equation_III}),
\begin{equation}
\Big(\bar u_0,\bar c_{110},\bar c_{120},\bar c_{220}\Big)=\Big({\ri{\bar k}^{-1}(1-\bar k\bar Y)},-4\ri\lambda  ,2\lambda ,2\Big)d_0\re^{-\bar k\bar Y}.
\end{equation}
It is obvious that the perturbations attenuate exponentially in the large-$\bar Y$ limit.
We can also estimate the asymptotic behaviours of these perturbations in the lower limit
\begin{equation}
\Big(\bar  u_0, \bar v_0, \bar p_0,\bar c_{110},\bar c_{120},\bar c_{220}\Big)\to\Big( {\ri }{\bar k}^{-1},\bar Y,  -2\ri\lambda \sigma,  -4\ri\lambda,  2\lambda , 2\Big)\quad\mbox{as }\bar Y\to 0.
\label{eq:upper_diffusive_asymp}
\end{equation}

Comparing the  expansions between (\ref{eq:expansion_UDL}) with  (\ref{eq:expansion_LDL}), we find that the magnitude of $\hat c_{110}$ in this layer is smaller than that in LDL, indicating that $\tilde c_{110}$ in LDL should approach zero in the large-$\tilde Y$ limit. The other perturbations should match in the overlapping region between the two layers.
Thus, we can obtain the upper boundary conditions for system (\ref{eq:equation_II}),
\begin{equation}
\Big(\tilde u_0', \tilde c_{110},\tilde c_{120}', \tilde c_{220}'\Big)\to 0,\quad \tilde p_0\to -2\sigma\bar k\lambda \tilde u_0\quad \mbox{as }\tilde Y\to \infty.
\label{eq:BC_LW_upper}
\end{equation}

\subsubsection{Numerical results for regime II}
\begin{figure}
\begin{center}
  \includegraphics[width = 0.5\textwidth]{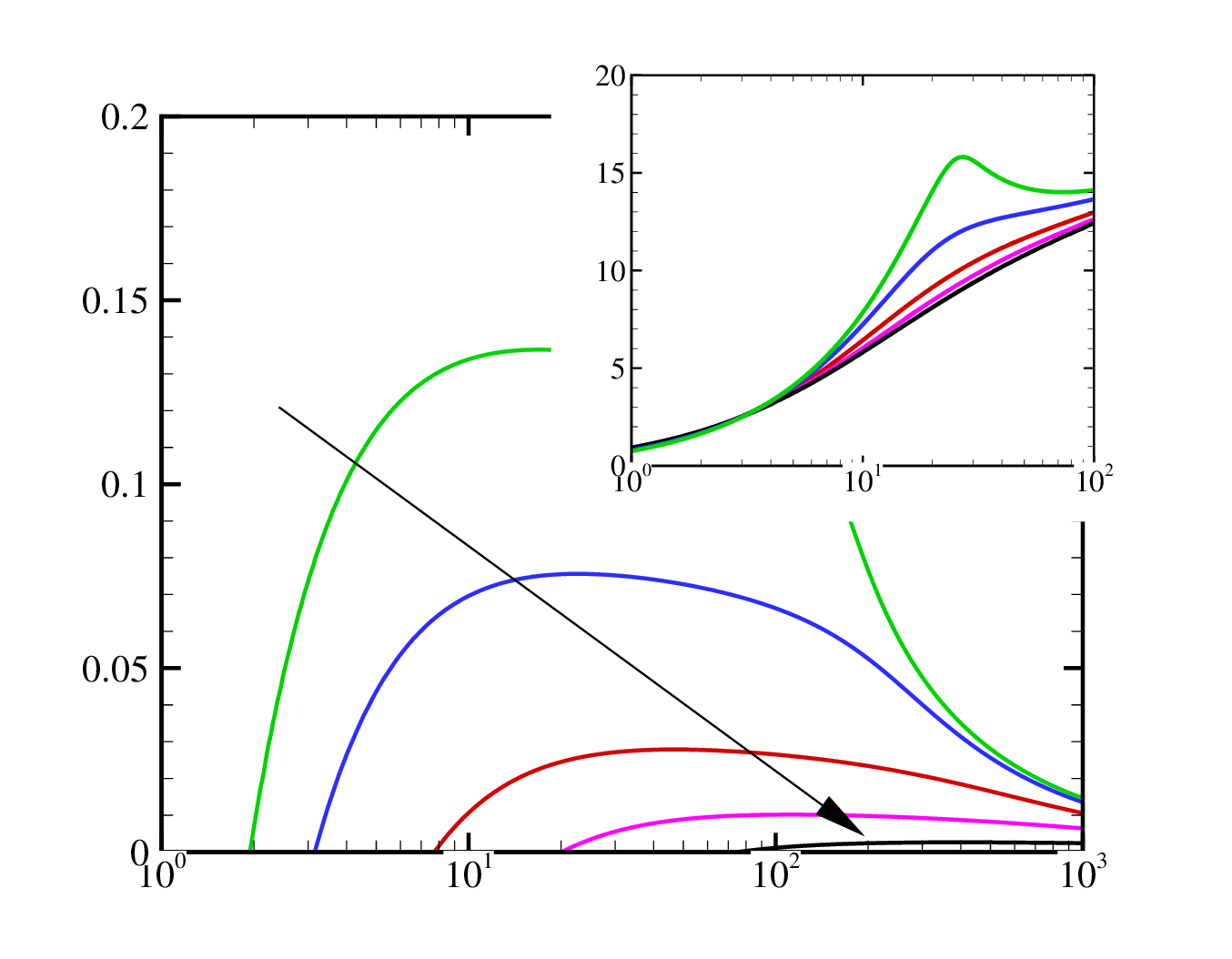}
  \put(-200,35){\rotatebox{90}{\large $\tilde c_{1i}=(\epsilon Wi)^{-1/2}c_i$}}
  \put(-130,0){\large$\tilde k=Wi^{3/2}\epsilon^{1/2}k$}\put(-200,130){(a)}
  \put(-130,40){Increasing $\beta$}
  \put(-78,65){$\tilde k(1-\beta)^{3/2}$}  \put(-109,95){\rotatebox{90}{$\tilde c_{1i}\tilde k$}}
    \includegraphics[width = 0.5\textwidth]{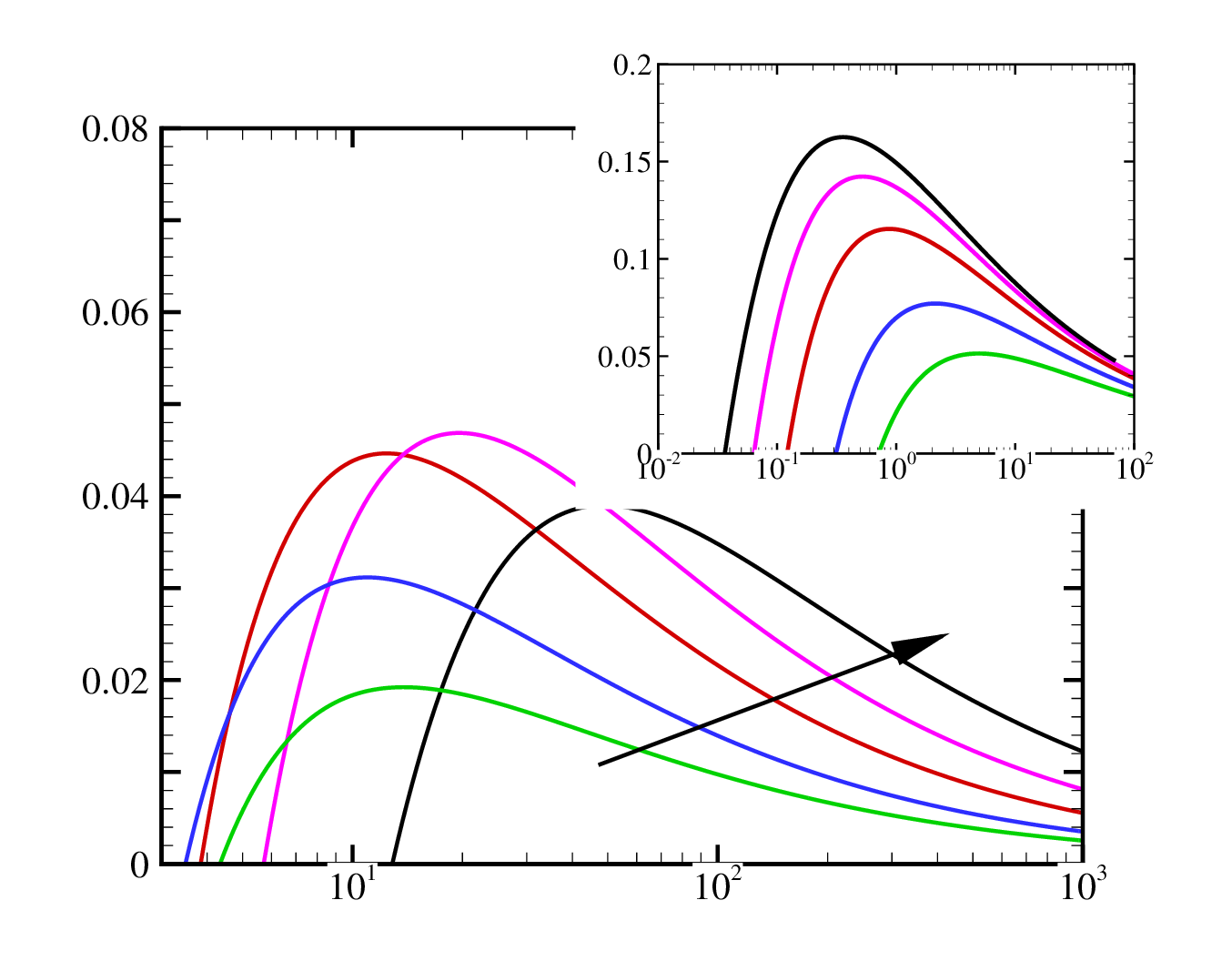}
      \put(-130,0){\large$\tilde k=Wi^{3/2}\epsilon^{1/2}k$}\put(-200,130){(b)}
        \put(-119,25){Increasing $\beta$}  \put(-70,70){$\tilde k(1-\beta)^{3/2}$}  \put(-109,95){\rotatebox{90}{$\tilde c_{1i}\tilde k^{1/3}$}}
  %\\

  %  \includegraphics[width = 0.8\textwidth]{growth_lower_branch2.eps}  \put(-160,6){\large$\sigma\tilde k^{2/3}$}  \put(-305,100){\rotatebox{90}{\large $\tilde c_{1,i}/\sigma$}}
  %  \put(-305,200){\large (b)}     \put(-190,155){{\color{black}$\beta=0.90$}}
   %  \put(-190,60){{\color{black}$\beta=0.98$}}
   %  \put(-190,110){{\color{black}$\beta=0.95$}}     \put(-190,33){{\color{orange}$\beta=0.99$}}
  \caption{Dependence of the rescaled growth rate $\tilde c_{1i}$ on the rescaled wavenumber $\tilde k$ with  $\beta=0.7$, 0.8, 0.9, 0.95 and 0.98  for regime-II PDI mode for a PPF. {\color{black}The inset in each panel shows the curves obtained after applying an appropriate rescaling.} (a): BC12; (b): BC3. }\label{fig:growth_lower}
\end{center}
\end{figure}
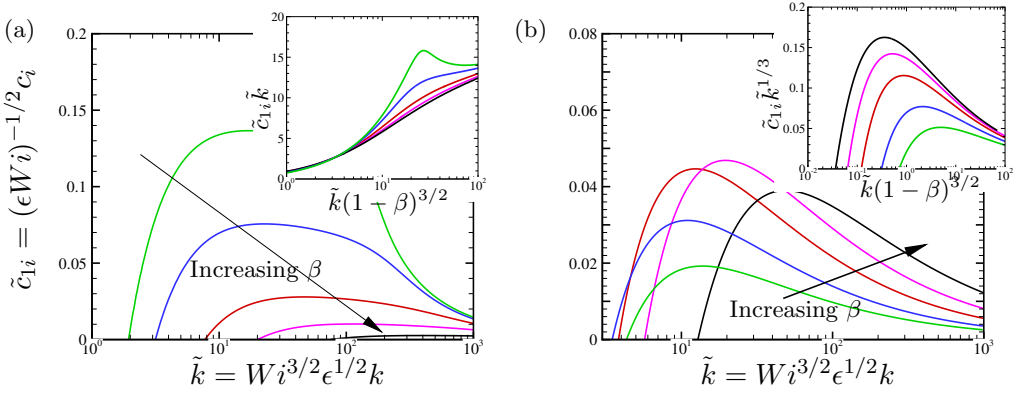
In regime II, the eigenvalue system  is reformulated as (\ref{eq:equation_II}) with boundary conditions (\ref{eq:LB_BC}) (or (\ref{eq:LB_BCiii})) and (\ref{eq:BC_LW_upper}), where $\tilde c_1$ is the eigenvalue. The number of the controlling parameters is further reduced to two: $\tilde k$ and $\beta$. Choosing a few representative $\beta$ values, we plot the dependence of the rescaled growth rate $\tilde c_{1i}$ on the rescaled wavenumber $\tilde k$ for a PPF ($\lambda=2$), as shown in figure \ref{fig:growth_lower}. The results for both boundary conditions  (\ref{eq:LB_BC}) and (\ref{eq:LB_BCiii}) are examined.
 For each curve, the PDI becomes unstable when the rescaled wavenumber $\tilde k$ exceeds a critical value, known as the neutral point. The rescaled growth rate $\tilde c_{1i}$ increases with $\tilde k$ until reaching its most unstable state. However, for further higher $\tilde k$ values, the rescaled growth rate decreases monotonically.
 For BC12, represented by  (\ref{eq:LB_BC}), the neutral wavenumber $\tilde k$ shifts to higher values as  $\beta$ approaches unity,  while $\tilde c_{1i}$ decreases notably. In contrast, for BC3, represented by (\ref{eq:LB_BCiii}), the variation of $\tilde k$ on $\beta$ is not monotonic.
  Notably, for $\tilde k\gg1$, the rescaled growth rate $\tilde c_{1i}$ for BC12 reduces as the dilute limit is approached, while the opposite is true for BC3. {\color{black}In the inset of each panel, we apply a suitable rescaling so that the curves tend to collapse in the dilute limit, $\beta\to 1$. Remarkably, for BC12, the growth rate $\tilde c_{1i}$ scales as $\tilde k^{-1}$, while for BC3, $\tilde c_{1i}$ scales as $\tilde k^{-1/3}$. On the other hand, both conditions share the same horizontal scaling parameter, $\tilde k(1-\beta)^{3/2}$, implying the fundamental scaling relation $\tilde k\sim(1-\beta)^{-3/2}$ in the dilute limit.
  Motivated  by \cite{Dong2022}, in which the centre-mode instability system is further reduced by taking the dilute limit, $\beta\to 1$, we explore PDI in the dilute limit in the following subsection, referring this specific scaling as regime III.}
  \subsection{Regime III analysis}
\label{sec:regime_II_dilute}
\subsubsection{Formulation for regime III}
Now we take $\beta\to 1$, and introduce $\sigma=(1-\beta)/\beta\ll 1$. {\color{black}Then, from the scaling relation revealed by figure \ref{fig:growth_lower}, we set $\tilde k\sim \sigma^{-3/2}$ and introduce}
\begin{equation}
\breve \sigma=\tilde k^{2/3}\sigma=O(1).
\end{equation}
From a scaling estimate, we find that the thickness of LDL becomes $\tilde Y\sim \tilde k^{-1/3}$. Thus,  we introduce a local coordinate,
\begin{equation}
\breve Y=\tilde k^{1/3}\tilde Y=O(1),
\end{equation}
and expand the perturbation field as
\refstepcounter{equation}
$$
(\hat u_0,\hat v_0,\hat p_0)=\Big(\breve U_0,Wi^{-1}\tilde k^{2/3}\breve V_0,Wi^{-3/2}\tilde k\breve P_0\Big)+\cdots.\eqno{(\theequation a)}
\label{eq:dilute_LDL}
$$
$$
 (\hat c_{110},\hat c_{120},\hat c_{220})=\Big(Wi^{-1/2}\tilde k^{1/3}\breve C_{110},Wi^{-3/2}\tilde k\breve C_{120},Wi^{-5/2}\tilde k\breve C_{220}\Big)
+\cdots.\eqno{(\theequation b)}
$$

Examining the boundary conditions (\ref{eq:LB_BC}c) for BC12, we find that only when $\tilde c_1\sim \tilde k^{-1}$ can these boundary conditions appear at leading-order impact. This scaling is selected as the magnitude of the phase speed correction for BC12. In contrast, for BC3, the boundary conditions (\ref{eq:LB_BCiii}) always appear in leading order, and to construct an eigenvalue system, the phase speed correction must appear in the governing equations with $\tilde c_1\sim \tilde k^{-1/3}$. These estimates also align with the numerical observations from figure \ref{fig:growth_lower} by applying $\beta\to 1$.
Therefore, we express the phase speed expansion in a unified form,
\begin{equation}
c=U(-1)+\epsilon^{1/2}Wi^{1/2}\tilde k^{\alpha}\breve c_1
+\cdots,
\label{eq:phase_speed2}
\end{equation}
where $\alpha=-1$ for BC12 and =-1/3 for BC3.

Substituting the above expansions into (\ref{eq:equation_II}) and collecting the leading-order terms, we obtain
\refstepcounter{equation}
$$
\breve U_0''+\breve \sigma (\ri\breve C_{110}+\breve C_{120}')=0,\quad \breve V_0''+\breve \sigma \ri\breve C_{120}-\breve P_0'=0,
\eqno{(\theequation a,b)}
$$
$$
 \ri\breve U_0+\breve V_0'=0,\quad \breve C_{110}''=\breve S\breve C_{110}-2(2\ri \lambda^2\breve U_0+\lambda \breve C_{120}),
\eqno{(\theequation c,d)}
\label{eq:equation_IV}
$$
$$
\breve C_{120}''= \breve S\breve C_{120}-2\ri \lambda^2\breve V_0,\quad\breve C_{220}''=\breve S\breve C_{220}-2\ri\lambda\breve V_0.\eqno{(\theequation e,f)}
$$
where $\breve S=\ri\lambda \breve Y$ for BC12 and $\breve S=\ri(\lambda \breve Y-\breve c_1)$ for BC3.

For BC12, the phase speed correction $\breve c_1$ does not appear in the equation system (\ref{eq:equation_IV}); it only appears in the  wall boundary conditions,
\refstepcounter{equation}
$$
\breve U_0(0)=\breve V_0(0)=\breve C_{120}(0)=\breve C_{220}(0)=0,\quad {\breve S_0} \breve C_{110}(0)={2\lambda\breve U_0'(0)}.
\eqno{(\theequation a,b,c,d,e)}
\label{eq:BC_LDL_dilute}
$$
where $\breve S_0=-\ri \breve c_1+1$. Note that the boundary condition for $\tilde c_{120}$, (\ref{eq:LB_BC}d), comes to the second order. In contrast, $\breve c_1$ appears in the leading-order equation system for BC3, and the boundary conditions are
\refstepcounter{equation}
$$
\breve U_0(0)=\breve V_0(0)=\breve C_{110}'(0)=\breve C_{120}'(0)=\breve C_{220}'(0)=0.
\eqno{(\theequation a,b,c,d,e)}
\label{eq:BC_LDL_dilute_BC3}
$$

From a scaling analysis, we obtain that the UDL thickness in this regime becomes $\bar Y\sim \tilde k^{-1/3}$. Following the same mathematical procedure, we obtain the upper boundary conditions of the LDL perturbations from matching with the UDL,
\begin{equation}
\Big(\breve U_0', \breve C_{110},\breve C_{120}', \breve C_{220}'\Big)\to 0,\quad \breve P_0\to -2\breve\sigma\lambda \breve U_0\quad \mbox{as }\breve Y\to \infty.
\label{eq:BC_LDL_upper_dilute}
\end{equation}

{\color{black}
Eliminating $\hat u$, $\hat p$ and $\hat{\bm c}$  governing equations (\ref{eq:equation_IV}), we obtain a seventh-order ordinary differential equation (ODE) for $\breve V_0$,
%\begin{equation}
%\breve V_0^{(7)}-2 \breve S \breve V_0^{(5)}-2\ri \lambda \breve V_0^{(4)}+(\breve S^2+2\lambda^2\breve \sigma)\breve V_0'''-2\breve S\lambda^2\breve\sigma\breve V_0'+2\ri\lambda^3\breve \sigma\breve V_0=0.
%\label{eq:ODE_7}
%\end{equation}
%It can be recast to
\begin{equation}
\Big[{\cal D}^4-\breve S{\cal D}^2+\ri\lambda {\cal D}+2\lambda^2\breve\sigma\Big][{\cal D}^3-\breve S{\cal D}+\ri\lambda]\breve V_0=0,
\label{eq:ODE_7_1}
\end{equation}
where ${\cal D}:=d/d\breve Y$. The no-slip, non-penetration boundary conditions are readily rewritten as
\begin{equation}
\breve V_0(0)=\breve V_0'(0)=0.\label{eq:regime_III_BC0}
 \end{equation}
 To convert the boundary conditions of the conformation tensor, we need to observe the near-wall behaviours of the perturbation field.

Let us expand the perturbation field in the near-wall region,
\refstepcounter{equation}
 $$
 \breve C_{110}=\bar a_0+\bar a_1\breve Y+\bar a_2\frac{\breve Y^2}2+\bar a_3\frac{\breve Y^3}6+\cdots,\eqno{(\theequation a)}$$
 $$\breve C_{120}=\bar b_0+\bar b_1\breve Y+\bar b_2\frac{\breve Y^2}2+\bar b_3\frac{\breve Y^3}6+\bar b_4\frac{\breve Y^4}{24}\cdots,\eqno{(\theequation b)}\label{eq:near_wall_perturbation}
 $$
 where $(\bar a_0,\bar a_1,\bar a_2,\bar a_3,\bar b_0,\bar b_1,\bar b_2,\bar b_3,\bar b_4)$ are constants.

 For BC12, we have $\breve S(0)=\breve C_{120}(0)=0$, then, it is easy to derive that $\bar a_2=\bar b_0=\bar b_2=\bar b_3=0$. Substituting (\ref{eq:near_wall_perturbation}) into (\ref{eq:equation_IV}) and (\ref{eq:BC_LDL_dilute}e), we obtain a set of algebraic equations,
 \refstepcounter{equation}
 $$
 \ri\breve V_0^{(3)}(0)+\breve \sigma(\ri\bar  a_0+\bar b_1)=0,\quad \ri \breve V_0^{(4)}(0)+\ri\breve \sigma \bar a_1=0,\quad  \ri\breve V_0^{(5)}(0))=0,\eqno{(\theequation a,b,c)}
 $$
 $$
  \ri\breve V_0^{(6)}(0)+\breve \sigma(\ri \bar a_3+\bar b_4)=0,\quad \bar a_3=\ri\lambda \bar a_0+4\lambda^2   \ri\breve V_0''(0)-2\lambda \bar b_1,\eqno{(\theequation d,e)}
 $$
 $$
 \bar b_4=2\ri\lambda\bar  b_1-2\ri\lambda^2 \breve V_0''(0),\quad (1-\ri\breve c_1)\bar a_0=2\ri\lambda \breve V_0''(0).   \eqno{(\theequation f)}
 $$
Eliminating the $(\bar a_0,\bar a_1,\bar a_3,\bar b_1,\bar b_4)$, we obtain the conformation boundary conditions to (\ref{eq:ODE_7_1}) for BC1,
\begin{equation}
\breve V_0^{(5)}(0)=0,\quad (1-\ri\breve c_1)\breve V_0^{(6)}(0)-2\ri\breve c_1 \lambda^2\breve \sigma \breve V_0''(0)=0.\label{eq:regime_III_BC1}
\end{equation}

For BC3, following the same procedure, we derive
\begin{equation}
\breve V_0^{(5)}(0)+\ri\breve c_1\breve V_0^{(3)}(0)-\frac{\lambda\breve V_0^{(4)}(0)}{\breve c_1}=0,\quad \breve V_0^{(6)}(0)-\ri\lambda \breve V_0^{(3)}(0)+\ri\breve c_1\breve V_0^{(4)}(0)+2\lambda^2\breve \sigma\breve V_0''(0)=0.\label{eq:regime_III_BC2}
\end{equation}

Additionally, the upper boundary conditions (\ref{eq:BC_LDL_upper_dilute}) can be converted to
\begin{equation}
(\breve V_0'',\breve V_0^{(3)},\breve V_0^{(4)})\to 0\quad\mbox{as }\breve Y\to \infty.
\label{eq:regime_III_BC4}
\end{equation}
}

\subsubsection{Numerical results for regime III}

\begin{figure}
\begin{center}
  \includegraphics[width = 0.48\textwidth]{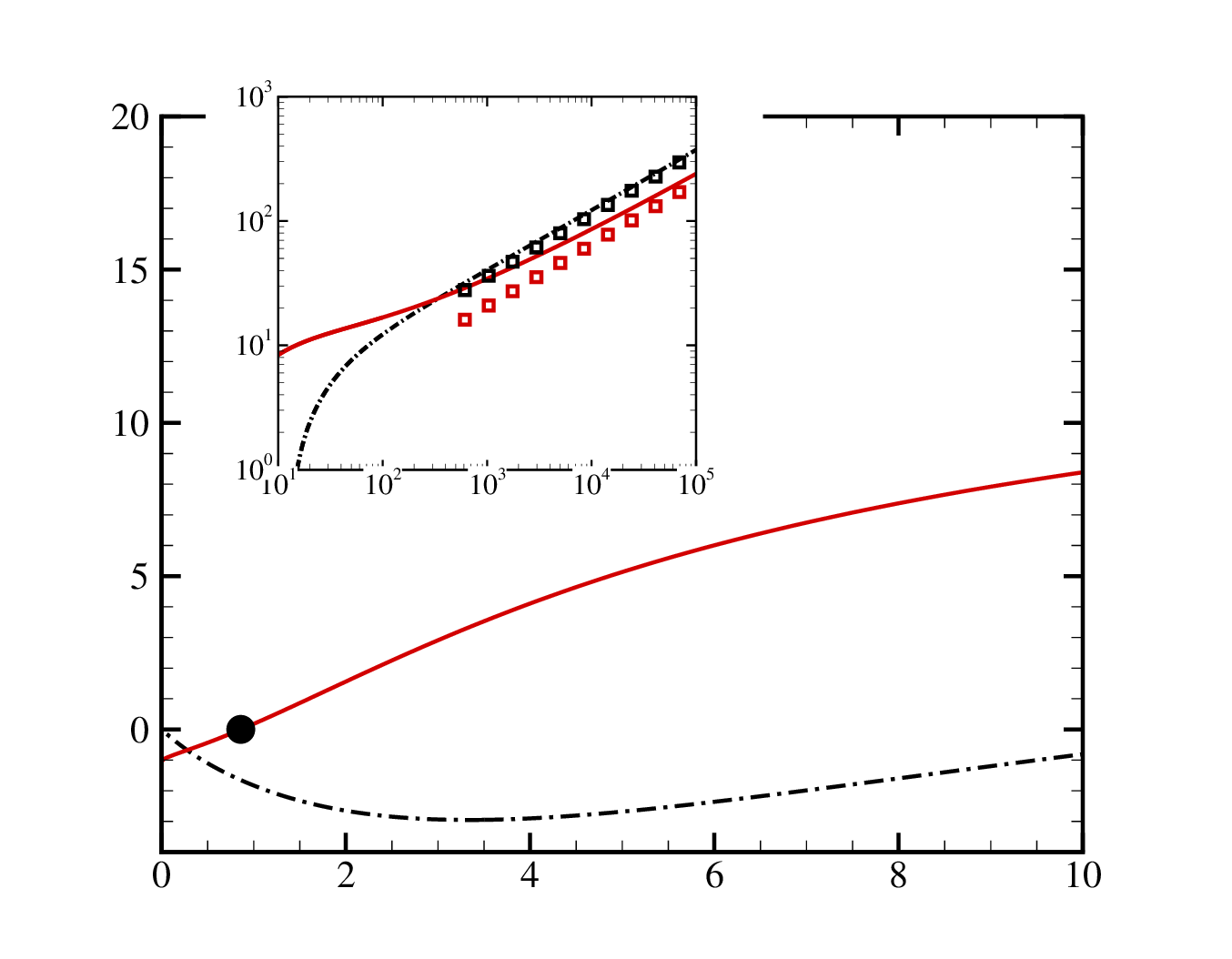}
  \put(-182,50){\rotatebox{90}{$\breve c_1=\tilde k\tilde c_1$}}  \put(-100,2){$\breve \sigma$}\put(-182,120){(a)}
  \includegraphics[width = 0.48\textwidth]{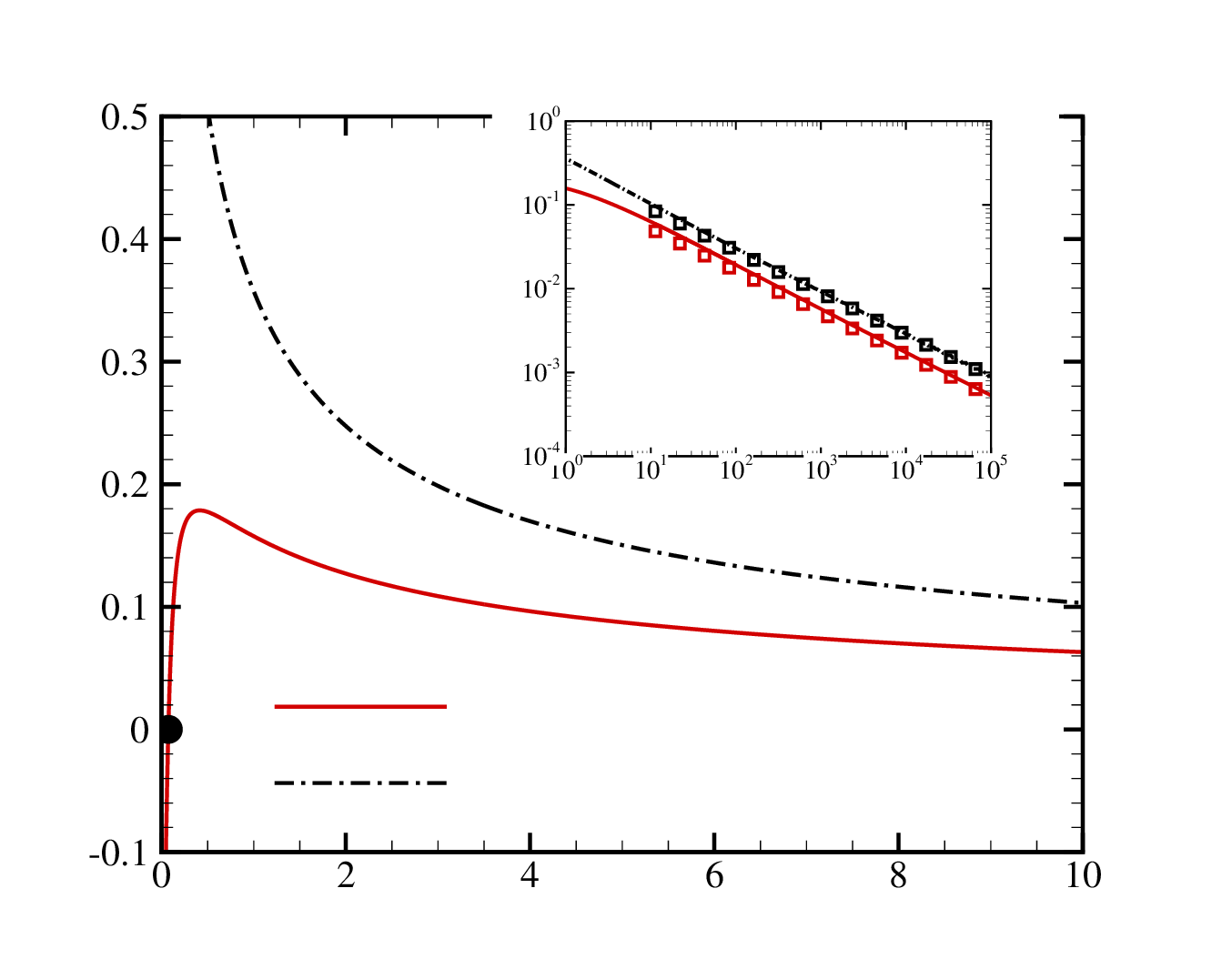}
  \put(-182,50){\rotatebox{90}{$\breve c_1=\tilde k^{1/3}\tilde c_1$}} \put(-100,2){$\breve \sigma$} \put(-115,35){$\breve c_{1i}$}\put(-115,25){$\breve c_{1r}$}\put(-182,120){(b)}
  \caption{Dependence on $\breve \sigma$ of the real and imaginary parts of $\breve c_{1}$ for regime III with BC12 (a) and BC3 (b), where $\lambda=2$ (PPF). The circle in each panel marks the neutral point. {\color{black}The inset shows the log-log plot to illustrate  the large-$\breve\sigma$ asymptote, with the squares in (a) and (b) denoting the asymptotic predictions by (\ref{eq:dispersion_RegimeIV_BC12}) and (\ref{eq:dispersion_RegimeIV_BC3}), respectively}.}\label{fig:dilute_growth-0}
\end{center}
\end{figure}
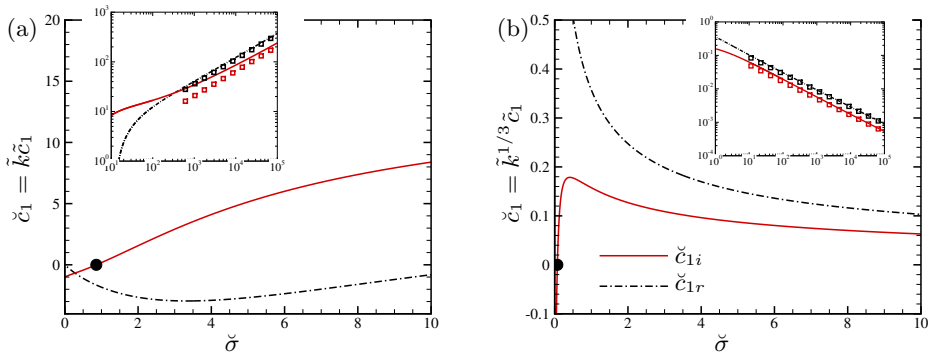
In this regime, the eigenvalue system is characterised by (\ref{eq:ODE_7_1})), complemented by the   boundary conditions (\ref{eq:regime_III_BC0}), (\ref{eq:regime_III_BC1}) (or (\ref{eq:regime_III_BC2})) and (\ref{eq:regime_III_BC4}). The eigenvalue $\breve c_1$ is solely governed by a single parameter, $\breve \sigma$, whose real and imaginary parts for PPF are plotted in  figure \ref{fig:dilute_growth-0}. For BC12, the PDI mode becomes unstable when $\breve \sigma$ exceeds 0.86, with its rescaled growth rate $\breve c_{1i}$ increasing monotonically with $\breve \sigma$.  The rescaled phase speed correction $\breve c_{1r}$ initially declines with increasing $\breve \sigma$, although a reversal occurs when $\breve \sigma>3.3$. In contrast, for BC3, the neutral point emerges at $\breve \sigma=0.0735$, with the rescaled growth rate peaking at $\breve \sigma=0.416$; the rescaled phase speed decreases monotonically with increasing $\breve \sigma$. It is crucial to emphasise that despite  the magnitude of the rescaled $\breve c_1$ for BC12 is greater than that for BC3, it does not mean that the PDI for BC12 is more unstable, due to the differing scaling laws for the two phase speed corrections, varying by a   factor $\tilde k^{-2/3}$.
{\color{black}The insets in figures \ref{fig:dilute_growth-0}-(a) and (b) further present log-log plots, clearly revealing an excellent scaling relationship between $\breve c_1$ and $\breve \sigma$ in the limit of $\breve\sigma\to \infty$. Motivated by this observation, we proceed to simplify the instability system further by taking the asymptotic limit of removing the only remaining controlling parameter. This approach leads to a parameter-free system, allowing the analytical derivation of solutions. We will investigate this simplified system in the following subsection, referring to it as Regime IV.}

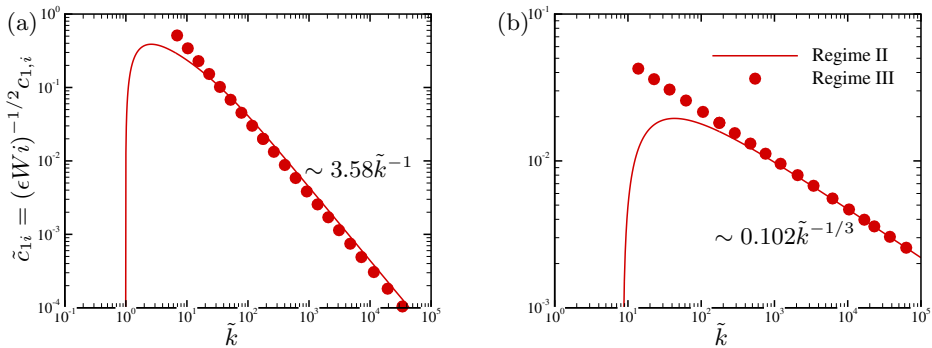
\begin{figure}
\begin{center}
  \includegraphics[width = 0.48\textwidth]{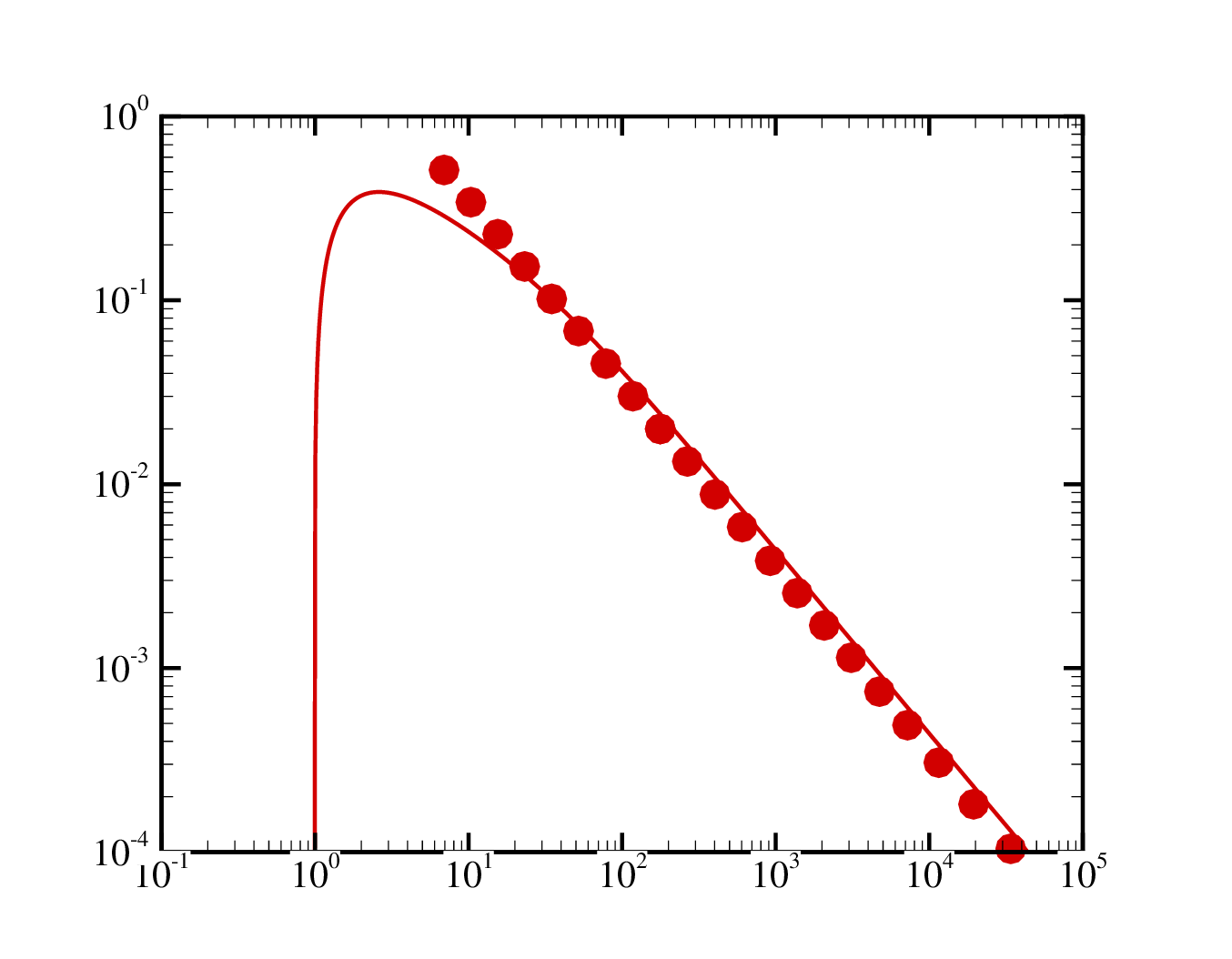}
  \put(-182,30){\rotatebox{90}{ $\tilde c_{1i}=(\epsilon Wi)^{-1/2}c_{1,i}$}}
  \put(-100,2){$\tilde k$}\put(-182,120){(a)}\put(-70,65){$\sim3.58\tilde k^{-1}$}
    \includegraphics[width = 0.48\textwidth]{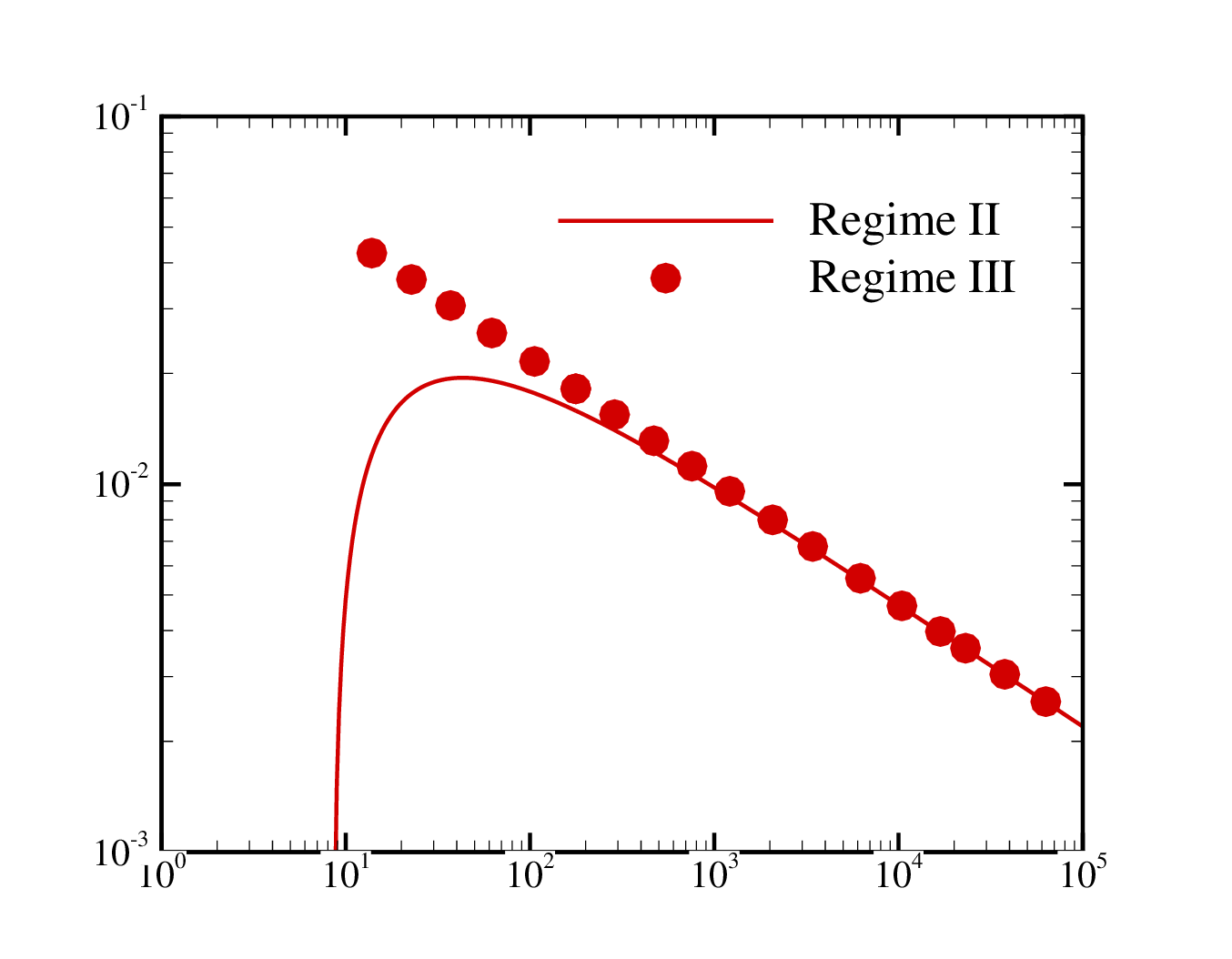}\put(-182,120){(b)}
     \put(-100,2){$\tilde k$}\put(-100,40){$\sim0.102\tilde k^{-1/3}$}
  \caption{Comparison of the rescaled growth rate $\tilde c_{1i}$ in the dilute and high-$\tilde k$ limits  obtained by regime II and III calculations, where $\breve \sigma=\tilde k^{2/3}\sigma=3.5$ and $\lambda=2$ (PPF). (a): BC12; (b): BC3. .}\label{fig:dilute_growth}
\end{center}
\end{figure}
The curves in  figure \ref{fig:dilute_growth} illustrate  the variation with $\tilde k$ of the rescaled growth rate $\tilde c_{1i}$ predicted by the regular-concentration regime II discussed in $\S$\ref{sec:regime_II}. Specifically, we choose $\breve\sigma=\sigma/\tilde k^{2/3}=3.5$  to match the dilute-limit scaling, where $\tilde k\to \infty$ signifies the dilute limit $\sigma\to 0$. For both BC12 and BC3, a sharp increase in $\tilde c_{1i}$ with $\tilde k$ is evident near the lower-branch neutral point, followed by a gradual decrease with a scaling $\tilde c_{1i}\sim \tilde k^{-1}$,  which represents the dilute-limit scaling. The   neutral wavenumber $\tilde k$ is found to be greater for BC3. In the high-$\tilde k$ limit, the regime-II calculations of $\tilde c_{1i}$ converge toward the predictions for regime III, as indicated by the circles, suggesting a natural connection between the two regimes. When comparing the results of the two boundary conditions, it is evident that, although the growth rate for BC12 at a moderate $\tilde k$   is higher than that for BC3, it decreases at a much higher rate as $\tilde k$ increases.

\begin{figure}
\begin{center}
  \includegraphics[width = 0.48\textwidth]{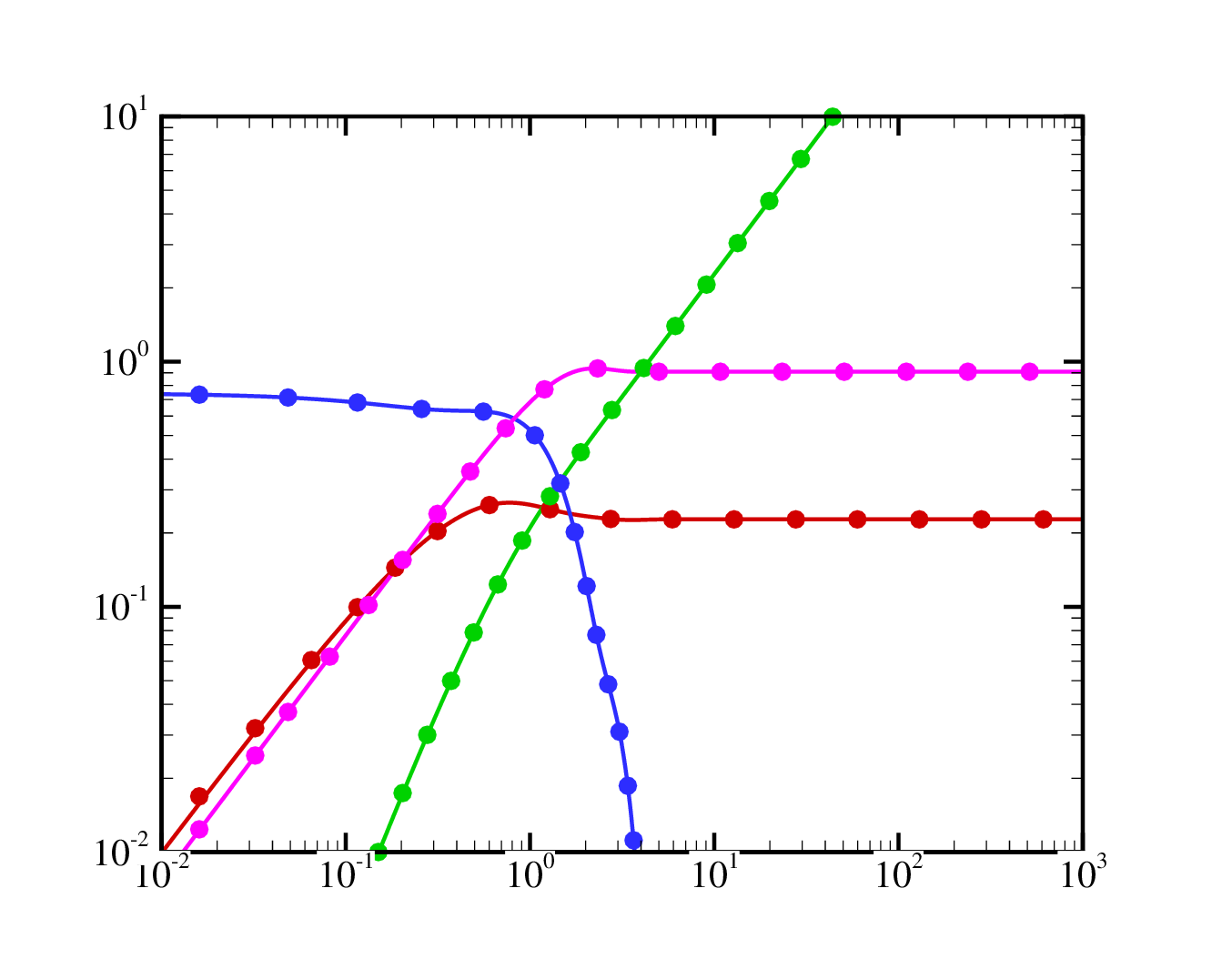}
  \put(-182,30){\rotatebox{90}{ $\breve U_0,\ \breve V_0,\ \breve C_{110},\ \breve C_{120}$}}
  \put(-95,2){$\breve Y$}\put(-182,120){(a)}
   \put(-120,30){$|\breve V_0|$}    \put(-150,90){$|\breve C_{110}|$}
      \put(-60,55){$|\breve U_0|$}    \put(-60,95){$|\breve C_{120}|$}
    \includegraphics[width = 0.48\textwidth]{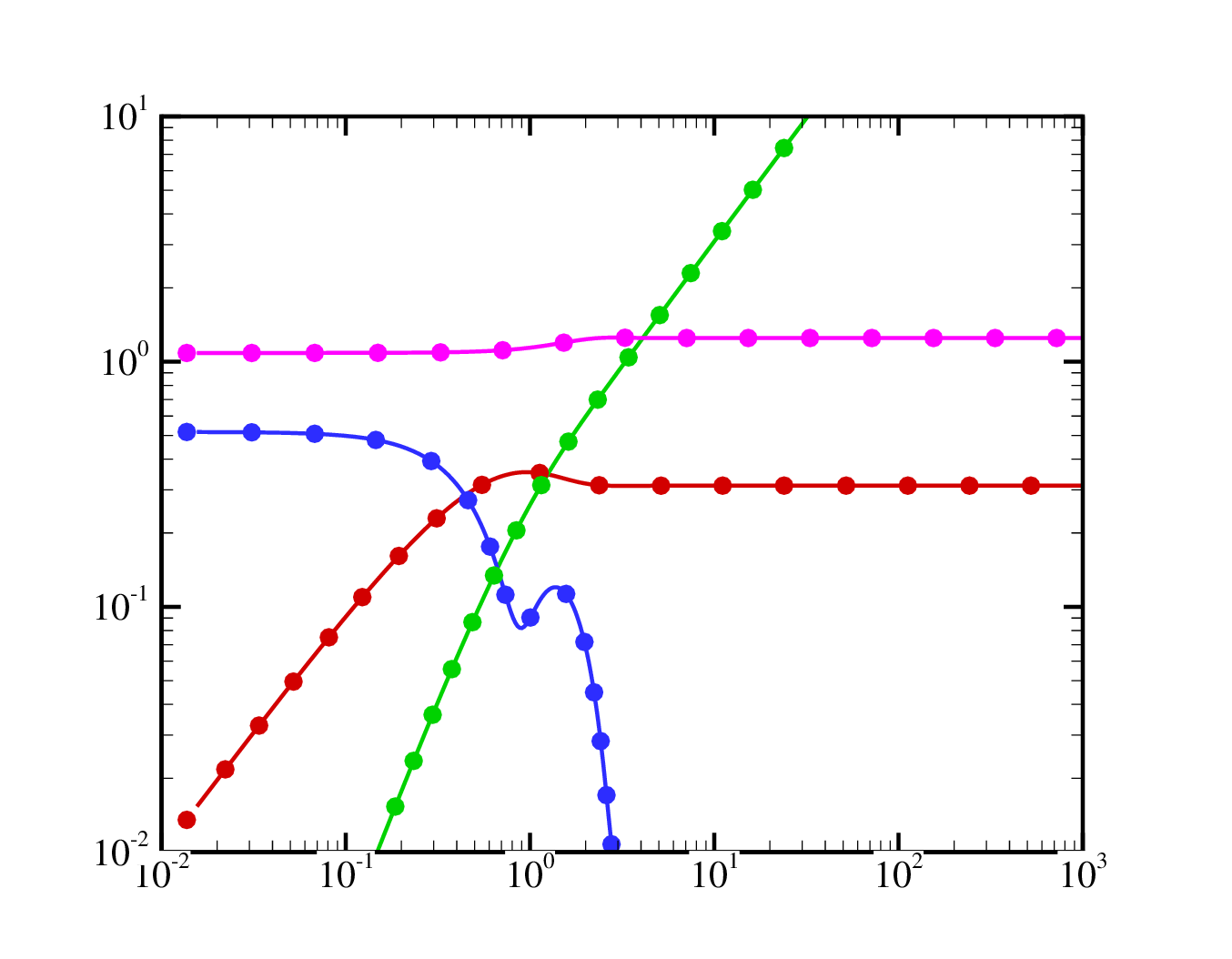}\put(-182,120){(b)}
   \put(-95,2){$\breve Y$}
    \put(-118,30){$|\breve V_0|$}    \put(-150,70){$|\breve C_{110}|$}
      \put(-60,75){$|\breve U_0|$}    \put(-60,100){$|\breve C_{120}|$}
  \caption{  Comparison of  the eigenfunctions between the regime-II and III calculations in the dilute limit, where $\breve \sigma=3.5$ and $\lambda=2$ (PPF). For regime II, we choose  $\beta=0.997$ ($\sigma=0.003$) and $\tilde k=3.97\times 10^4$. Solid lines: regime II; circles: regime III. (a): BC12; (b): BC3. We choose $\breve U_0'(0)=1$ for normalisation.}\label{fig:dilute_eigf}
\end{center}
\end{figure}
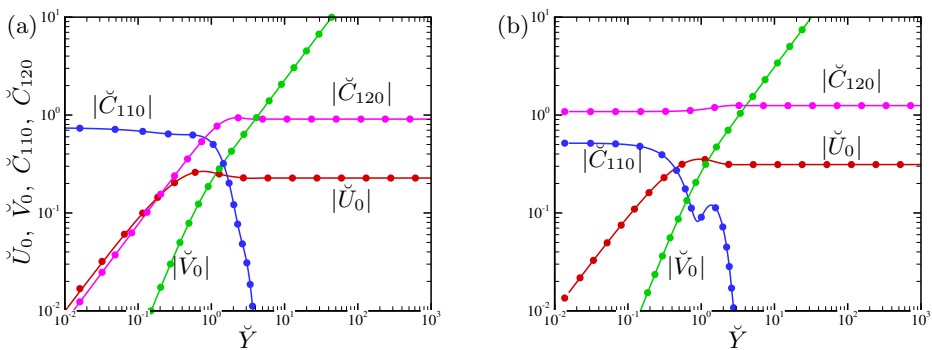
In figure \ref{fig:dilute_eigf}, we compare the norms of the eigenfunctions obtained from calculations based on Regimes II and III. The dilute limit with $\beta=0.997$ (or $\sigma=0.003$) and $\breve \sigma=3.5$ is considered.  Both boundary conditions, BC12 and BC3, exhibit similar near-wall behaviors of the perturbation velocity field, specifically: $\breve V_0\sim \breve Y^2$, $\breve U_0\sim \breve Y$. However, the near-wall conformation perturbations show evident discrepancy: $\breve C_{120}$ for BC12 is almost proportional to $\breve Y$,  while for BC3, it exhibits a zero gradient.
The alignment of the calculations from the two regimes further validates the accuracy of the predictions made by our asymptotic analyses.

{\color{black}
\subsection{Regime IV analysis}
\label{sec:regime_IV}}
As shown in {\color{black}figure \ref{fig:dilute_growth-0}}, the system (\ref{eq:ODE_7_1}) becomes unstable when  $\breve\sigma$ exceeds a certain threshold, {\color{black} marked by a dot, and a clear scaling relation between $\breve c_{1}$ and $\breve\sigma$ is evident as $\breve\sigma\to \infty$.
To facilitate an analytical derivation of the dispersion relation by further simplifying the instability formulation, we perform an asymptotic analysis in this subsection by taking $\breve \sigma$ to be asymptotically large. In this regime, no controlling parameter remains.
 A scaling estimate determines that the LDL further splits into two distinct layers, a bulk sublayer with $\breve Y=O(1)$ and a wall sublayer with $\breve Y\sim \breve \sigma^{-1/4}$.
\subsubsection{Regime-IV solution for BC12}
\label{sec:regime_IV_BC12}
For BC12, figure \ref{fig:dilute_growth-0}(a) reveals that both the real and imaginary parts of $\breve c_1$ increase with $\breve\sigma$, and fitting the data shows that $\breve c_{1}\sim \breve \sigma^{1/2}$ as $\breve\sigma\to \infty$. Therefore, we expand the phase speed as
\begin{equation}
\breve c_{1}=\breve\sigma^{1/2}(\breve c_{10}+\breve\sigma^{-1/4}\breve c_{11}+\cdots).
\label{eq:expansion_c1}
\end{equation}

In the bulk sublayer, the solution is expanded as
\begin{equation}
\breve V_0=\sum_{l=0}^{\cdots}\breve\sigma^{-l/4}\breve V_{0l}.
\label{eq:V0_expansion_bulk}
\end{equation}
Substituting into (\ref{eq:ODE_7_1}), we obtain the governing equations to the first three orders,
\refstepcounter{equation}
$$
{\cal L}\breve V_{00}=0,\quad {\cal L}\breve V_{01}=0,\quad {\cal L}\breve V_{02}=0,
\eqno{(\theequation a,b,c)}
\label{eq:bulk_sublayer_EQ}
$$
where ${\cal L}:=[{\cal D}^3-\ri\lambda \breve Y{\cal D}+\ri\lambda]$.
Excluding the exponentially growing component, we derive the solutions:
\begin{equation}
\vspace{.2cm}\breve V_{0l}= d_{1l}\tilde\xi+d_{2l}\Big(\tilde\xi\int_\infty^{\tilde\xi}\mbox{Ai}(\tilde\xi)d\tilde\xi-\mbox{Ai}'(\tilde\xi)\Big)
,\quad\mbox{with }l=0,1,2,
\label{eq:solution_ODE7}
\end{equation}
where Ai denotes the Airy function of the first  kind \citep{Abramowitz1964}, $\tilde \xi=(\ri\lambda)^{1/3}\breve Y$, and $(d_{1l},d_{2l})$ are constants to be determined. Here, the exponentially growing part is dropped to satisfy the matching condition with the UDL solution. As the wall is approached, the asymptotic behaviour of $\breve V_{0l}$ is
\begin{equation}
\breve V_{0l}\to {\cal V}_{l0}+{\cal V}_{l1}\breve Y+{\cal V}_{l2}\breve Y^2/2+\cdots\quad\mbox{as }\breve Y\to 0,
\label{eq:bulk_sublayer_V0}
\end{equation}
where
\begin{equation}
{\cal V}_{l0}=-d_{2l}\mbox{Ai}'(0),\quad{\cal V}_{l1}=(d_{1l}-d_{2l}/3)(\ri\lambda)^{1/3},\quad {\cal V}_{l2}=d_{2l}\mbox{Ai}(0)(\ri\lambda)^{2/3}.\label{eq:coefficient_Vl0}
\end{equation}
 For a non-trivial expansion, we assume that $d_{10}$ and $d_{20}$ cannot both be zero.
Since the conditions in (\ref{eq:regime_III_BC0}) are not satisfied for this case, we need to consider the underneath wall sublayer.

In the wall sublayer, we introduce $\eta=\breve\sigma^{1/4}\breve Y=O(1)$, and expand the solution as
\begin{equation}
\breve V_0=\sum_{l=0}^{\cdots}\breve\sigma^{-l/4}\breve v_l.
\label{eq:wall_sublayer_v0}
\end{equation}
The governing equations to the first three orders read
\refstepcounter{equation}
 $$
(\bar{\cal D}^4+2\lambda^2)\bar{\cal D}^3\breve v_{l}=0\quad\mbox{with }l=0,1,2,\eqno{(\theequation)}\label{eq:Eq_v_layer_II}
$$
where $\bar{\cal D}:=d/d\eta$.
The wall boundary conditions read
\refstepcounter{equation}
$$
\breve v_{l}(0)=\breve v_{l}'(0)=\breve v_{l}^{(5)}(0)=0\quad\mbox{with }l=0,1,2,
\eqno{(\theequation a,b,c)}\label{eq:BC_wall1}
$$
\refstepcounter{equation}
 $$
\breve v_{0,1}^{(6)}(0)+2\lambda^2\breve v_{0,1}''(0)=0,\quad \breve v_{2}^{(6)}(0)+2\lambda^2\breve v_{2}''(0)=\breve v_0^{(6)}(0)/(\ri\breve c_{10}).
\eqno{(\theequation a,b)}\label{eq:BC_wall}
$$

The leading-order solution is
\begin{equation}
\breve v_0={\cal V}_{00}+a_0\re^{\bar\lambda_1\eta}+b_0\re^{\bar\lambda_2\eta},
\label{eq:wall_v0}
\end{equation}
where $\bar\lambda_{1,2}=(-1\pm\ri)(2\lambda^2)^{1/4}/\sqrt{2}$, and $(a_0,b_0)$ are constants to be determined. To match the bulk sublayer, the exponentially growing complementary solution is excluded. Applying the boundary conditions at the wall, it is determined that
\refstepcounter{equation}
$$
{\cal V}_{00}+a_0+b_0=0,\quad \bar\lambda_1a_0+\bar\lambda_2b_0=0,\quad \bar\lambda_1^5 a_0+\bar\lambda_2^5b_0=0,\quad \bar\lambda_1^6 a_0+\bar\lambda_2^6b_0+2\lambda^2(\bar\lambda_1^2 a_0+\bar\lambda_2^2b_0)=0.
\eqno{(\theequation a,b,c,d)}\label{eq:algebraic_0}
$$
The equation (\ref{eq:algebraic_0}d) is satisfied automatically, and the conditions (\ref{eq:algebraic_0}b) and (\ref{eq:algebraic_0}c) are equivalent. Then, the solutions to (\ref{eq:algebraic_0}a,b) are
\begin{equation}
a_0=\frac{-\bar\lambda_2{\cal V}_{00}}{\bar\lambda_2-\bar\lambda_1},\quad b_0=\frac{\bar\lambda_1{\cal V}_{00}}{\bar\lambda_2-\bar\lambda_1}.
\label{eq:a0b0}
\end{equation}

Considering the matching condition, the second-order solution is
\begin{equation}
\breve v_1={\cal V}_{10}+{\cal V}_{01}\eta+a_1\re^{\bar\lambda_1\eta}+b_1\re^{\bar\lambda_2\eta}.
\label{eq:wall_v1}
\end{equation}
 Applying the wall boundary conditions, we obtain that ${\cal V}_{01}$ must be zero (the solutions for $a_1$ and $b_1$ are omitted for brevity because they do not affect the determination of the dispersion relation).

The third-order solution is
\begin{equation}
\breve v_2={\cal V}_{20}+{\cal V}_{11}\eta+{\cal V}_{02}\eta^2/2+a_2\re^{\bar\lambda_1\eta}+b_2\re^{\bar\lambda_2\eta}.
\label{eq:wall_v2}
\end{equation}
Applying the wall boundary condition (\ref{eq:BC_wall}b), we obtain
\refstepcounter{equation}
$$
{\cal V}_{20}+a_2+b_2=0,\quad {\cal V}_{11}+\bar\lambda_1a_2+\bar\lambda_2b_2=0,\quad \bar\lambda_1^5 a_2+\bar\lambda_2^5b_2=0,\eqno{(\theequation a,b,c)}$$
 $$\bar\lambda_1^6 a_2+\bar\lambda_2^6b_2+2\lambda^2(\bar\lambda_1^2 a_2+\bar\lambda_2^2b_2+{\cal V}_{02})=(\bar\lambda_1^6a_0+\bar\lambda_2^6b_0)/(\ri\breve c_{10}).
\eqno{(\theequation d)}\label{eq:algebraic_2}
$$
Combining (\ref{eq:a0b0}) and (\ref{eq:algebraic_2}d), and considering  ${\cal V}_{02}/{\cal V}_{00}=-(\ri\lambda)^{2/3}\mbox{Ai}(0)/\mbox{Ai}'(0)$, we  derive
\begin{equation}
\breve c_{10}  =-\frac{a_0\bar\lambda_1^2+b_0\bar\lambda_2^2}{\ri {\cal V}_{02}}=\left\{\begin{array}{ll}
1.1249+0.6495\ri&\mbox{for }\lambda=2;\\
 0.8929 + 0.5155\ri&\mbox{for }\lambda=1.   \end{array}
\right.
\label{eq:dispersion_RegimeIV_BC12}
\end{equation}
Here, the positive imaginary part corresponds to the unstable PDI. The asymptotic prediction (\ref{eq:dispersion_RegimeIV_BC12}) is represented by the squares in figure \ref{fig:dilute_growth-0}-(a), and excellent agreement with the numerical solutions is achieved in the limit of $\breve\sigma\to \infty$.

\subsubsection{Regime-IV solution for BC3}
\label{sec:regimeIV_BC3}
For BC3, on the contrary, the complex phase speed $\breve c_1$ decreases with $\breve \sigma$ at a rate of $\breve c_1\sim \breve \sigma^{-1/2}$. Consequently, the expansion in (\ref{eq:expansion_c1}) is changed to
\begin{equation}
\breve c_{1}=\breve\sigma^{-1/2}(\breve c_{10}+\breve\sigma^{-1/4}\breve c_{11}+\cdots).
\label{eq:expansion_c1_BC3}
\end{equation}

In the bulk sublayer, the  expansion of $\breve V_0$ remains  identical  to that given by (\ref{eq:V0_expansion_bulk}), and the first two orders of the governing equations   are unchanged from  (\ref{eq:bulk_sublayer_EQ}a,b). Therefore, the solutions for $\breve V_{00}$ and $\breve V_{01}$ are the same as those presented in  (\ref{eq:solution_ODE7}), with the asymptotic behaviours in the limit of $\breve Y\to 0$ following directly from (\ref{eq:bulk_sublayer_V0}) with $l=0,1$.

In the wall sublayer, the leading-order solution takes the same form as in (\ref{eq:wall_v0}). Applying  boundary conditions (\ref{eq:algebraic_0}a,b), $\breve v_0^{(4)}=0$ and $\breve v_0^{(6)}+2\lambda^2\breve v_0''=0$,  we find that ${\cal V}_{00}=0=a_0=b_0$, implying $\breve v_0=0$. Consequently, using (\ref{eq:coefficient_Vl0}), we obtain $d_{20}=0$, ${\cal V}_{01}=d_{10}(\ri\lambda)^{1/3}$, and thus, $\breve V_{00}=d_{10}(\ri\lambda)^{1/3}\breve Y$.

Following the same procedure, we can determine the coefficients to second-order solution (\ref{eq:wall_v1}):
${\cal V}_{10}=0$ and
\begin{equation}
a_1=-b_1={\cal V}_{01}/(\bar\lambda_2-\bar\lambda_1).
\label{eq:a1b1_BC3}
\end{equation}
This also implies $d_{21}=0$ and $\breve V_{01}=d_{11}(\ri\lambda)^{1/3}\breve Y$.

However, due to the presence of $-\ri\breve c_1$ in $\breve S$ for BC3, the third-order governing equations include the impact of the rescaled phase speed, expressed as
\begin{equation}
{\cal L}\breve V_{02}=-\ri\breve c_{10}{\cal D}\breve V_{00}=-(\ri\lambda)^{1/3}\ri\breve c_{10}d_{10}.
\end{equation}
The solution can be expressed as
\begin{equation}
\breve V_{02}=-\frac{\ri\breve c_{10}d_{10}}{(\ri\lambda)^{2/3}}+
d_{12}\tilde\xi+d_{22}\Big(\tilde\xi\int_\infty^{\tilde\xi}\mbox{Ai}
(\tilde\xi)d\tilde\xi-\mbox{Ai}'(\tilde\xi)\Big).
\end{equation}
Thus, the coefficients in the asymptotic behaviour (\ref{eq:bulk_sublayer_V0}) for $l=2$ read
\begin{equation}
{\cal V}_{20}=-\frac{\ri\breve c_{10}d_{10}}{(\ri\lambda)^{2/3}}-d_{22}\mbox{Ai}'(0),\quad{\cal V}_{21}=\Big(d_{12}-d_{22}/3\Big)(\ri\lambda)^{1/3},\quad
 {\cal V}_{22}=d_{22}\mbox{Ai}(0)(\ri\lambda)^{2/3}.\label{eq:asymptotic_V2}
\end{equation}

 In the wall sublayer, the third-order governing equation remains identical to  (\ref{eq:Eq_v_layer_II}), and so the solution can be expressed in the same form as   (\ref{eq:wall_v2}). However, the coefficients $a_2$ and $b_2$ change, since the lower boundary conditions  are now modified to
 \refstepcounter{equation}
$$
{\cal V}_{20}+a_2+b_2=0,\quad {\cal V}_{11}+\bar\lambda_1a_2+\bar\lambda_2 b_2=0,\quad \lambda {\cal V}_{20}=\breve c_{10}{\cal V}_{01}.\eqno{(\theequation a,b,c)}\label{eq:BC3_BC3}
$$

The fourth-order solution $\breve v_3$ does not provide valuable information for determination of the dispersion relation, and so we directly move to the fifth order. The governing equation is
\begin{equation}
(\bar{\cal D}^4+2\lambda^2)\bar{\cal D}^3\breve v_{4}=-2\ri\lambda^3{\cal V}_{10}-2\ri\lambda^3\Big[a_1\re^{\bar\lambda_1\eta}(3+\bar\lambda_1\eta)
+b_1\re^{\bar\lambda_2\eta}(3+\bar\lambda_2\eta)\Big],
\end{equation}
which is subject to the wall boundary conditions
\begin{equation}
\breve v_4(0)=\breve v_4'(0)=0,\quad \lambda \breve v_4^{(4)}(0)=\breve c_{10}\breve v_3^{(5)}(0)+\cdots,\quad \breve v_4^{(6)}(0)+2\lambda^2\breve v_4''(0)=\ri\lambda \breve v_1^{(3)},\label{eq:BC_v3}
\end{equation}
where $\cdots$ denotes terms associating with high-order expansions of $\breve c_{1}$.
The solution can be expressed as
\begin{equation}
\begin{array}{rl}\vspace{.2cm}\displaystyle
\breve v_4=&\displaystyle{\cal V}_{40}+{\cal V}_{31}\eta+{\cal V}_{22}\frac{\eta^2}2+a_4\re^{\bar\lambda_1\eta}+b_4\re^{\bar\lambda_2\eta} \\
&\displaystyle+\ri\lambda
\Big[-{\cal V}_{10}\frac{\eta^3}{6}+\frac{a_1\re^{\bar\lambda_1\eta}}{8\bar\lambda_1^2}
(\bar\lambda_1\eta^2-3\eta)
+\frac{b_1\re^{\bar\lambda_2\eta}}{8\bar\lambda_2^2}(\bar\lambda_2\eta^2-3\eta)\Big].
\end{array}
\end{equation}
Substituting into the fourth relation in (\ref{eq:BC_v3}), and considering the solutions (\ref{eq:a1b1_BC3}), we obtain
%$$
%\ri\lambda\frac{-60(a_1\bar\lambda_1^3+b_1\bar\lambda_2^3)}{8}+2\lambda^2{\cal V}_{22}-2\ri\lambda^3\frac{44(a_1\bar\lambda_1^3+b_1\bar\lambda_2^3)}{8\bar\lambda_1^4}=
%\ri\lambda(a_1\bar\lambda_1^3+b_1\bar\lambda_2^3)
%$$
\begin{equation}
{\cal V}_{22}=\frac{\ri}{\sqrt{2}}{\cal V}_{01}.
\end{equation}
Combining with (\ref{eq:BC3_BC3}c) and (\ref{eq:asymptotic_V2}), we can derive
\begin{equation}
\breve c_{10}=-\frac{(\ri\lambda)^{1/3}\mbox{Ai}'(0)}{2\sqrt{2}\mbox{Ai}(0)}=\left\{\begin{array}{ll}
0.2812+0.1624\ri&\mbox{for }\lambda=2;\\
 0.2232 + 0.1289\ri&\mbox{for }\lambda=1.   \end{array}
\right.\label{eq:dispersion_RegimeIV_BC3}
\end{equation}

Again the positive imaginary part of $\breve c_{10}$ implies an unstable PDI. This asymptotic prediction is plotted in \ref{fig:dilute_growth-0}-(b), which aligns with the numerical curves of regime III when  $\breve\sigma$ is asymptotically large.

}

\section{Development of new conformation boundary conditions to eliminate unstable PDI}
\label{sec:BC}
\subsection{Derivation of the new conformation boundary conditions based on regime {\color{black}IV}}
\label{sec:proper_III}
{\color{black}Thanks to the previous asymptotic analyses, the original system (\ref{eq:instability}) and (\ref{eq:instability_OB}), initially characterised  by five controlling parameters, has been substantially simplified into regime IV, which features no remaining controlling parameters.} In this subsection, we aim to construct a stable asymptotic solution based on regime IV, and derive a new set of conformation boundary conditions from the near-wall behaviours of this asymptotic solution.

\subsubsection{Mathematical details}
\label{sec:ASY_sigma_infty}
{\color{black}
We first summarise the results obtained in $\S$\ref{sec:regime_IV}. For BC12, the phase speed $\breve c_1$ does not appear in the leading-order equations within the bulk sublayer, and consequently, the dispersion relation is largely determined by the wall sublayer, subject to the relation between the constants ${\cal V}_{00}$ and ${\cal V}_{02}$ imposed by the bulk-sublayer solution. In principle, many choices are possible for the set of conformation boundary conditions at the wall; however, the present set promotes an unstable PDI and is therefore unfavourable. In contrast, for BC3, the rescaled phase speed contributes to both the bulk and wall sublayers, although it emerges only at third order in the bulk sublayer. Given that both sets of boundary conditions yield unstable solutions, we now seek an alternative set of boundary conditions that ensures a negative growth rate, thus guaranteeing stability.

In fact, setting $\breve S=\ri\lambda \breve Y-\ri \breve c_{1}$ represents a more generic form. For BC1, one may assume that $\breve c_{1}$ appears with a small prefactor. Guided by this asymptotic structure, we aim to construct a set of conformation boundary conditions that ensures $\breve c_1=O(1)$ and yields a negative imaginary part, $\breve c_{1i}<0$. Considering that  the wall sublayer is not expected to generate active perturbations, we expand the phase speed according to the bulk-sublayer property, where the second-order correction scales as $O(\breve \sigma^{-1})$. Therefore, the phase speed expansion differs from the earlier expansions (\ref{eq:expansion_c1}) and (\ref{eq:expansion_c1_BC3}), expressed as}
\begin{equation}
\breve c_1=\breve c_{10}+\breve \sigma^{-1}\breve c_{11}+\cdots.
\end{equation}
Correspondingly, the perturbation in the bulk region should be expanded as
\begin{equation}
\breve V_0=\breve V_{00}+\breve\sigma^{-1}\breve V_{01}+\cdots.
\end{equation}
{\color{black}Comparing with the expansions for the BC12 and BC3, we note that the terms at $O(\breve \sigma^{-1/4},\breve \sigma^{-1/2},\breve \sigma^{-3/4})$ become trivial in the solution to be constructed.
The governing equations are
\refstepcounter{equation}
$$
{\cal L}_1\breve V_{00}=0,\quad {\cal L}_1\breve V_{01}=-\ri\breve c_{11}{\cal D}\breve V_{00},
\eqno{(\theequation a,b)}
\label{eq:bulk_sublayer_EQ}
$$
where ${\cal L}_1:=[{\cal D}^3-\ri(\lambda \breve Y-\breve c_{10}){\cal D}+\ri\lambda]$.

The leading-order solution takes the same form as  (\ref{eq:solution_ODE7}), with the definition of $\tilde \xi$ changed to $\tilde \xi=(\ri\lambda)^{1/3}\breve Y+\tilde\xi_0$, where  $\xi_0=-(\ri\lambda)^{1/3}\breve c_{10}/\lambda$.
The second-order solution can be expressed as
\begin{equation}
\breve V_{01}= \frac{-\ri\breve c_{11}}{(\ri\lambda)^{2/3}}\Big(d_{10}+d_{20}\int_\infty^{\tilde\xi}\mbox{Ai}(\tilde\xi)d\tilde\xi\Big) +d_{11}\tilde\xi+d_{21}\Big(\tilde\xi\int_\infty^{\tilde\xi}\mbox{Ai}(\tilde\xi)d\tilde\xi-\mbox{Ai}'(\tilde\xi)\Big)
\end{equation}
The lower-limit asymptotic behaviours defined in (\ref{eq:bulk_sublayer_V0}) are expressed as
\refstepcounter{equation}
$$
{\cal V}_{00}=d_{10}\tilde\xi_0+d_{20}\Big(\tilde \xi_0\int_\infty^{\tilde \xi_0}\mbox{Ai}(\tilde \xi)d\xi-\mbox{Ai}'(\tilde\xi_0)\Big),\eqno{(\theequation a)}$$
$${\cal V}_{01}=\Big(d_{10}+d_{20}\int_\infty^{\tilde \xi_0}\mbox{Ai}(\tilde \xi)d\xi\Big)(\ri\lambda)^{1/3},\quad {\cal V}_{02}=d_{20}\mbox{Ai}(\xi_0)(\ri\lambda)^{2/3}, \eqno{(\theequation b,c)}\label{eq:coefficient_V00_new}
$$
$$
{\cal V}_{10}=\frac{-\ri\breve c_{11}}{(\ri\lambda)^{2/3}}\Big(d_{10}+d_{20}\int_\infty^{\tilde\xi_0}\mbox{Ai}(\tilde\xi)d\tilde\xi\Big)+ d_{11}\tilde\xi_0+d_{21}\Big(\tilde \xi_0\int_\infty^{\tilde \xi_0}\mbox{Ai}(\tilde \xi)d\xi-\mbox{Ai}'(\tilde\xi_0)\Big), \eqno{(\theequation d)}
$$
$$
{\cal V}_{11}=\frac{-\ri\breve c_{11}d_{20}}{(\ri\lambda)^{1/3}}\mbox{Ai}(\tilde\xi_0)+ (\ri\lambda)^{1/3}\Big( d_{11}+d_{21}\int_\infty^{\tilde \xi_0}\mbox{Ai}(\tilde \xi)d\xi\Big), \eqno{(\theequation e)}$$
$${\cal V}_{12}={-\ri\breve c_{11}d_{20}}\mbox{Ai}(\tilde\xi_0)+ d_{21}\mbox{Ai}(\tilde \xi_0)(\ri\lambda)^{2/3}. \eqno{(\theequation f)}
$$

Focusing now on the wall sublayer, where $\eta=O(1)$, we adopt the same expansion for the solution $\breve V_0$ as in (\ref{eq:wall_sublayer_v0}). The leading-order solution remains identical in form to (\ref{eq:wall_v0}), while the second-order solution takes the form $\breve v_1={\cal V}_{01}\eta+a_1\re^{\bar\lambda_1\eta}+b_1\re^{\bar\lambda_2\eta}$. Our strategy is to minimise the perturbation in the wall sublayer. Being similar to the case for BC3, if we take}
\begin{equation}
d_{10}\tilde\xi_0+d_{20}\Big(\tilde \xi_0\int_\infty^{\tilde \xi_0}\mbox{Ai}(\tilde \xi)d\xi-\mbox{Ai}'(\tilde\xi_0)\Big)=0,\label{eq:condition1}
 \end{equation}
{\color{black} then the leading-order perturbation $\breve v_0$ vanishes. Next, we further seek to reduce the magnitude of $\breve v_{1}$. Notably, if we take ${\cal V}_{01}=0$ and ${\cal V}_{10}=0$, then the solution is simplified to $\breve v_1=0$. These conditions can indeed be satisfied, leading to the following required condition:}
\refstepcounter{equation}
$$
\Big(d_{10}+d_{20}\int_\infty^{\tilde \xi_0}\mbox{Ai}(\tilde \xi)d\xi\Big)(\ri\lambda)^{1/3}=0.\eqno{(\theequation)}\label{eq:condition2}
$$
Combining conditions (\ref{eq:condition1}) and (\ref{eq:condition2}), we can directly obtain the dispersion relation,
\begin{equation}
\mbox{Ai}'(\tilde\xi_0)=0.
\label{eq:dispersion}
\end{equation}
The derivative of the Airy function Ai$'$ has zero values only in the negative real axis, which corresponds to a negative value of $\tilde c_{1i}$, indicating a stable discrete mode.
The first zero point of Ai$'(\tilde\xi_0)$ appears at $\tilde \xi_0=-1.0188$. Thus, for $\lambda=2$ (the PPF base flow), the numerical solution of the rescaled phase speed is
\begin{equation}
\breve c_{10}=1.40-0.809\ri,
\label{eq:c_1}
 \end{equation}
 with the relation between $d_{10}$ and $d_{20}$ expressed as
$
 d_{20}=d_{10}\int_{\tilde\xi_0}^\infty\mbox{Ai}(\tilde\xi)d\tilde\xi
$.

{\color{black}
Solving the third- and fourth-order equations in the wall sublayer, and applying the matching condition and the no-slip conditions, we derive
\begin{equation}
\breve v_2={\cal V}_{02}\eta^2/2,\quad\breve v_3=0.
\end{equation}

The fifth- and sixth-order governing equations are
$$
(\bar{\cal D}^4+2\lambda^2)\bar{\cal D}^3\breve v_4=-\ri\breve c_{10}2\lambda^2{\cal V}_{02}\eta,\quad (\bar{\cal D}^4+2\lambda^2)\bar{\cal D}^3\breve v_5=\ri\lambda^3{\cal V}_{02}\eta^2,
$$
The solutions are
\begin{equation}
\breve v_4={\cal V}_{10}-\frac{\ri\breve c_{10}{\cal V}_{02}\eta^4}{24}+a_4\re^{\bar\lambda_1\eta}+b_4\re^{\bar\lambda_2\eta},\quad
\breve v_5={\cal V}_{11}\eta+\frac{\ri\lambda{\cal V}_{02}\eta^5}{120}+a_5\re^{\bar\lambda_1\eta}+b_5\re^{\bar\lambda_2\eta}.
\end{equation}
Applying the no-slip and non-penetration conditions at the wall, we obtain
\refstepcounter{equation}$$
{\cal V}_{10}+a_4+b_4=0,\quad \bar\lambda_1a_4+\bar\lambda_2b_4=0,\eqno{(\theequation a,b)}
$$
$$
a_5+b_5=0,\quad \bar\lambda_1a_5+\bar\lambda_2b_5+{\cal V}_{11}=0.\eqno{(\theequation c,d)}
$$

The seventh-order solution reads
\begin{equation}
\breve v_6=\frac{{\cal V}_{12}\eta^2}{2}-\frac{\ri\breve c_{10}{\cal V}_{04}\eta^6}{6!}-\frac{\ri\breve c_{10}\eta}{4}\Big(\frac{a_4\re^{\bar\lambda_1\eta}}{\lambda_1}+\frac{b_4\re^{\bar\lambda_2\eta}}{\lambda_2}\Big)+a_6\re^{\bar\lambda_1\eta}+b_6\re^{\bar\lambda_2\eta},
\end{equation}
The no-slip, non-penetration conditions determine
\refstepcounter{equation}$$
a_6+b_6=0,\quad \bar\lambda_1a_6+\bar\lambda_2b_6-\frac{\ri\breve c_{10}}{4}\Big(\frac{a_4}{\bar\lambda_1}+\frac{b_4}{\bar\lambda_2}\Big)=0.\eqno{(\theequation a,b)}
$$

The next task is to find boundary conditions from high-order derivatives of $\breve v_{4,5,6}$. A few representative quantities are listed as follows:
\refstepcounter{equation}$$
\breve v_4^{(4)}(0) =2\lambda^2{\cal V}_{10}-\ri\breve c_{10}{\cal V}_{02},\quad\breve v_5^{(4)}(0) =-2\lambda^2\Big(a_5+b_5\Big)=0,\eqno{(\theequation a,b)}
$$
$$
 \breve v_4^{(5)}(0) =-2\lambda^2(\bar\lambda_1a_4+\bar\lambda_2b_4)=0,\quad  \breve v_5^{(5)}(0) =2\lambda^2{\cal V}_{11}
+\ri\lambda{\cal V}_{02},\eqno{(\theequation c,d)}
$$
$$
\breve v_4^{(3)}(0)=-2\lambda^2\Big(\frac{a_4}{\bar\lambda_1}+\frac{b_4}{\bar\lambda_2}\Big),\quad \breve v_6^{(5)}(0) =-2\lambda^2(\bar\lambda_1a_6+\bar\lambda_2b_6)
+2\lambda^2\frac{5\ri\breve c_{10}}{4}\Big(\frac{a_4}{\bar\lambda_1}+\frac{b_4}{\bar\lambda_2}\Big),\eqno{(\theequation e,f)}\label{eq:v456}
$$
By adjusting ${\cal V}_{10}$ and ${\cal V}_{11}$, it is possible to obtain $\breve v_{4}^{(4)}(0)=\breve v_{5}^{(5)}(0)=0$. Remarkably, performing $\ri\breve c_{10}\times$(\ref{eq:v456}e)+(\ref{eq:v456}f), we can derive
\begin{equation}
\breve v_6^{(5)}(0)+\ri\breve c_{10}\breve v_4^{(3)}(0)=0.
\end{equation}
Combining with $\breve v_l^{(5)}(0)=\breve v_l^{(4)}(0)=0$  for $l\in[0,5]$, we may construct the boundary condition to the perturbation $\breve V_0$,
}
\begin{equation}
\breve V_0^{(4)}(0)=0,\quad \breve V_0^{(5)}(0)+\ri\breve c_{10}\breve V_0'''(0)=0.
\label{eq:BC_proper_V}
\end{equation}

To convert (\ref{eq:BC_proper_V}) to the conformation boundary conditions for   system (\ref{eq:equation_IV}), we expand $\breve C_{110}$ and $\breve C_{120}$ in the near-wall region. This expansion takes the same form as in (\ref{eq:near_wall_perturbation}), but with a change in the coordinate from $\breve Y$ to $\eta$.
Substituting into (\ref{eq:equation_IV}) leads to
\refstepcounter{equation}
$$\ri \breve V_0^{(3)}+\breve \sigma(\ri \bar a_0+\bar b_1)=0,\quad \bar a_1+\bar b_2=0,\quad \ri \breve V_0^{(5)}+\breve \sigma(\ri \bar a_2+\bar b_3)=0,
\eqno{(\theequation a)}
$$
$$
\bar a_2=-\ri\breve c_{10} \bar a_0-2\lambda \bar b_0,\quad \bar b_2=-\ri\breve c_{10} \bar b_0,\quad \bar b_3=-\ri\breve c_{10} \bar b_1+\ri\lambda \bar b_0.\eqno{(\theequation  b)}
$$
Using the relation (\ref{eq:BC_proper_V}), we can derive
\begin{equation}
\bar  b_0=\bar b_2=\bar a_1=0,
\label{eq:b0b2a1}
 \end{equation}
 indicating $\breve C_{110}'(0)=\breve C_{120}(0)=\breve C_{120}''(0)=0$. Therefore, the new conformation boundary conditions can be expressed as
\begin{equation}
\breve C_{110}'(0)=\breve C_{120}(0)=0,
\label{eq:BC_III_proper}
\end{equation}
which is precisely the focus of this paper.
\subsubsection{Discussion and validation}
{\color{black}The physical interpretation of boundary conditions (\ref{eq:BC_III_proper}) is that at the wall, the streamwise stretching of the polymer $\hat c_{11}$ remains with zero gradient in the wall-normal direction, while the shear deformation $\hat c_{12}$ is completely suppressed. {This is consistent with the analysis of polymer conformations in dilute solutions reported in \cite{Mavrantzas1999}, which shows that the polymer chains become flat near the wall, with negligible dimension in the perpendicular direction.}
 On the contrary, the BC12 conditions impose an abrupt change of the conformation diffusion  (from zero at the wall to a finite value in the internal flow region, as analysed in $\S$\ref{sec:regime_IV_BC12}), which drives additional instability mechanisms. Similarly, for BC3, even though the conformation tensor remains continuous in the near-wall region, the mismatch of the perturbations in the wall sublayer and those in the bulk sublayer also leads to instability, as analysed in $\S$\ref{sec:regimeIV_BC3}. In fact, from the physical point of view, we are unable to determine which set of conformation boundary conditions is more appropriate. However, as demonstrated by the asymptotic analysis in   $\S$\ref{sec:ASY_sigma_infty},    the boundary conditions (\ref{eq:BC_III_proper}) minimise the perturbation in the wall sublayer, avoiding additional instability induced from the wall, which is favourable in numerical simulations.
}

In Section 2.7 of \cite{Lewy2024}, the mechanism of PDI was discussed based on the leading-order terms  derived from the asymptotic analysis in the limit of $\beta\to 0$. Through an energy estimate, the growth rate was expressed as (equation (28) in that paper)
\begin{equation}
kc_i\overline{|\hat Q|^2}=\cdots+\frac{1}{2}\Big[\epsilon\partial_y|\hat Q|^2+\frac{\beta k^2T_{11}}{1-\beta}\partial_y|\hat\omega|^2\Big]_{-1}^1,
\label{eq:cite}
\end{equation}
where $\hat Q=[-\ri k \partial_y \hat c_{11}-(\partial_y^2+k^2)\hat c_{12}]/Wi$, $\hat\omega=\ri k\hat v-\partial_y \hat u$, $T_{11}=(C_{11}-1)/Wi$, and the terms denoted by '$\cdots$' indicate the volume terms with negative signs. To achieve an unstable PDI with a positive $c_i$, the contributions from the boundary must also be positive. If $\epsilon$ is set to be zero (indicating no ACD is introduced),  the first term of the boundary contribution vanishes. While the second term may still remain positive in this situation, the negative volume terms could overwhelm its contribution, resulting in a negative growth rate.

Although  the concentration levels assumed in \cite{Lewy2024} do not align with those in  the regime-III analysis, the growth rate estimate in (\ref{eq:cite})  still remains an insightful observation, {\color{black}because the dominant contributors in the instability equations remain the same.}
Based on the near-wall behaviour described in (\ref{eq:near_wall_perturbation}) with (\ref{eq:b0b2a1}), we can estimate that $\hat Q(0)\sim y$ in the near-wall region, indicating that $\partial_y|\hat Q|^2=0$ at $y=\pm 1$. The implication is that the perturbation field governed by the boundary conditions  (\ref{eq:BC_III_proper}) ensures that the contribution of the first term on the right-hand side of (\ref{eq:cite}) is zero, thereby highlighting the effectiveness of this set of boundary conditions in eliminating unstable PDI.

\subsubsection{Application of (\ref{eq:BC_III_proper}) to regime III}
\begin{figure}
\begin{center}
   \includegraphics[width = 0.48\textwidth]{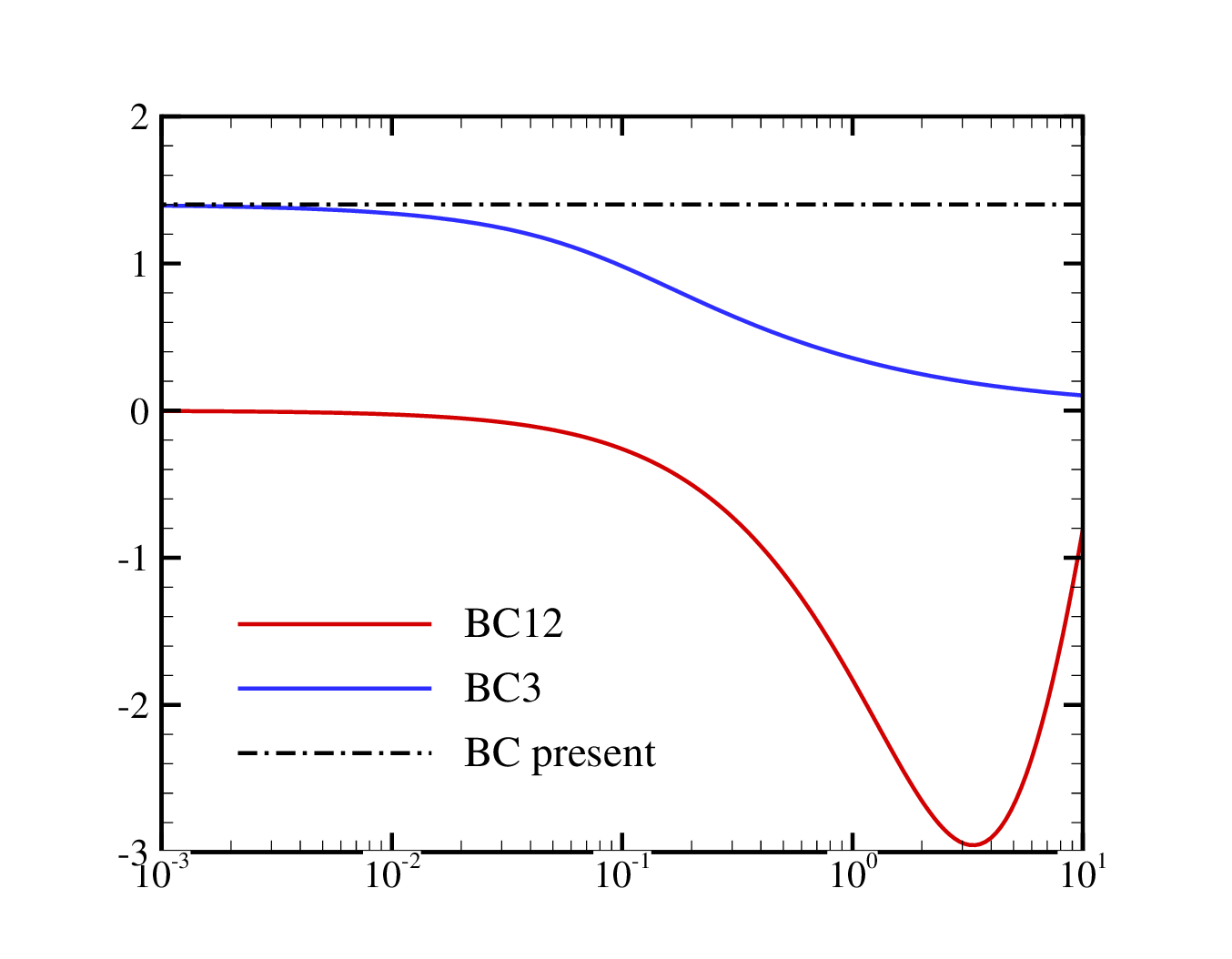}
  \put(-182,70){\rotatebox{0}{ $\breve c_{1r}$}}
  \put(-100,2){$\breve\sigma$}\put(-182,120){(a)} \includegraphics[width = 0.48\textwidth]{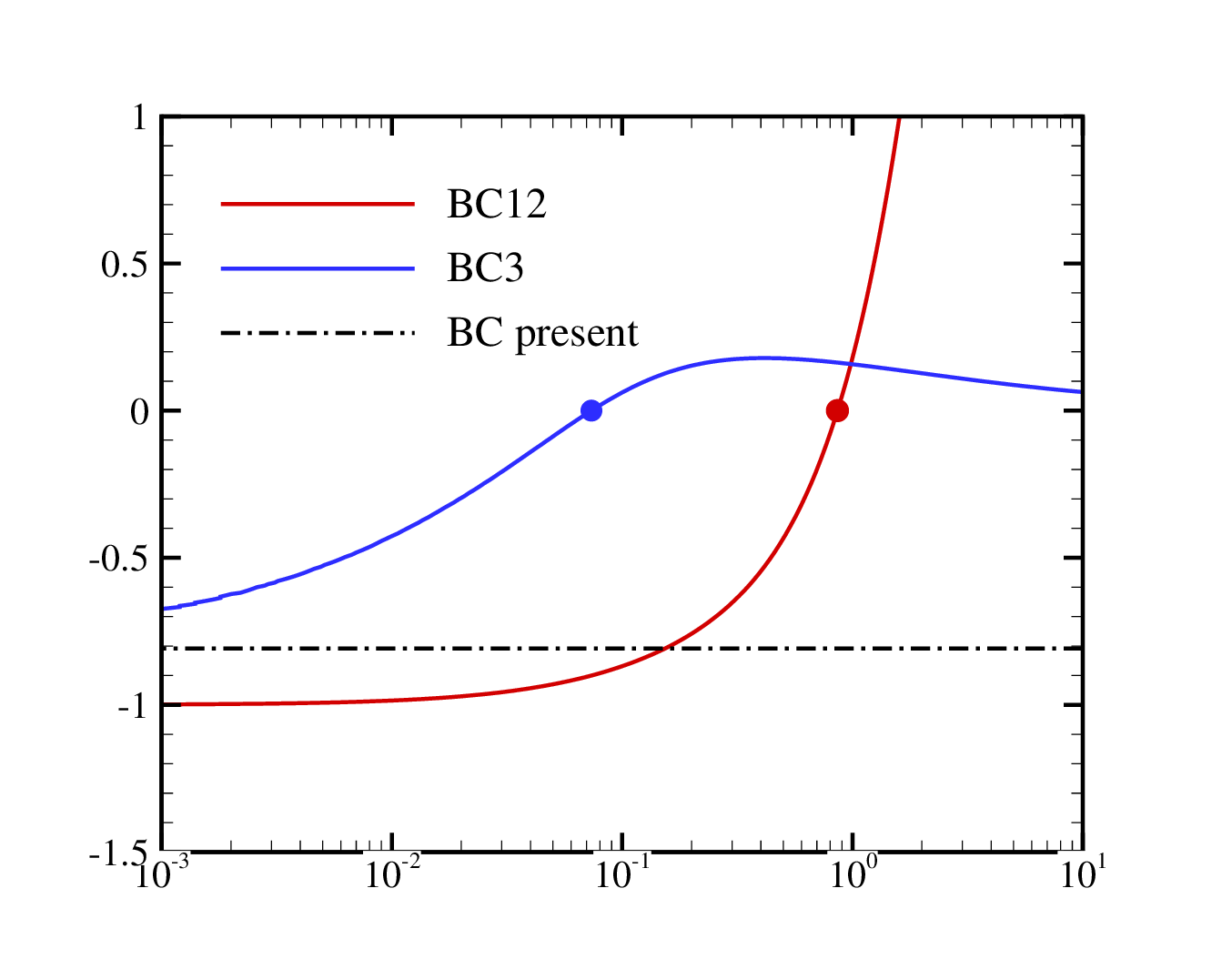}
  \put(-182,70){\rotatebox{0}{ $\breve c_{1i}$}}
  \put(-100,2){$\breve\sigma$}\put(-182,120){(b)}
    \caption{Variation with $\breve\sigma$ of the rescaled phase speed $\breve c_{1r}$ (a) and growth rate $\breve c_{1i}$ (b)  for different boundary conditions, where $\lambda=2$ (PPF). The circle in (b) marks the neutral point for BC12 and BC3. }\label{fig:iota_try1}
\end{center}
\end{figure}
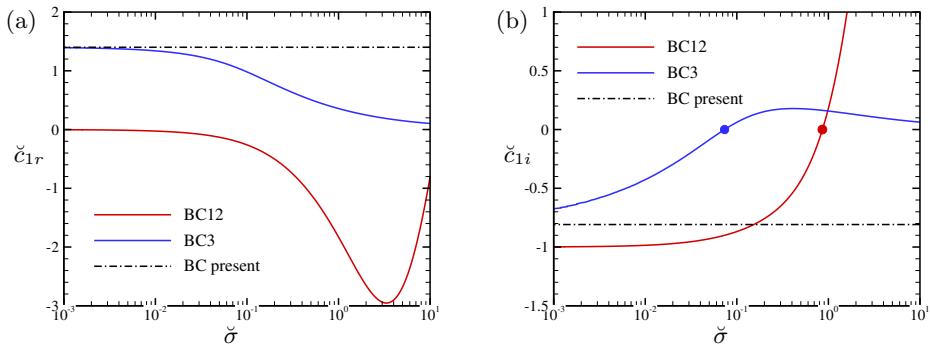

To confirm the effectiveness of the boundary conditions (\ref{eq:BC_III_proper}) for finite values of $\breve \sigma$, we perform numerical calculations of the regime-III system over the entire range of $\breve \sigma$, as shown in figure \ref{fig:iota_try1}.
The black dot-dashed lines  demonstrate the rescaled phase speed and growth rate obtained by solving (\ref{eq:equation_IV}) with the new boundary conditions (\ref{eq:BC_III_proper}). Notably, these results do not vary with $\breve\sigma$, and the values agree precisely with the analytical prediction (\ref{eq:c_1}). For comparison, we also present the curves obtained based on BC12 and BC3. It is confirmed that our new boundary conditions effectively eliminate the unstable PDI.

\subsection{Effectiveness of the  new boundary conditions in regimes I and II}
\label{sec:proper_II}
\begin{figure}
\begin{center}
   \includegraphics[width = 0.48\textwidth]{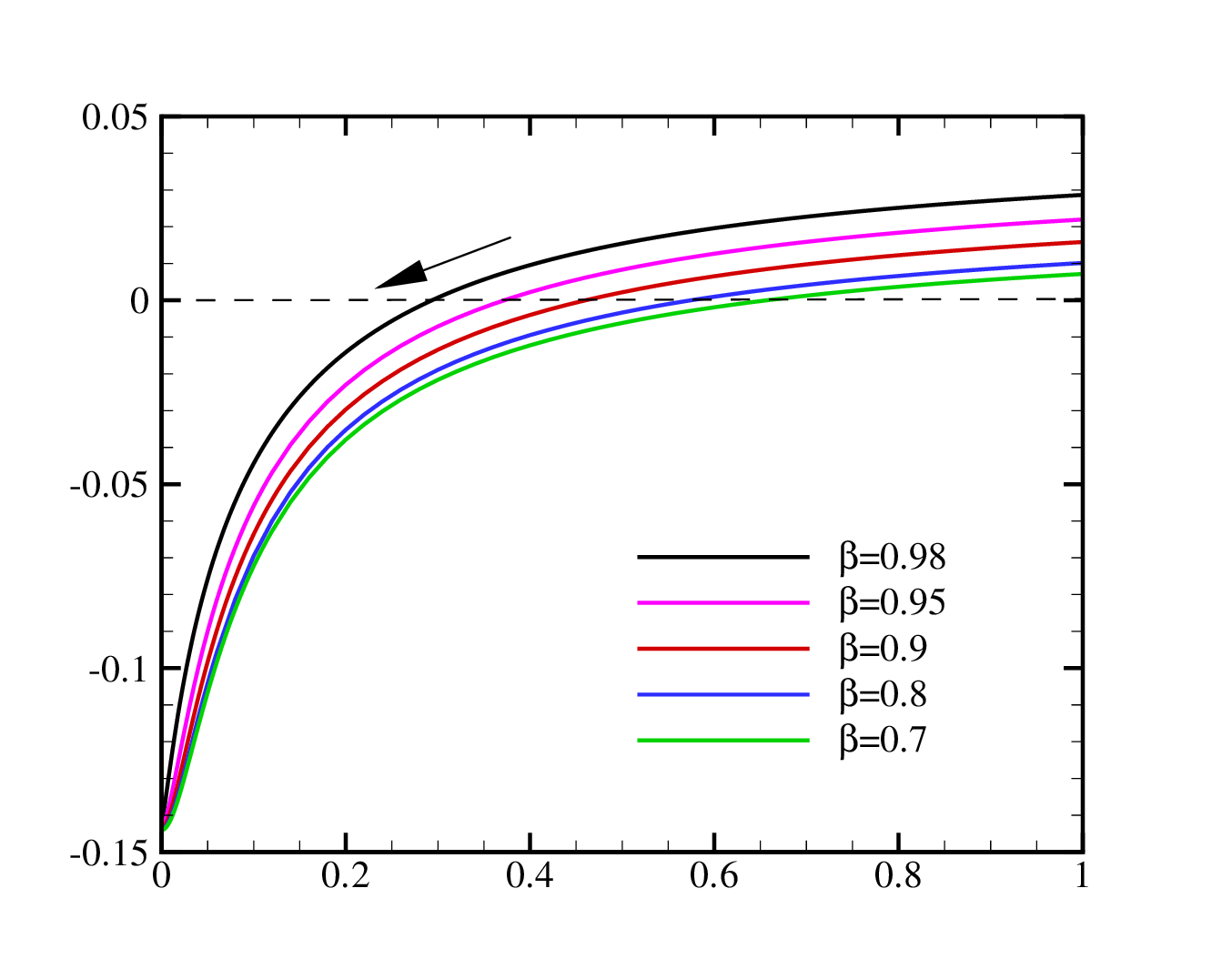}
  \put(-185,75){\rotatebox{0}{ $\tilde c_{1i}$}}\put(-140,110){Reducing $\kappa$}
  \put(-100,2){$\kappa$}\put(-185,120){(a)}
  \includegraphics[width = 0.48\textwidth]{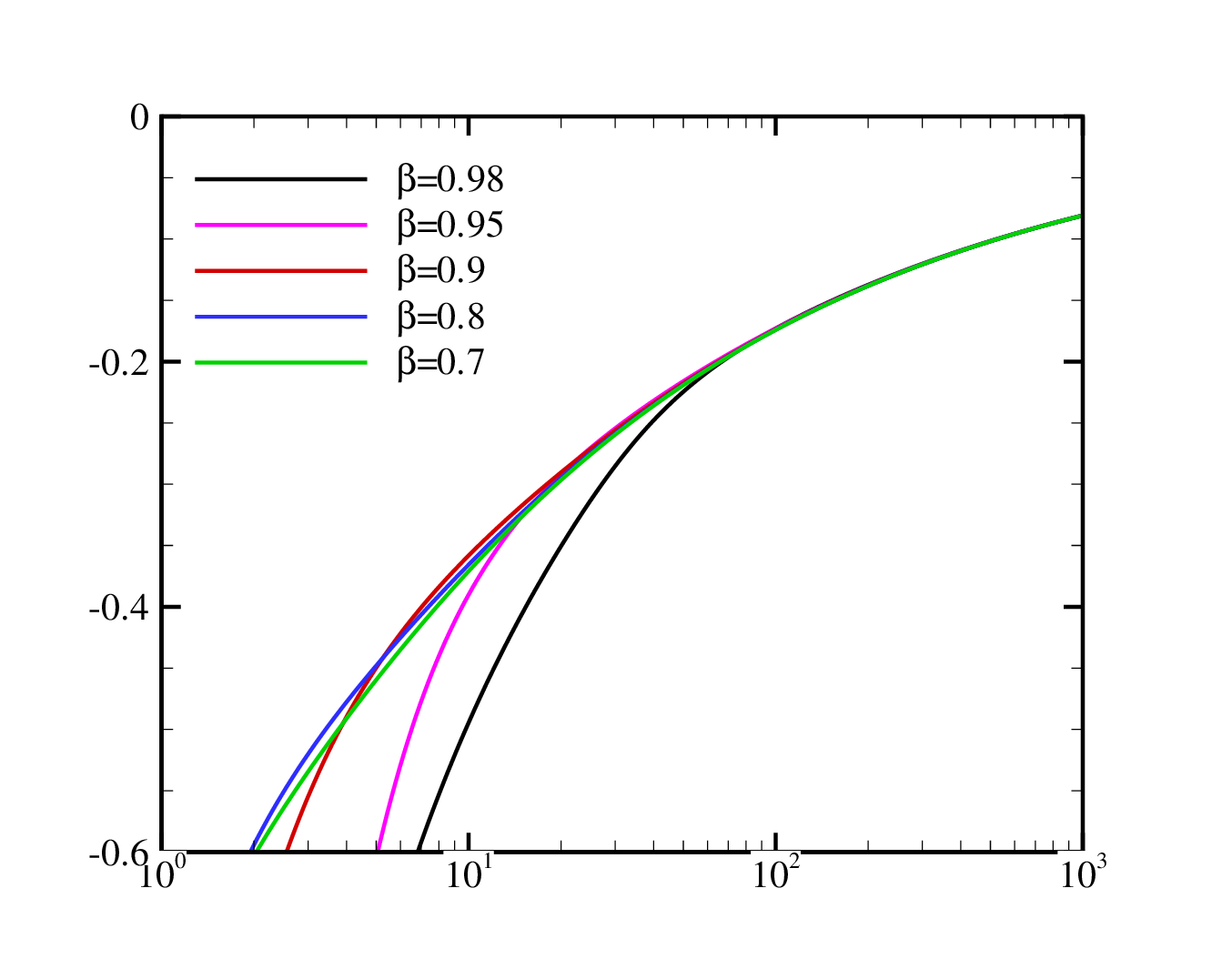}
  \put(-182,70){\rotatebox{0}{ $\tilde c_{1i}$}}
  \put(-100,2){$\tilde k$}\put(-185,120){(b)}
    \caption{(a): Variation of the rescaled growth rate $\tilde c_{1i}$ with $\kappa$ for boundary conditions (\ref{eq:adjusting_BC}), where $\tilde k=177$ and $\lambda=2$ (PPF). The horizontal dashed line represent the zero growth line. (b): Variation of $\breve c_{1i}$ with $\tilde k$ for boundary conditions (\ref{eq:proper_improve_II}).  The same $\beta$ values as in figure \ref{fig:growth_lower} are selected.  }\label{fig:proper_regime_II}
\end{center}
\end{figure}
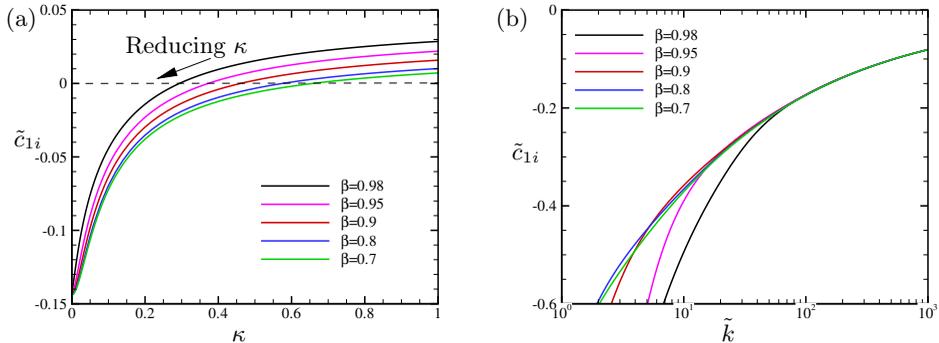

For regime II, the new boundary conditions can be readily translated to
\begin{equation}
\tilde c_{110}'(0)=\tilde c_{120}(0)=\tilde c_{220}(0)=0.
\label{eq:proper_improve_II}
\end{equation}
Therefore, the new eigenvalue system is represented by (\ref{eq:equation_II}) with boundary conditions given by (\ref{eq:BC_LW_upper}), (\ref{eq:LB_BC}a,b) and  (\ref{eq:proper_improve_II}).
 Given the existence of multiple eigenvalue solutions within this eigenvalue system, we identify the instability branch through a continuation method, which guarantees that all solutions originate from the same branch.
To achieve this goal, we express the wall boundary conditions for the conformation tensor in terms of
\begin{equation}
\tilde c_{110}'(0)=0, \quad (1-\kappa)\Big(\tilde c_{120}(0),\tilde c_{220}(0)\Big)=\kappa \Big(\tilde c_{120}'(0),\tilde c_{220}'(0)\Big),
\label{eq:adjusting_BC}
\end{equation}
where $\kappa=1$ corresponds to the boundary conditions (\ref{eq:BC_LDL_dilute_BC3}) (denoted by BC3) and $\kappa=0$ indicates the newly-developed  boundary conditions (\ref{eq:proper_improve_II}). Gradually reducing $\kappa$ from 1 to 0, we can trace the eigenvalue solutions along the same branch, as illustrated in \ref{fig:proper_regime_II}-(a). For all selected $\beta$ values, the rescaled growth rate decreases with reducing $\kappa$, and for $\kappa=0$, all the PDI modes become stable. This figure shows results for one specific value of  $\tilde k$, while  figure \ref{fig:proper_regime_II}-(b) further explore the rescaled growth rate obtained by boundary conditions (\ref{eq:proper_improve_II}) (or (\ref{eq:adjusting_BC}) with $\kappa=0$) across a wide range of $\tilde k$. It is confirmed that the unstable PDI is effectively suppressed by the  new boundary conditions throughout the entire  parameter space, and the eigensolutions are more stable for small $\tilde k$ values.

\begin{figure}
\begin{center}
   \includegraphics[width = 0.48\textwidth]{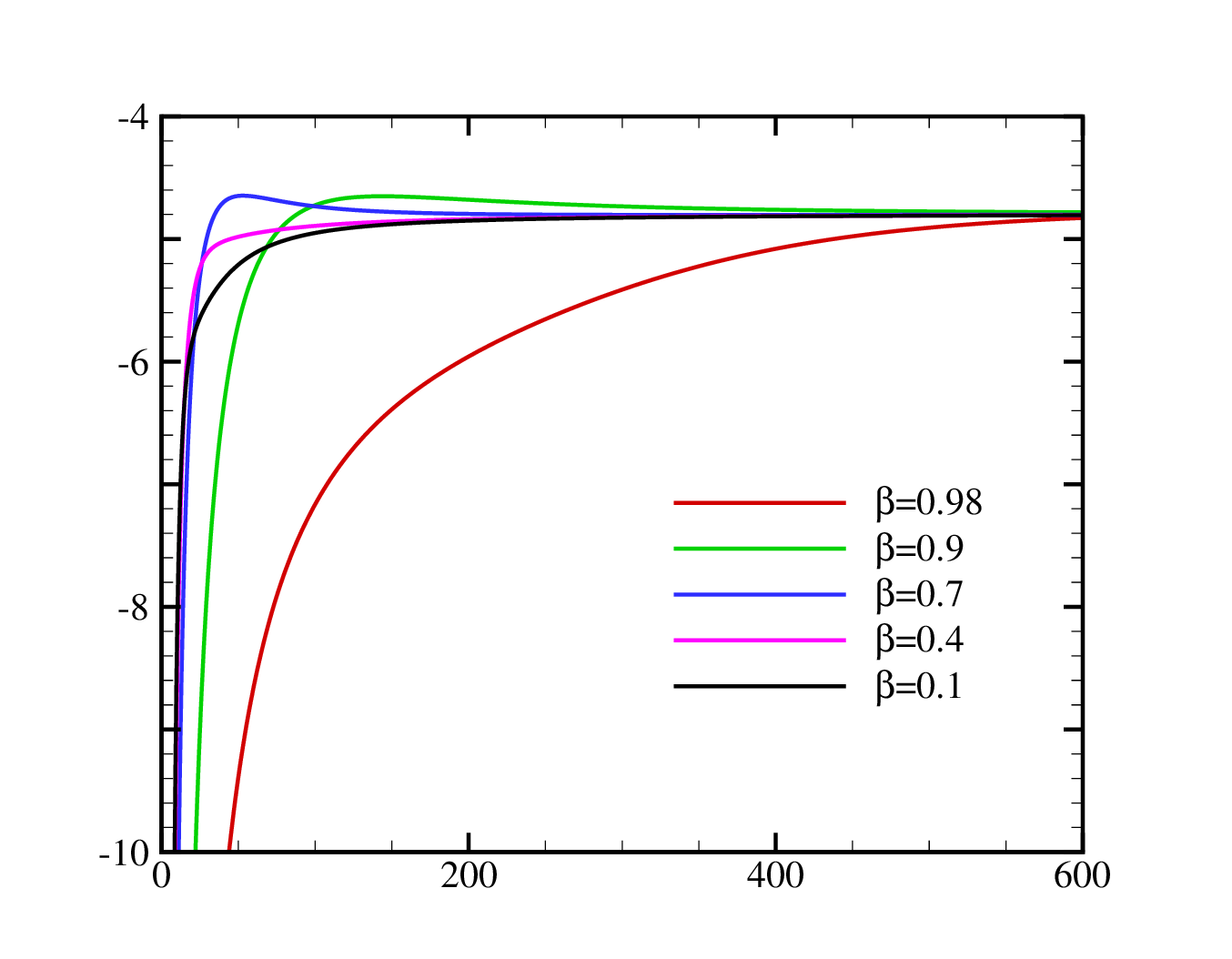}
  \put(-185,70){\rotatebox{0}{ $ c_{1i}$}}
  \put(-100,2){$Wi$}\put(-185,120){(a)} \includegraphics[width = 0.48\textwidth]{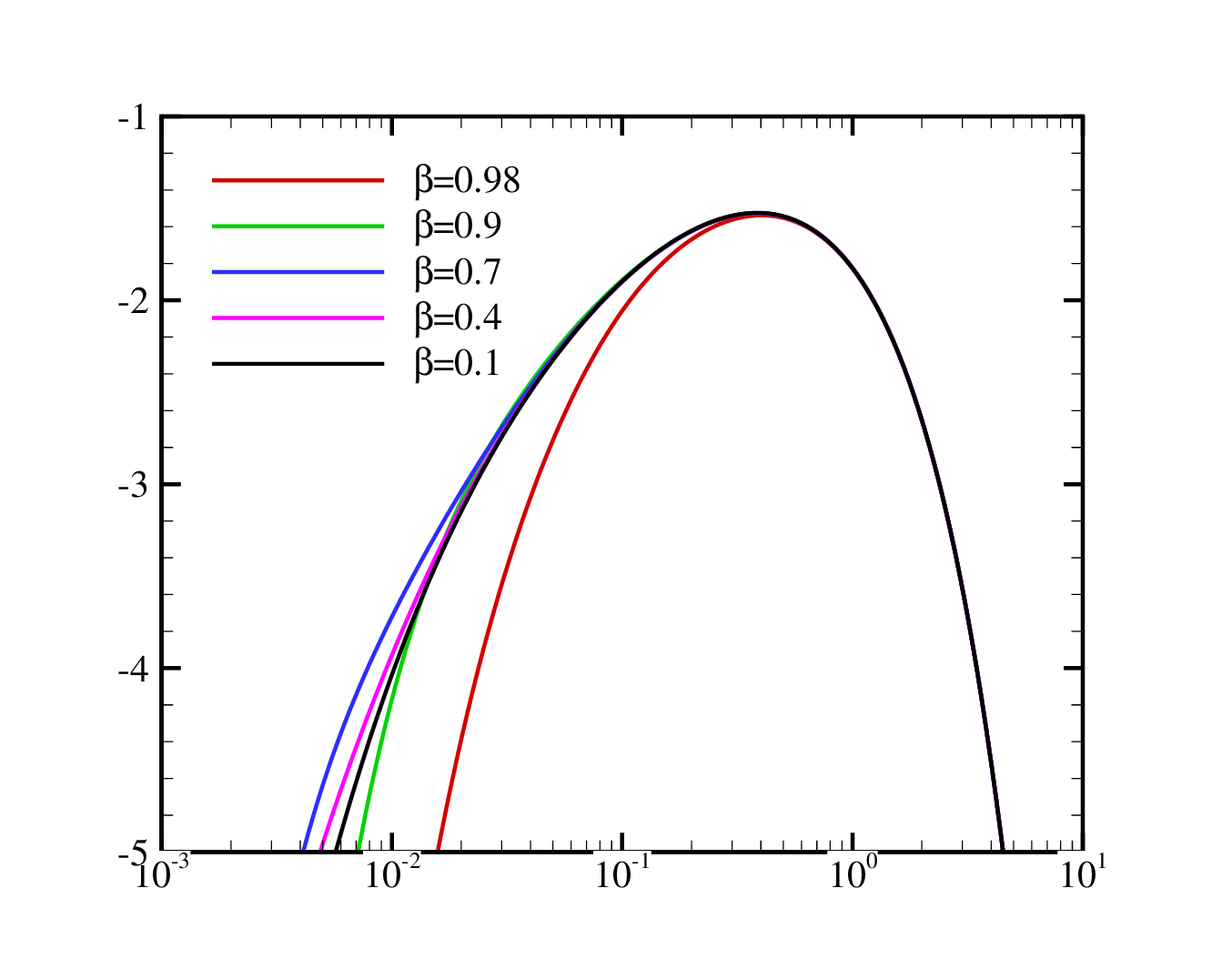}
  \put(-182,70){\rotatebox{0}{ $ c_{1i}$}}
  \put(-100,2){$\bar k$}\put(-185,120){(b)}
    \caption{The rescaled growth rate $c_{1i}$ for regime I with the new boundary conditions for various $\beta$ values, where $\lambda=2$ (PPF). (a): Dependence on   $Wi$ with $\bar k=0.00482$; (b): dependence on $\bar k$ with $Wi=40$. }\label{fig:proper_regime_I}
\end{center}
\end{figure}
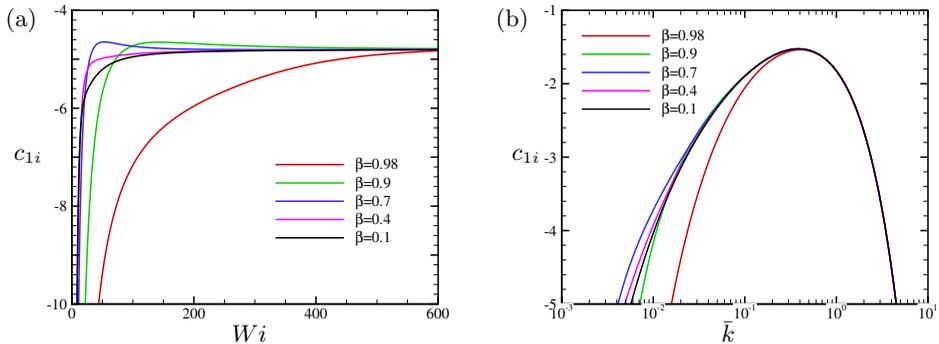

For regime I, the eigenvalue system is described by (\ref{eq:equation_I}), along with boundary conditions (\ref{eq:BC_upper}), (\ref{eq:BC_I}a,b) and the conformation boundary conditions
\begin{equation}
\hat c_{110}'(0)=\hat c_{120}(0)=\hat c_{220}(0)=0.
\end{equation}
 Applying the same continuation method as for regime II, we obtain the rescaled growth rate $c_{1i}$ for various $Wi$, $\bar k$ and $\beta$ values, as shown in figure \ref{fig:proper_regime_I}.  Once again, the eigenvalue calculations for the PPFs indicate that all modes become stable.

\subsection{Application of  the new boundary conditions for  system (\ref{eq:instability}) with Oldroyd-B model}
\label{sec:proper_original}
The validations of our new conformation boundary conditions, as outlined above, are conducted based on the asymptotic systems that consider the asymptotic structures of PDI. To confirm the robustness of this set of boundary conditions, it is essential to also examine its performance within the original system (\ref{eq:instability}), to rule out the emergence of new types of instabilities. At this stage, the eigenvalue system is represented by (\ref{eq:instability}) with boundary conditions given by (\ref{eq:BCuv}) and
\refstepcounter{equation}
$$
\hat c_{11}'(\pm 1)=0,\quad \hat c_{12}(\pm 1)=\hat c_{22}(\pm 1)=\hat c_{33}(\pm 1)=0.
\label{eq:BC_new}
\eqno{(\theequation a,b,c,d)}
$$
To obtain a broad range of the eigenspectra, the SC method is employed. In this subsection, the base flow is selected as the steady solution of the equation system (\ref{eq:GE}) with $\epsilon$ set to be zero; the impact of the ACD on the base flow will be discussed in $\S$\ref{sec:DNS}.

\begin{figure}
\begin{center}
   \includegraphics[width = 0.48\textwidth]{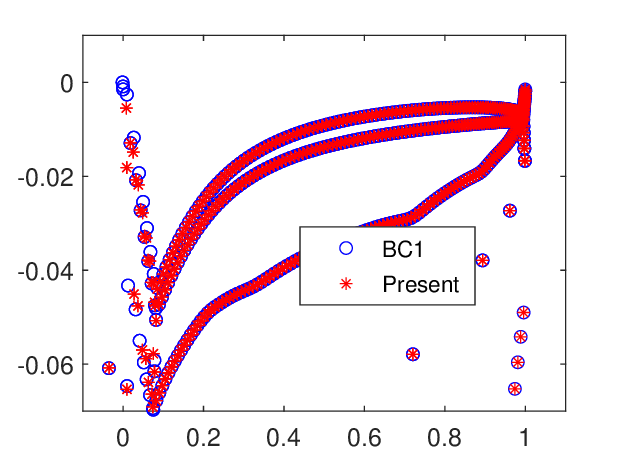}
  \put(-185,50){\rotatebox{0}{ \large$ c_{i}$}}
  \put(-90,-7){\large$c_r$}\put(-185,100){(a)}\put(-140,100){PDI}
     \includegraphics[width = 0.48\textwidth]{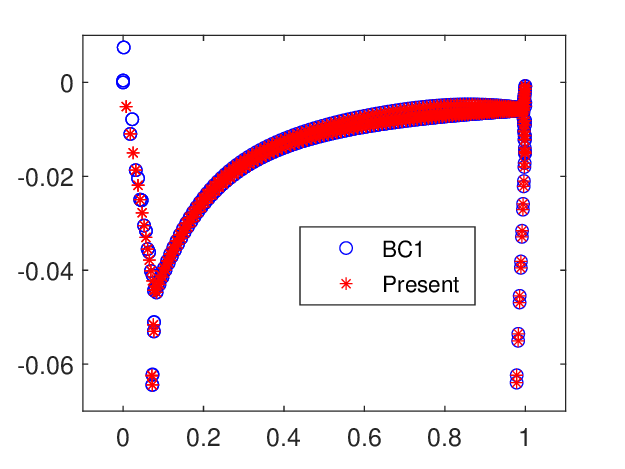}
  \put(-185,50){\rotatebox{0}{ \large$ c_{i}$}}
  \put(-90,-7){\large$c_r$}\put(-185,100){(b)}\put(-140,100){PDI}
    \caption{Comparison of the Oldroyd-B PPF eigenspectra obtained based on the boundary condition BC1 (\ref{eq:BC1}) and the present boundary condition (\ref{eq:BC_new}), where 301 collection points are employed. (a): $Re=1000$, $Wi=22.1$,  $k=52$, $\beta=0.9$ and $\epsilon=10^{-5}$; (b): $Re=1$, $Wi=2000$, $k=56$, $\beta=0.9$ and $\epsilon=10^{-5}$. The results labeled by 'BC1' and 'Present' are derived from boundary conditions (\ref{eq:BC1}) and (\ref{eq:BC_new}), respectively.
}\label{fig:QZ_Wi_2000}
\end{center}
\end{figure}

Given the five independent controlling parameters of the instability system, demonstrating the effectiveness of the boundary conditions (\ref{eq:BC_new}) across the entire parameter space is not possible. Choosing two representative parameter sets, we compare the eigenspectra obtained using both boundary conditions of BC1 (\ref{eq:BC1}) and our boundary conditions (\ref{eq:BC_new}), as illustrated by the circles and stars in figure \ref{fig:QZ_Wi_2000}, respectively. The parameters in panel (a) correspond to those in figure 1-(b) of \cite{Couchman2024}, showing good alignment in the results.  Both calculations reveal a branch of discrete modes with small $c_r$ values, marked by PDI in the figure. Notably, the least stable PDI under BC1 demonstrates a neutral nature, with  $c=c_r+\ri c_i=-0.00139+0\ri$. When the wall conformation boundary conditions are switched from BC1 to (\ref{eq:BC_new}), the PDI growth rates reduce,
 ensuring  all the discrete modes becoming stable.
 For a higher $Wi$ value, the eigenvalue calculation based on BC1 reveals an unstable discrete PDI mode with $c=0.00168+0.00744\ri$, as illustrated in panel (b). Again, our new boundary conditions effectively eliminates this unstable PDI.

 \begin{figure}
\begin{center}
   \includegraphics[width = 0.80\textwidth]{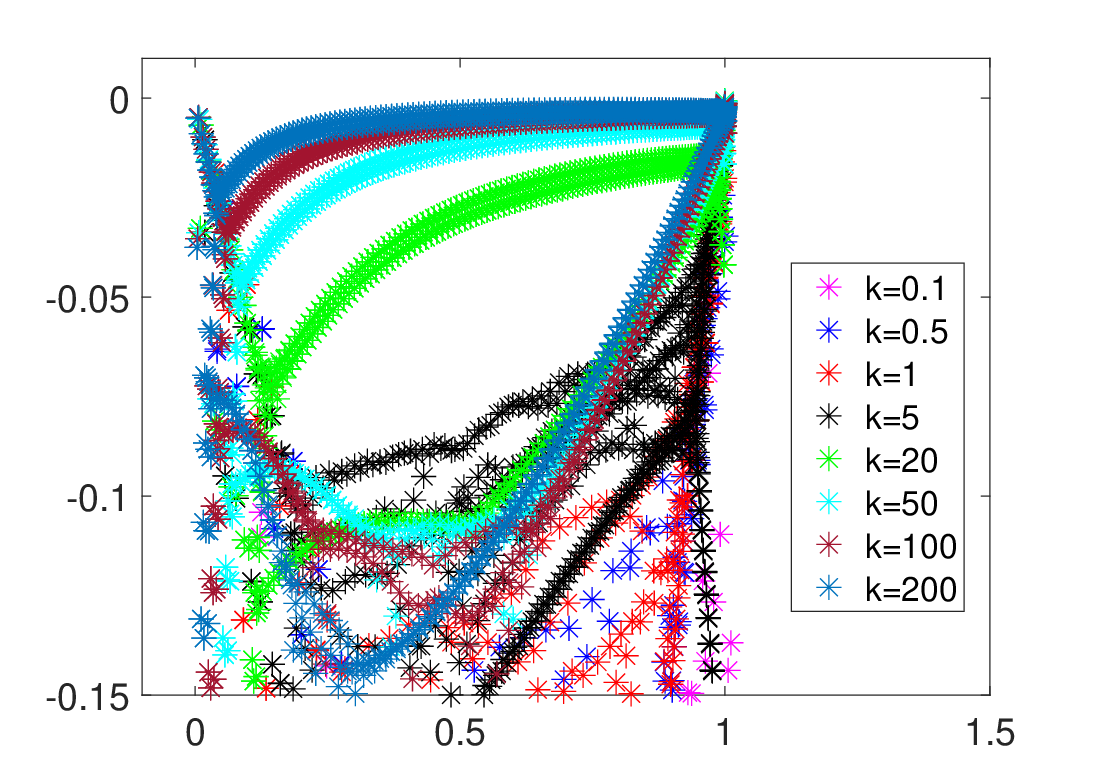}
  \put(-300,85){\rotatebox{0}{\large $ c_{i}$}}
  \put(-160,0){\large $c_r$}
    \caption{Eigenspectra of system (\ref{eq:instability}) for an Oldroyd-B PPF with various $k$, with $Re=1$, $Wi=100$,  $\beta=0.9$ and $\epsilon=10^{-5}$. For all the calculations, 301 collocation points are employed. }\label{fig:QZ_vary_k}
\end{center}
\end{figure}
Additionally, we have also selected  $\epsilon=10^{-5}$, $\beta=0.9$, $Wi=100$ and $Re=1$ for a PPF, and examined the eigenspectra for the Oldroyd-B instability system with our present boundary conditions by adjusting $k$ across values of 0.1, 0.5, 1.0, 5.0, 20, 50, 100 and 200, as shown by the stars in figure \ref{fig:QZ_vary_k}. It has been confirmed  that no unstable PDI emerges in all these calculations.
Although the validation of the effectiveness of our present boundary conditions is not exhaustive across the entire parameter space, the selected cases provide a representative span across perturbations with all possible length scales. Hence, we are confident in the efficacy of our  conformation boundary conditions in eliminating unstable PDI.
\subsection{Application of  the new boundary conditions for system (\ref{eq:instability}) with FENE-P model}
\label{sec:proper_FENE_P}
FENE-P model is another widely used constitutive model for simulating  viscoelastic fluids. In this model, the relation between the conformation stress and  tensor (\ref{eq:OB}) is modified by
\begin{equation}
\bm \tau=\frac{f(tr\bm c)\bm c-\bm I}{Wi},\quad f(s)=\Big(1-\frac{s}{L^2}\Big)^{-1},
\label{eq:FENE_P}
\end{equation}
where $L$ denotes the maximum extensibility of the polymer chains. In the limit of $L\to \infty$, the FENE-P model converges to the Oldroyd-B model. Since the asymptotic structures of the instabilities  for both constitutive models are identical, the PDI mode also appears in the FENE-P fluids \citep{Couchman2024} when the ACD is introduced. The instability system for the FENE-P fluids is described by (\ref{eq:instability}) with
\begin{equation}
(\hat \tau_{11},\hat \tau_{12},\hat \tau_{22},\hat \tau_{33})=\frac{f(tr \bm{C})}{Wi}(\hat c_{11},\hat c_{12},\hat c_{22},\hat c_{33}).
\end{equation}
At each wall, in addition to the no-slip, non-penetration conditions (\ref{eq:BCuv}) for velocity perturbations, we also require the boundary conditions for the conformation tensor. For BC1, the boundary conditions are
\refstepcounter{equation}
$$
\Big[({-\ri k c+Wi^{-1}f(tr\bm {C})})\hat c_{11}-{2(C_{12}\hat u'+U'\hat c_{12})}\Big]_{y=\pm1}=0,
\label{eq:BC1_FENE_P}\eqno{(\theequation a)}
$$
$$
 \Big[({-\ri k c+Wi^{-1}f(tr\bm {C})})\hat c_{12}-{\hat u'}\Big]_{y=\pm1}=0,\quad\hat c_{22}(\pm 1)=0.
\eqno{(\theequation b, c)}
$$
In contrast, to eliminate the PDI, it is also necessary to apply our  boundary conditions (\ref{eq:BC_new}).

\begin{figure}
\begin{center}
   \includegraphics[width = 0.96\textwidth]{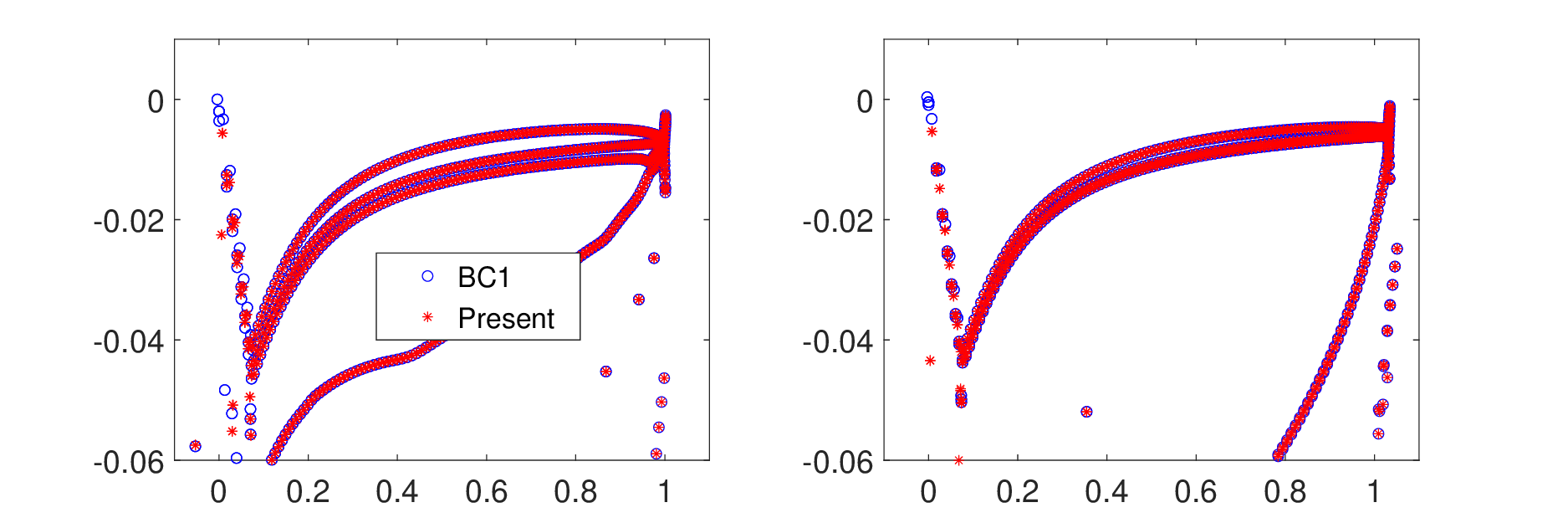}
  \put(-360,60){\rotatebox{0}{ \large$ c_{i}$}}\put(-352,100){(a)}
  \put(-103,-2){\large$c_r$}\put(-185,100){(b)}\put(-145,90){PDI}
  \put(-310,90){PDI}  \put(-270,-2){\large$c_r$}
    \caption{Comparison of the eigenspectra of FENE-P PPF instability system between BC1 (blue circles) and our boundary conditions (red stars), where $Re=1000$, $k=65$, $\beta=0.7$, $\epsilon=10^{-5}$ and $L=200$. The number of collocation points is $N=301$. (a):  $Wi=7.83$; (b): $Wi=40$.
    }\label{fig:FENE_P_1}
\end{center}
\end{figure}
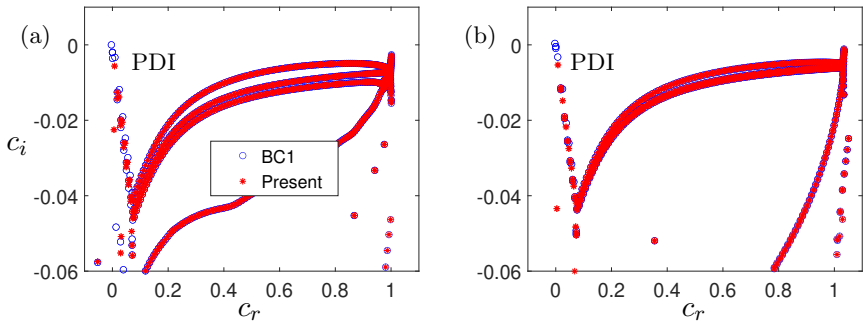

In figure \ref{fig:FENE_P_1}, we study the effect of our new boundary conditions on suppressing PDI in FENE-P fluids for two representative parameter sets with different $Wi$ values. The base flow is given by the analytical solution (2.14) of \cite{Zhang2013}. The blue circles denote the calculations based on  the traditional BC1 (\ref{eq:BC1_FENE_P}), which show a branch of discrete PDI mode  at small $c_r$ values. In panel (a), the least stable mode is neutral with $c=-0.00394+0\ri$, while in panel (b), it is unstable with $c=-0.00315+0.000386\ri$. However, when the conformation boundary conditions are replaced by  (\ref{eq:BC_new}), as shown by the red stars, the least stable PDI mode becomes stable. We have also examined the eigenvalue spectra for various values of $Wi$, $Re$ and $\beta$ under our present boundary conditions, and similar results were observed, although they are not shown here for brevity. Thus, we conclude that our present boundary conditions effectively eliminate the unstable PDI for both the Oldroyd-B and FENE-P models.

\subsection{Application in DNS}
\label{sec:DNS}
In the above calculations, the base flow was obtained by solving the equations without ACD, which can be expressed analytically. This choice is convenient for stability analysis and does not introduce significant errors, as the artificial diffusion is considered to be minimal.
However, for DNSs, we must account for the impact of ACD on the base flow, since both the base flow and the perturbations are governed by the same system of equations and boundary conditions. Given that the perturbations $(\hat c_{12},\hat c_{22},\hat c_{33})$ vanish at the wall, as indicated by (\ref{eq:BC_new}b,c,d), we can establish the boundary conditions for $(C_{12},C_{22},C_{33})$ to be close to the base-flow state in (\ref{eq:base_conformation}). A straightforward  choice is to set the boundary conditions for $(C_{12},C_{22},C_{33})$ to be the same as those obtained without ACD. For instance, using the Oldroyd-B constitutive model, we can simply set $(C_{12},C_{22},C_{33})=(Wi U',1,1)$ at $y=\pm 1$.
On the other hand, because the boundary condition for $\hat c_{11}$, as presented in (\ref{eq:BC_new}a), shows a zero gradient, we need to establish the boundary condition of $C_{11}$ in  Neumann form. A reasonable choice for this boundary condition is $C_{11}'(\pm 1)=0$. Although this does not satisfy the base flow  (\ref{eq:base_conformation}a) for a PPF, its impact on the base-flow velocity is minimal. In summary, we choose the wall boundary conditions for the Oldroyd-B mean conformation tensor as
\refstepcounter{equation}
$$
C_{11}'(\pm 1)=0,\quad C_{12}(\pm 1)=WiU'(\pm 1),\quad C_{22}(\pm 1)=C_{33}(\pm 1)=1.
\label{eq:BC_base}
\eqno{(\theequation a,b,c,d)}
$$

 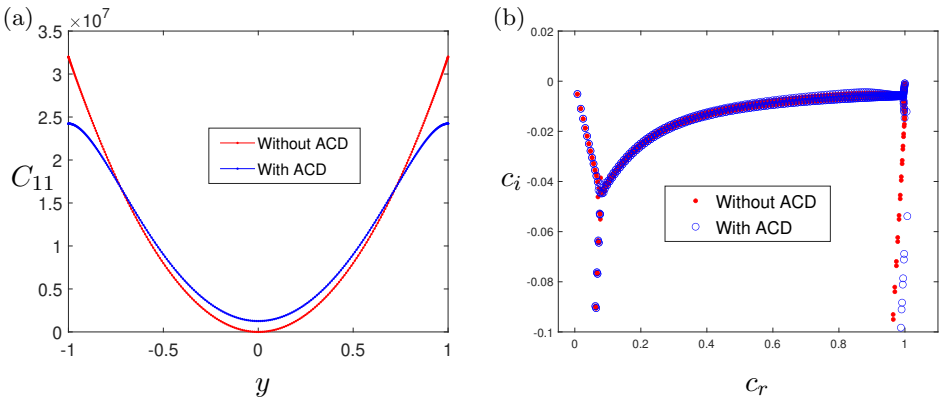
\begin{figure}
\begin{center}
   \includegraphics[width = 0.48\textwidth]{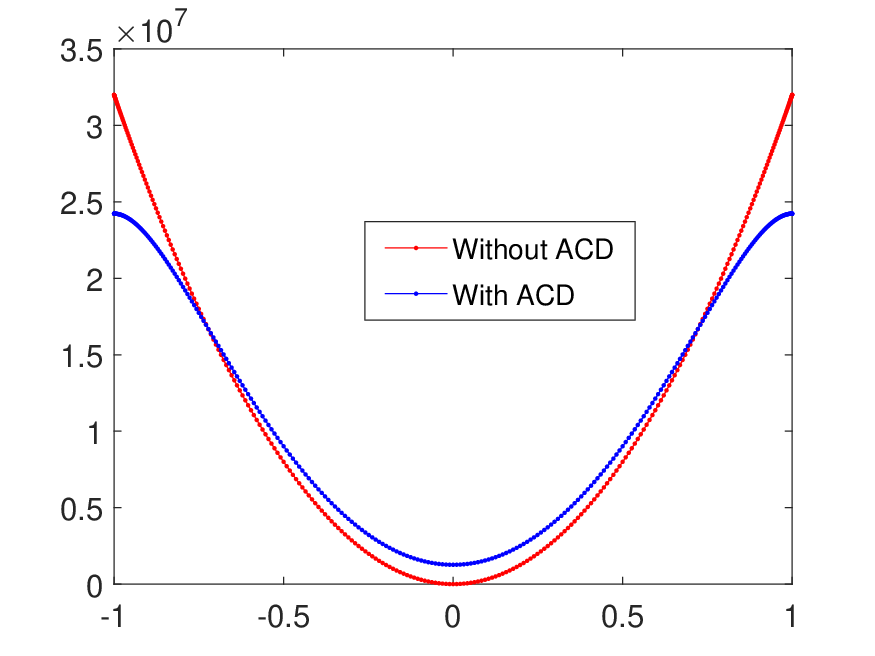}
  \put(-185,70){\rotatebox{0}{ \large$C_{11}$}}
  \put(-90,-7){\large$y$}\put(-185,130){(a)}
     \includegraphics[width = 0.48\textwidth]{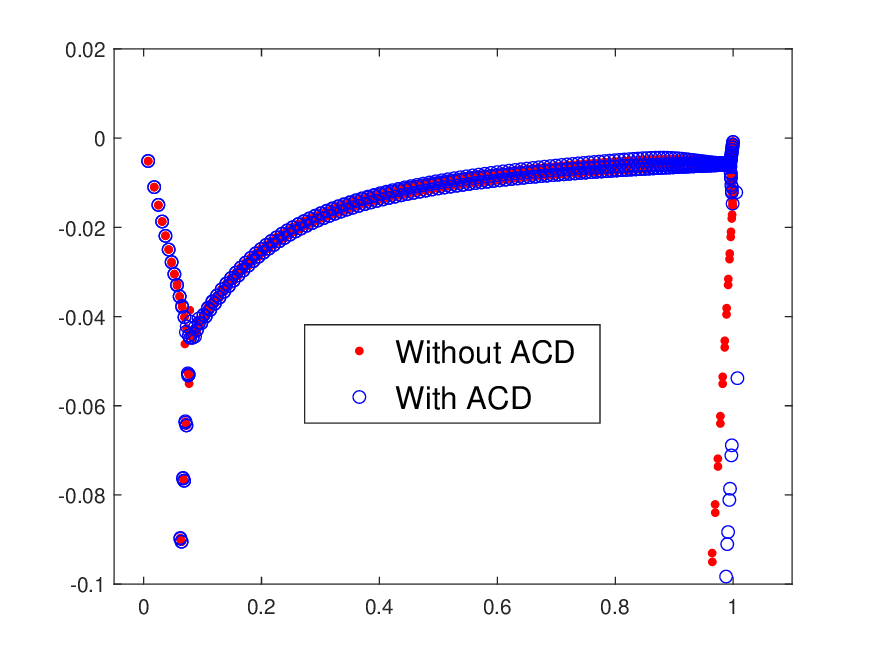}
  \put(-185,70){\rotatebox{0}{ \large$ c_{i}$}}
  \put(-90,-7){\large$c_r$}\put(-185,130){(b)}
    \caption{(a): Comparison of  $C_{11}$ of a PPF obtained from  the equations without ACD  versus those with ACD subject to boundary conditions (\ref{eq:BC_base}). (b): Comparison of the eigenspectra based on the base flows in (a), where the perturbation boundary conditions (\ref{eq:BC_new}) are applied for both calculations. The controlling parameters are     $Re=1$, $Wi=2000$, $k=56$, $\beta=0.9$ and $\epsilon=10^{-5}$. }\label{fig:base_flow_conformation}
\end{center}
\end{figure}
For the Oldroyd-B model, we compare the base flow obtained from the ACD equations using our boundary conditions (\ref{eq:BC_base}) with the analytical solutions given in (\ref{eq:base_velocity}) and (\ref{eq:base_conformation}). We find that a quantitative discrepancy arises only in the conformation component $C_{11}$, especially in the near-wall region, as displayed in figure \ref{fig:base_flow_conformation}-(a).
However, this discrepancy does not lead to noticeable differences in the eigenspectra calculations, as shown in panel (b).  This indicates that the application of the conformation boundary conditions for the base flow (\ref{eq:BC_base}) is indeed appropriate. The same test was also performed for the FENE-P model, yielding a similar conclusion.

In DNSs, {\color{black}we typically  have two options for selecting  the unknowns: (1) the instantaneous flow field $\bm\phi$, and (2) the perturbation field $\hat{\bm\phi}=\bm\phi-\bm\Phi$. For the latter, the base flow must be obtained analytically or numerically in advance, and the boundary conditions (\ref{eq:BC_new}) are applied to the perturbation equations. For the former, we denote the base-flow conformation tensor without ACD by $\bar{\bm C}$, and the boundary conditions for the total conformation tensor are
\refstepcounter{equation}
$$
\frac{\partial c_{11}}{\partial y}=0,\quad (c_{12},c_{22},c_{33})=(\bar C_{12},\bar C_{22},\bar C_{33})\quad\mbox{at }y=\pm 1.
\label{eq:BC_DNS}
\eqno{(\theequation a,b,c,d)}
$$
For instance, in the Oldroyd-B model, this set of boundary conditions is expressed as
\refstepcounter{equation}
$$
\frac{\partial c_{11}}{\partial y}=0,\quad (c_{12},c_{22},c_{33})=(Wi\frac{\partial U}{\partial y},1,1)\quad\mbox{at }y=\pm 1.
\label{eq:BC_DNS1}
\eqno{(\theequation a,b,c,d)}
$$
For three-dimensional cases, we can simply set $c_{13}=\bar C_{13}$ and $c_{23}=\bar C_{23}$ at the wall.

 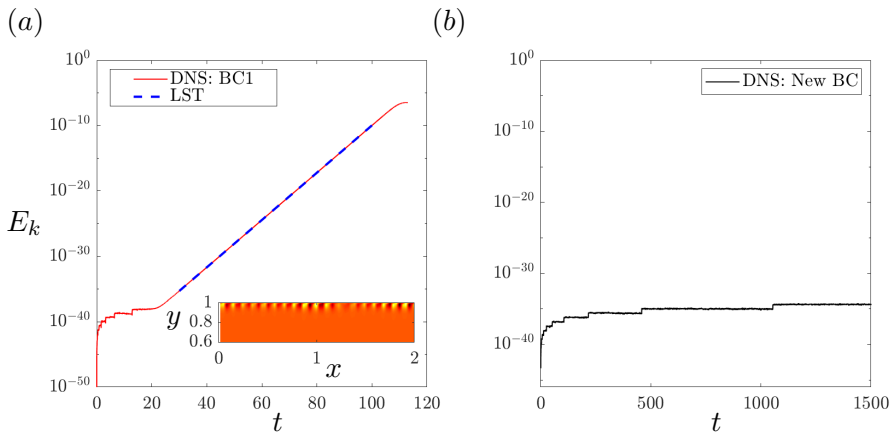
\begin{figure}
\begin{center}
   \includegraphics[width = 0.96\textwidth]{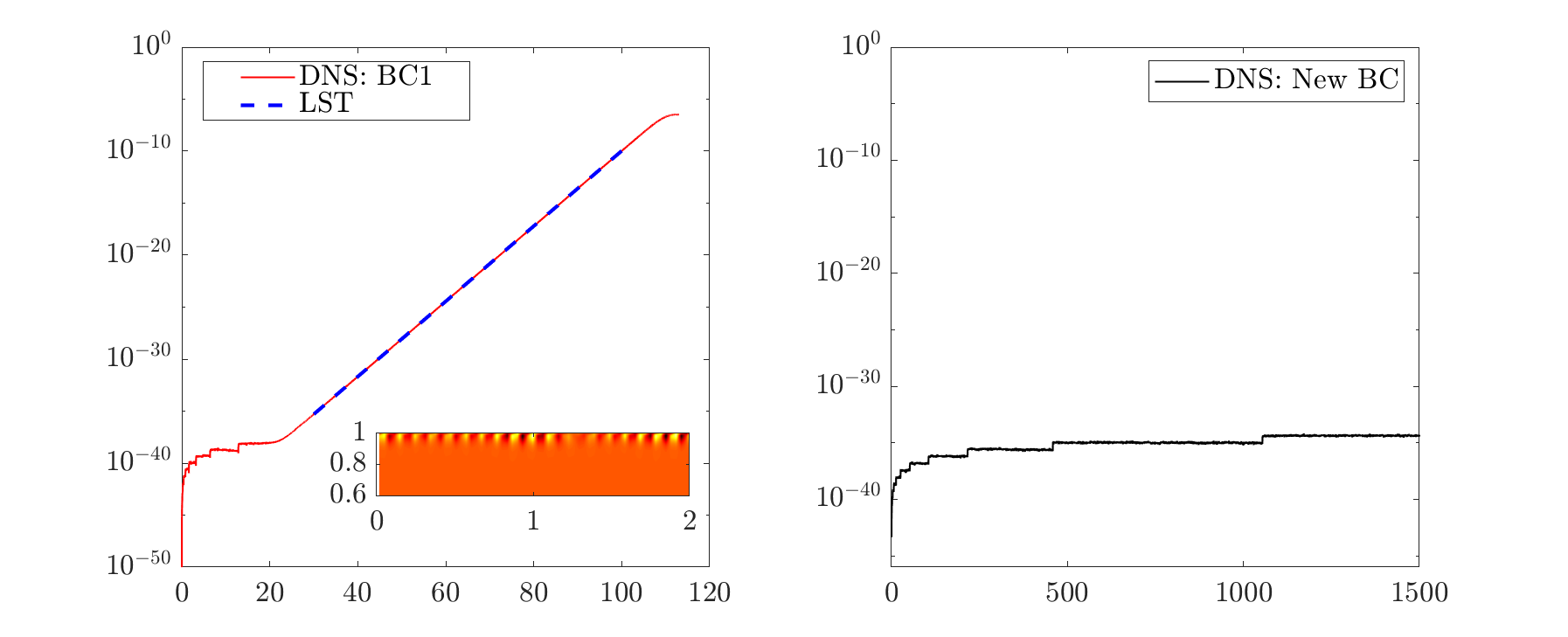}
  \put(-95,0){\large$t$}  \put(-260,0){\large$t$}\put(-360,75){\large$E_k$}
  \put(-240,20){\large$x$}\put(-300,40){\large$y$}
  \put(-360,150){\large$(a)$}\put(-200,150){\large$(b)$}
    \caption{Evolution of the perturbation kinetic energy of the Oldroyd-B PPF fluids for $k=56$ obtained by DNS, where $Re=1$, $Wi=2000$, $k=56$, $\beta=0.9$ and $\epsilon=10^{-5}$. (a): Boundary conditions BC1 by setting $\epsilon=0$ at the wall; (b): new boundary conditions (\ref{eq:BC_DNS}). The  inbox in (a) shows the contours of the perturbation wall-normal velocity of PDI at $t=100$.}\label{fig:DNS}
\end{center}
\end{figure}

We now perform  DNSs to investigate the evolution of perturbations in Oldroyd-B polymeric fluids initialised with minimal random perturbations.
{\color{black}In our DNS, the governing equations (\ref{eq:GE}) are discretised using a spectral method in the streamwise and spanwise directions, and a finite-difference method based on collocated Chebyshev-Lobatto notes in the wall-normal direction. A second-order predictor-corrector scheme based on the Crank-Nicolson method is employed for time advancing. Periodic boundary conditions are imposed in the streamwise and spanwise directions, while the no-slip, non-penetration conditions, combined with either the conventional BC1 or our new conformation conditions are employed at the wall.
Specifically, in the present two-dimensional simulations, we choose the domain length to be $L_x=2\pi$ and $L_y=2$, and $N_k=64$ Fourier modes in the streamwise direction and $N_y=400$ Chebyshev-Lobatto nodes in the wall-normal direction.  The time step is set as $\Delta_t=0.0001$. The DNS solver was previously validated and used in \cite{Sun2021Nonlinear} for simulating nonlinear perturbation evolution  in  viscoelastic pipe flows.
}

Figure \ref{fig:DNS}-(a) displays the temporal evolution of the perturbation kinetic energy $E_k$ at a specific wavenumber, $k=56$, using boundary conditions BC1, which reveals a clear exponential growth within the time interval
 $t\in[25,110]$. The growth rate  closely resembles the LST prediction, $\omega_i=kc_i=0.4166$, and the perturbation contours displayed in the inset highlight the PDI perturbation in the near-wall region, characterized by a small wavelength. In contrast, the perturbation evolution using our present boundary conditions (\ref{eq:BC_DNS}) does not exhibit any evident instability at all over a considerably extended time interval, confirming the effectiveness of our boundary conditions in suppressing the unstable PDI.

 {\color{black}
 \begin{figure}
\begin{center}
   \includegraphics[width = 0.48\textwidth]{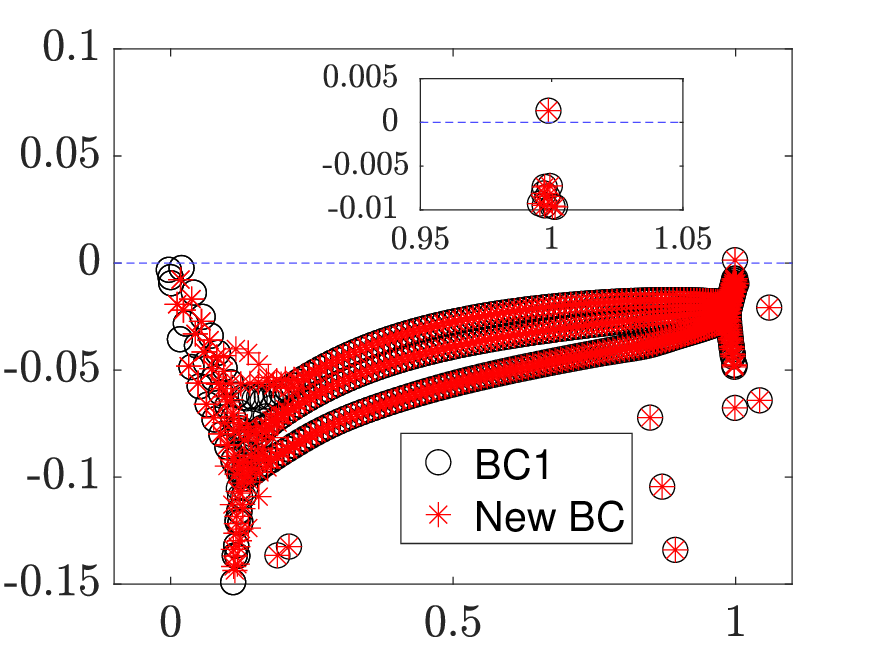}
  \put(-80,0){\large$c_r$}  \put(-195,120){\large$(a)$}\put(-185,70){\large$c_i$}\put(-90,108){\tiny$0.999+0.00134\ri$}
     \includegraphics[width = 0.48\textwidth]{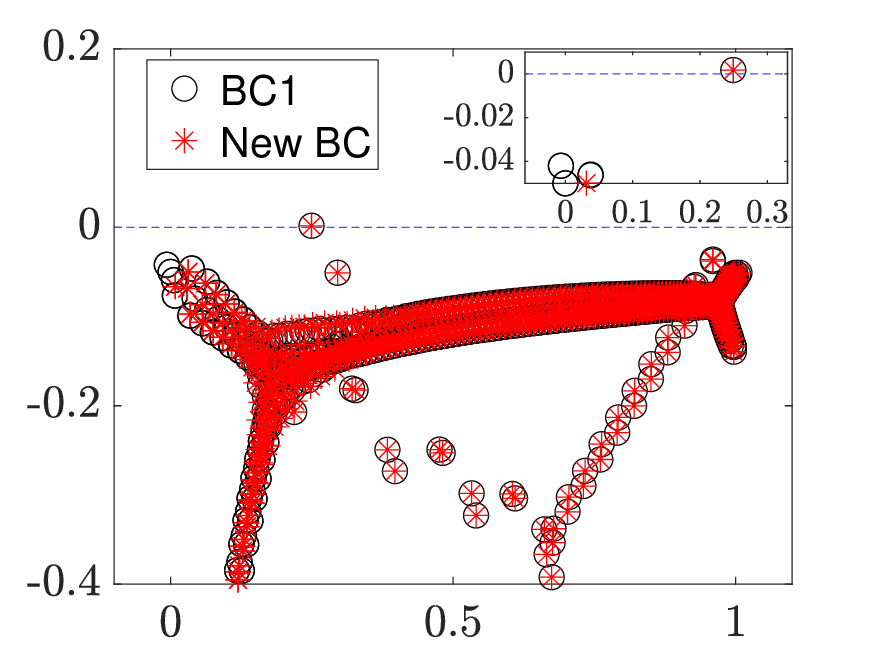}
       \put(-80,0){\large$c_r$}  \put(-195,120){\large$(b)$}\put(-185,70){\large$c_i$}\put(-70,115){\tiny$0.249+0.00173\ri$}\\
          \includegraphics[width = 0.48\textwidth]{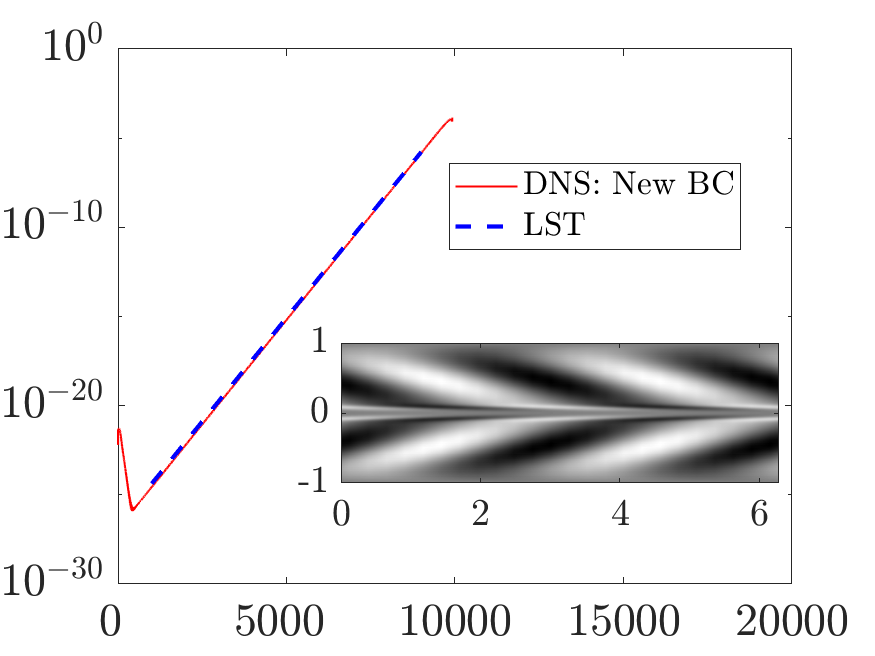}
               \put(-90,-8){\large$t$}  \put(-195,120){\large$(c)$}\put(-190,70){\large$E_k$}
     \put(-70,24){\large$x$}\put(-130,50){\large$y$}
     \includegraphics[width = 0.48\textwidth]{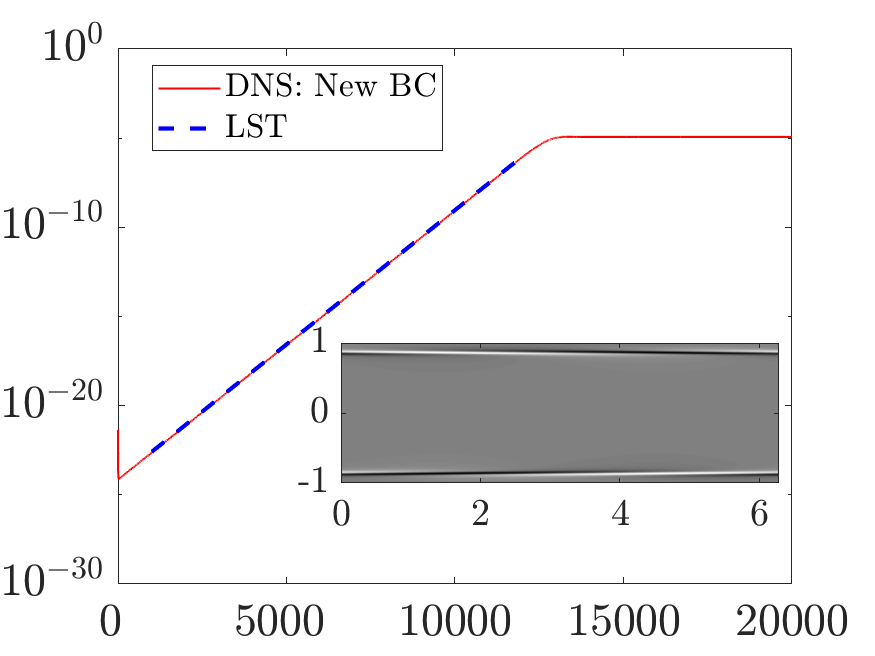}
     \put(-90,-8){\large$t$}  \put(-195,120){\large$(d)$}\put(-190,70){\large$E_k$}
     \put(-70,24){\large$x$}\put(-130,50){\large$y$}
      \caption{(a): Eigenspectra for $Re=190$, $Wi=76$, $\beta=0.9$, $k=2$ and $\epsilon=10^{-6}$. (b) Eigenspectra for $Re=8000$, $Wi=20$, $\beta=0.9$, $k=1$ and $\epsilon=10^{-6}$. The insets in (a) and (b) provide zoomed-in views of the unstable modes. (c,d): Temporal evolution of perturbation kinetic energy $E_k$ of the most unstable modes in (a,b). The insets in (c,d) present the contours of trace of the perturbation conformation tensor at $t=5000$.}\label{fig:DNS1}
\end{center}
\end{figure}
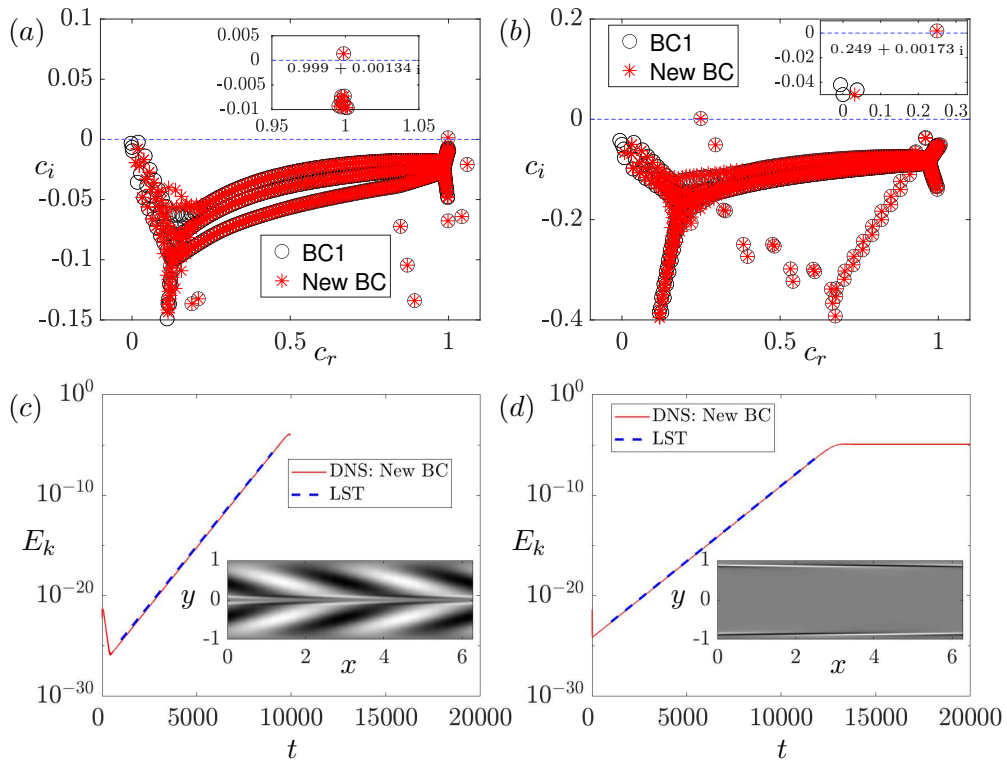
We now evaluate whether the new set of conformation boundary conditions affects the calculations of other instability modes, including centre modes \citep{Garg2018,Dong2022} and wall modes \citep{Shenkar2021}. Figure \ref{fig:DNS1}-(a)  presents the eigenspectra in the $c_r$-$c_i$ plane for $Re=190$, $Wi=76$, $\beta=0.9$, $k=2$ and $\epsilon=10^{-6}$, where an unstable centre mode, with $c=0.999+0.00134\ri$, emerges. The results obtained using both  BC1 and the new boundary conditions demonstrate excellent agreement. Having chosen these parameters, we carry out DNS based on the new boundary conditions initialized from minimal random perturbations, with the evolution of the perturbation kinetic energy shown in panel (c). After a certain period, the centre mode becomes apparent, with the energy growth  $2\omega_i$ aligning perfectly with the LST prediction. The inset displays the contours of the trace of the perturbation conformation tensor, exhibiting the characteristic arrow shape of centre modes.  Similarly, in panel (b), we plot the eigenspectra  for the parameters featuring an unstable wall mode, indicated by $c=0.249+0.00173\ri$. Using the new boundary conditions, we perform DNS for this parameter set in panel (d), observing that the exponential growth of the wall mode aligns well with the LST prediction. The perturbation field is illustrated in the inset. These results confirm that the new set of conformation boundary conditions does not impact the computations of other instability modes, and thus is expected to be an effective means for simulating transition routes in viscoelastic flows.
}
\section{Concluding remarks}
\label{sec:conclusion}
Although artificial conformation diffusion is regarded as an effective tool for suppressing numerical instability in simulations of transitional and turbulent polymer-solution flows, the recent discovery of PDI presents a significant challenge to its application in wall-shear flows. In this paper, we aim to tackle this issue by proposing a set of new boundary conditions for the conformation tensor to avoid the emergence of PDI.

To achieve this task, we first employ an asymptotic technique that reveals {\color{black}four} distinct regimes across different parameter ranges. Regime I occurs at small conformation diffusivity,
 $\epsilon\ll 1$, with the streamwise wavenumber scaling as $k\sim \epsilon^{-1/2}$. For both PPFs and PCFs, unstable PDI modes emerge when the Weissenberg number  $Wi$ exceeds a threshold that is dependent on the concentration level. As $Wi$ increases, the peak of the PDI growth rate is observed at lower $\bar k=k\epsilon^{1/2}$ values. This observation prompts us to explore Regime II, where  $k\epsilon^{1/2}\sim Wi^{-3/2}$ with $Wi\gg 1$. In this regime, the near-wall diffusive layer splits into two sublayers, with the dispersion relation determined within the lower-diffusive layer.
 Numerical solutions illustrate the emergence of unstable PDI for $\tilde k=Wi^{3/2}\epsilon^{1/2}k$ ranging from a critical value to infinity.
 Furthermore, different choices of conformation boundary conditions result in an inverse dependence of  the growth rate on polymer concentration. As a result, further investigation into PDI in the dilute limit, referred to as Regime III, becomes of particular interest. This regime specifically focuses on the scaling relation $k\epsilon^{1/2}Wi^{3/2}\sim (1-\beta)^{-3/2}$, leading to a reduction in the number of controlling parameters from five (in the original instability system) to just one ($\breve\sigma$). {\color{black}To simplify the instability system even further, we subsequently consider the limit $\breve\sigma\to \infty$, leading to a parameter-free system (referred to as regime IV) that facilitates an analytical solution of the dispersion relation.}

Then, we aim to derive the new conformation boundary conditions based on the  simplified instability system inferred from {\color{black}regime IV}.
{\color{black}Based on the concept that the perturbations in the wall sublayer should be minimised}, we construct an analytical solution with a negative growth rate.
 By examining the near-wall behaviour of this solution, we derive a set of conformation boundary conditions designed for eliminating unstable PDI. For the streamwise stretch component $\hat c_{11}$, we impose a Neumann condition $\hat c_{11}'=0$, so that near the wall $\hat c_{11}$ does not vary in the wall-normal direction. This prevents any artificial buildup or depletion of polymer stretching at the boundary and lets the bulk flow set its value. For all other conformation components ($\hat c_{12},\hat c_{22}$), we impose Direchlet conditions, which enforces  zero shear or normal stretching of the polymer in the wall-normal direction. There conditions are readily carried over to the governing systems in regimes I, II and III.

 The effectiveness of our new conformation boundary conditions can also be interpreted through the analytical model for the onset of unstable PDI developed in \cite{Lewy2024}. In this model, the PDI growth rate is primarily influenced by the shear of the conformation tensor and the shear of the vorticity at the boundaries (see (\ref{eq:cite})), in addition to negative volume contributions. Notably, the boundary conformation shear emerges as the dominant factor that ensures a positive PDI growth rate. By applying our new conformation boundary conditions, this term favorably becomes zero, indicating the elimination of any unstable PDI modes.

For representative configurations, this set of conformation boundary conditions has been validated as effective in eliminating unstable PDI within the  ACD instability system, using both the Oldroyd-B and FENE-P constitutive models.
Therefore, we can expect their effectiveness throughout the entire parameter space.
Moreover, the new conformation boundary conditions can be conveniently integrated into the DNS code with ACD, as discussed in $\S$\ref{sec:DNS}.
Comparing the DNS results based on the original boundary conditions in \cite{Beneitez2023} and our present boundary conditions, we confirm that the latter does ensure stable simulations of polymer fluids without unstable PDI. {\color{black}Remarkably, the new set of conformation boundary conditions  does not affect the calculations of other types of instabilities, including the centre and wall modes.}

It needs to be emphasised that the new set of conformation boundary conditions proposed in this paper is not the unique way to eliminate unstable PDI, and we can expect more types of effective boundary conditions. With this set of boundary conditions ensuring the absence of unstable PDI throughout the parameter space, using ACD in simulations of transitional and turbulent polymer-solution flows can be considered as a reliable and efficient approach for  future studies.

\vspace{.4cm}
\noindent\textbf{Acknowledgements}{
This research is supported by National Natural Science Foundation of China (92371104,12588201), CAS project for Young Scientists in Basic Research (YSBR-087), and Strategic Priority Research Program of Chinese Academy of Sciences (XDB0620102).
}

  \vspace{.4cm}

\noindent\textbf{Declaration of Interests}{
The authors report no conflict of interest.
}

\bibliographystyle{jfm}

\bibliography{jfm-instructions}

\end{document}